%% file: article.tex
\documentclass[DIV=calc, fleqn, 10pt, twocolumn]{scrartcl}

\usepackage{amsmath}
\usepackage{iftex}
\ifLuaTeX
  \usepackage{fontspec, luaotfload}
  
  \usepackage{unicode-math}
\else
  \usepackage[utf8]{inputenc}
  \usepackage[T1]{fontenc}
\fi
\usepackage{libertine}
\recalctypearea

\newcommand{\picsize}{0.48\textwidth}
\newcommand{\halfpicsize}{0.24\textwidth}
\newcommand{\clusterpicsize}{0.24\textwidth}

\usepackage[english]{babel}
\usepackage[babel=true]{microtype}
\usepackage[]{siunitx}
\usepackage[pdfauthor={Reint Hieronimus, Andreas Heuer}, pdftitle={How to model the interaction of charged Janus particles}, colorlinks=false, pdfborder={0 0 0}, hidelinks]{hyperref}

\usepackage[noblocks]{authblk}

\usepackage{booktabs, csquotes, fancybox, graphicx, tabularx, wrapfig}
\graphicspath{{./}}

\usepackage[font={small, sf}, format=plain, labelfont=bf, labelsep=period, skip=4pt, textformat=simple, width=\picsize]{caption}
\usepackage[margin=5pt]{subcaption}

\usepackage[backend=bibtexu, autopunct=false, sorting=none, sortcites=true, maxnames=99, url=true, doi=false, isbn=false, style=chem-angew]{biblatex}
\bibliography{article}

\usepackage{balance}

\newcommand{\eg}{e.\,g. } % Abkürzung "e. g."
\newcommand{\ie}{i.\,e. } % Abkürzung "i. e."
\newcommand{\etal}{\textit{et\,al.} } % Abkürzung "et al."
\newcommand{\D}{\,\mathrm{d}} % Differentialoperator
\newcommand{\uvec}[1]{\hat{\mathbf{#1}}} % Einheitsvektor
\renewcommand{\vec}[1]{\mathbf{#1}} % Vektoren fett statt mit Pfeil
\renewcommand{\exp}[1]{\mathrm{exp}\left(#1\right)} % e-Funktion

\ifLuaTeX
  
  \renewcommand{\textoverline}[1]{$\overline{\mbox{#1}}$}
\else
  
  \newcommand{\textoverline}[1]{$\overline{\mbox{#1}}$}
\fi

\begin{document}
% Autoren
\title{How to model the interaction of\\charged Janus particles}
\author[1,*]{Reint Hieronimus}
\author[1]{Simon Raschke}
\author[1, **]{Andreas Heuer}
\affil[1]{Westfälische Wilhelms-Universität Münster, Institut für physikalische Chemie, Corrensstraße 28/30, 48149 Münster, Germany}
\date{\vspace*{-0.4em}\small Dated: 18 July 2016}

% Abstract
\twocolumn[
  \begin{@twocolumnfalse}
    \maketitle
    \begin{abstract}
      \noindent We analyse the interaction of charged Janus particles including screening effects. The explicit interaction is mapped via a least square method on a variable number $n$ of systematically generated tensors that reflect the angular dependence of the potential. For $n = 2$ we show that the interaction is equivalent to a model previously described by Erdmann, Kröger and Hess (EKH). Interestingly, this mapping is not able to capture the subtleties of the interaction for small screening lengths. Rather, a larger number of tensors has to be used. We find that the characteristics of the Janus type interaction plays an important role for the aggregation behaviour. We obtained cluster structures up to the size of $13$ particles for $n = 2$ and $36$ and screening lengths $\kappa^{-1} = 0.1$ and $1.0$ via Monte Carlo simulations. The influence of the screening length is analysed and the structures are compared to results for an electrostatic-type potential and for multipole-expanded Derjaguin-Landau-Verwey-Overbeek (DLVO) theory. We find that a dipole-like potential (EKH or dipole DLVO approximation) is not able to sufficiently reproduce the anisotropy effects of the potential. Instead, a higher order expansion has to be used to obtain clusters structures that are identical to experimental results for up to $N = 8$ particles. The resulting minimum-energy clusters are compared to those of sticky hard sphere systems. Janus particles with a short-range screened interaction resemble sticky hard sphere clusters for all considered particle numbers, whereas for long-range screening even very small clusters are structurally different.
    \end{abstract}
    \vspace{0.9em}
  \end{@twocolumnfalse}
]

% E-Mail der Autoren
\let\oldthefootnote\thefootnote
\renewcommand{\thefootnote}{\fnsymbol{footnote}}
\footnotetext[1]{Electronic mail: r.hieronimus@uni-muenster.de}
\footnotetext[7]{Electronic mail: andheuer@uni-muenster.de}
\let\thefootnote\oldthefootnote

% Text
\section{Introduction}
Janus particles are known for more than two decades, after Veyssié and coworkers were able to prepare spherical molecules with both a hydrophilic and a hydrophobic hemisphere\autocite{Casagrande1989}. Since then, they have been of large interest due to their anisotropic character, and today not only particles with hydrophilic-hydrophobic interaction\autocite{Granick2008, Granick2011} are synthetically accessible but also magnetic\autocite{Velev2009, Sacanna2012, Granick2012b} or charged\autocite{Granick2006, Ravoo2013} particles. For a better understanding of the particle properties, effort has been put into theoretical investigations: the self-assembly behaviour and phase diagrams of hard spheres\autocite{Cacciuto2009, Sciortino2010, Sciortino2013} and soft spheres\autocite{Li2012, Hagy2012} have been studied, for example, as well as the structure formation on surfaces\autocite{Vanakaras2006, Kahl2010} and the interaction with surfaces \autocite{Klapp2011}.

\begin{figure}[h]
  \captionsetup[subfigure]{labelformat=empty}
  \centering
  \begin{subfigure}{0.24\textwidth}
    \includegraphics[width=0.9\textwidth]{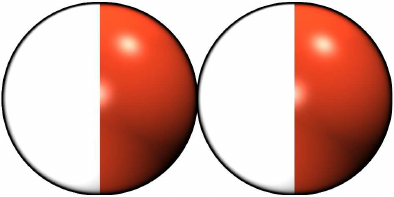}
    \caption{\textit{ns}}
    \label{fig:configurations_parallel_good}
  \end{subfigure}
  \begin{subfigure}{0.24\textwidth}
    \includegraphics[width=0.9\textwidth]{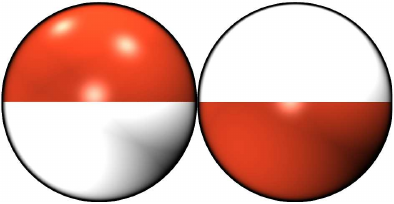}
    \caption{\textoverline{\textit{ee}}}
    \label{fig:configuration_perpendicular_good}
  \end{subfigure}
  \begin{subfigure}{0.24\textwidth}
    \includegraphics[width=0.9\textwidth]{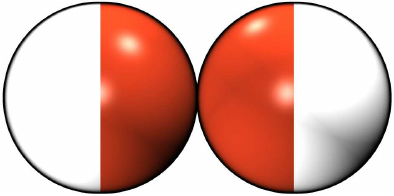}
    \caption{\textit{nn}}
    \label{fig:configurations_parallel_bad}
  \end{subfigure}
  \begin{subfigure}{0.24\textwidth}
    \includegraphics[width=0.9\textwidth]{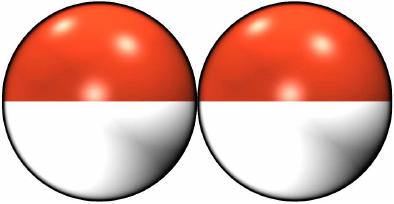}
    \caption{\textit{ee}}
    \label{fig:configuration_perpendicular_bad}
  \end{subfigure}
  \caption{Examples of configurations of two Janus particles including nomenclature. Depending on the type of interaction, like patches either attract or repel each other.}
  \label{fig:configurations}
\end{figure}

Depending on the type of interaction, like patches either attract or repel each other. For an electrostatic interaction, which in this paper we are interested in, oppositely charged patches are attractive. Basic configurations are shown in Fig.~\ref{fig:configurations}. The \textit{ns} configuration with the \enquote{north pole} of one particle pointing towards the \enquote{south pole} of another particle is the preferred configuration. By rotating the particles antiparallely towards their \enquote{equators} as in the \textoverline{\textit{ee}} configuration the interaction becomes gradually weaker. For the \textit{ee} configuration the interaction is repulsive, and finally the \textit{nn} configuration with touching \enquote{north poles} is the most repulsive. The potential of the not shown configurations \textit{ne} or \textit{se} is zero as here the interactions of the two \enquote{equators} cancel out each other.

For the simulation of Janus particles in various systems a wide number of models reproducing their anisotropic properties are known. The most straightforward approach is to model the particles as hard spheres and to use a square-well potential that is sensitive to the orientation of the patches, as introduced by Kern and Frenkel\autocite{Kern2003}. This potential can easily be implemented and is used quite often\autocite{Glotzer2005, Sciortino2009, Cacciuto2010, Sciortino2010, Sciortino2010b, Sciortino2011, Sciortino2012, Sciortino2012b, Sciortino2013}. Also variants with distance-dependent potentials, such as Lennard-Jones or Yukawa, have been employed for the simulation of hard \autocite{Goegelein2008} and soft spheres\autocite{Doye2007, Doye2007b, Doye2007c, Cacciuto2009, Doye2009, Doye2010, Kahl2010, Glotzer2014}. For all these potentials an explicit expression, albeit slightly empirical, is given for the orientation dependence.

An improved approach for charged particles is to distribute elementary charges on the patch surfaces and to calculate the electrostatic interaction of all pairs of charges. Ideally one would take an infinite number of charges into account, \ie integrate over the surfaces, but the integral is analytically not to be solved and needs to be approximated by a large number of charges to achieve sufficient accuracy. Due to the computational expensiveness this approach is only rarely used\autocite{Granick2006, Granick2008, Hagy2012}. In the reported implementations the sphere interaction and the interaction of the charges were treated independently. It was claimed that the cluster formation was independent of the range or shape of the potential, as long as the range is less than \SI{30}{\percent} of the sphere diameter, and that the clusters obtained via Monte Carlo simulation agreed with the experimental findings\autocite{Granick2006}. The finite range of the potential reflects the presence of screening effects.

A natural choice would be to express the screening effects between the elementary charges via a Yukawa potential. However, this approach implicitly assumes that the electrolyte can penetrate the particle in order to compensate the surface charge. It has been shown that on a microscopic scale the electrolyte needs to be considered and thus the arrangement of ions around a charged Janus particle\autocite{Dijkstra2010, Dijkstra2012}. In the limit of low charge densities and large interparticle distances compared to the interaction range, the Derjaguin-Landau-Verwey-Overbeek (DLVO) approximation describes the electrostatic field around a homogeneously charged particle correctly. A multipole-expanded approximation has been derived for the anisotropic Janus particles where the dipolar term is the most dominant term, overestimating the potential in axial direction and underestimating it in perpendicular direction of the orientational vector of the particle, but capturing the general physics of interacting Janus particles\autocite{Dijkstra2012}.

A class of particles related to charged Janus particles is referred to as Inverse Patchy Colloids (IPCs): charged particles that are decorated with oppositely charged patches, \ie equally charged regions are repulsive and only the interaction of particle and patch is attractive\autocite{Bianchi2011, Bianchi2015, Bianchi2015b}. While in general these particles can have an arbitrary patch size or number of patches, one-patch IPCs with a hemispherical patch again describe charged Janus particles. Recently a system using this one-patch model with an exponential screening of the potential, but without explicit charges, has been studied\autocite{Dempster2016}. The influence of the patch size on the phase diagram as well as low-temperature phases have been examined, the results allow in case of the hemispherical patch a direct comparison to our findings.

In this paper we investigate how the choice of the potential and subsequent approximations influence the aggregation behaviour of Janus particles. We start with an approach similar to Hong \etal\autocite{Granick2006} but choose a Yukawa rather than a square-well potential. The key idea is to perform a numerical multipole-expansion and to examine the dependence of the potential on the termination of this expansion. In this way we can check the importance for the structure formation of the error made due to the dipolar approximation, as used in the DLVO approach. Furthermore, we check the impact of the hard-core properties of Janus particles on the screened potential. We observe that the termination on the dipolar level has severe consequences whereas the additional accuracy of the DLVO approach has a minor effect.

The structure of the paper is as follows. In the first part we describe possible models for the interaction of charged Janus particles. An electrostatic Yukawa-based potential including screening effects due to an electrolyte is described in Sec.~\ref{sec:electrostatic_potential} and serves us as a reference. We employ a model from Erdmann, Kröger and Hess described in Sec.~\ref{sec:EKH_potential} to introduce a fit potential in Sec.~\ref{sec:fit_potential}. It uses a set of basis tensors that capture the anisotropic properties of Janus particles. By mapping it to the electrostatic potential we can profit from high accuracy at low computational costs. In the following Sec.~\ref{sec:DLVO_potential} we compare the potential derived for DLVO theory to the tensor model. In the second part of the paper we discuss the structure formation with respect to the previously described models. We use Monte Carlo simulations together with the parallel tempering technique to obtain minimum potential energy structures. The resulting structures are studied in Sec.~\ref{sec:cluster_formation} to discuss the effects of the electrostatic and the DLVO potential. Furthermore, we analyse the influence of the screening length. The structures are also compared to sticky hard sphere clusters and aggregates of Janus-like IPCs. Finally, we conclude our results in Sec.~\ref{sec:conclusion}.

\section{Electrostatic potential}
\label{sec:electrostatic_potential}
Due to the symmetry of Janus spheres, the configurational space of two particles is described by only four parameters giving the position and orientation of the second particle relative to the first one: the distance $r$ between the centres of mass of the particles and the Euler angles $\vartheta_1$ for the position and $\vartheta_2$ and $\varphi_2$ for the orientation. This configurational space is, for our purposes, discretised to a grid with $K$ points in each angular dimension and a smaller number of points in the spatial dimension $r$. The energies of all these points need to be determined so that the tensor-based model can be mapped to them. The Janus spheres are considered in this paper to be of diameter $\sigma = \num{1}$ which is equivalent to the minimum distance between particles.

The interaction potential of two Janus particles
\begin{equation}
  u_\mathrm{ES} = u_\mathrm{iso}(r) + u_\mathrm{aniso}(\uvec{n}_k, \uvec{n}_l, \vec{r})
\end{equation}
with the interparticle vector $\vec{r}$ and the normal vectors $\uvec{n}_{k,l}$ of the patches $k$ and $l$ of the two particles can be split into an isotropic part $u_\mathrm{iso}$ for the cores and an anisotropic part $u_\mathrm{aniso}$ for the patches. For the former we use a hard sphere term
\begin{equation}
  u_\mathrm{HS}(r) = \begin{cases}
                       \infty &\text{if $r < \sigma$}\\
                       0      &\text{if $\sigma \leq r$}
                     \end{cases}
\end{equation}
where $\sigma$ is the particle diameter and $r$ the interparticle distance, for the latter we consider the patch surfaces uniformly covered by point charges. Each pair of point charges of different Janus particles is interacting via the Yukawa potential
\begin{equation}
\label{eq:potential_yukawa}
  u_\mathrm{Yu}(r_{kl}) = \frac{q_k q_l}{4\pi\epsilon r_{kl}} \, \mathrm{exp}\left(-\kappa r_{kl}\right)
\end{equation}
with the screening length $\kappa^{-1}$ for the interaction of the charges $q_{k,l}$ which are located at $\vec{s}_k$ and $\vec{s}_l$ separated by the distance $r_{kl} = \left| \vec{r} + \vec{s}_l - \vec{s}_k \right|$. Expressing the point charges by local densities $q(s_k)$ and $q(s_l)$ respectively, one can write the total interaction for every pair of patches after integration over both patch surfaces as
\begin{equation}
  u_\mathrm{Yu}(\uvec{n}_k, \uvec{n}_l, \vec{r}) = \iint q(s_k) q(s_l) \, u_\mathrm{Yu}(r_{kl}) \D \vec{s}_k \D \vec{s}_l \,.
\end{equation}
The integration is carried out numerically using the trapezoidal rule and with a sufficiently large number $L$ of subintervals in $\varphi$-direction, while in \enquote{longitudinal} $\vartheta$-direction the number is scaled by $\sin{\vartheta}$ to achieve a uniform coverage. The actual total number of charges per patch, as shown in Table~\ref{tab:number_charges}, is approximately given by $\num{0.64} L^2$.

As Janus particles have two hemispheres, summation over all four possible combinations of pairs of patches gives the total potential $u_\mathrm{ES}$ for each configuration of two particles.

\begin{table}
  \centering
  \caption{Actual number of charges on the surface of a patch corresponding to the parameter $L$}
  \label{tab:number_charges}
  \begin{tabular}{c | S[table-format=2.0] S[table-format=3.0] S[table-format=3.0] S[table-format=4.0] S[table-format=4.0] S[table-format=4.0] S[table-format=5.0]}
  \toprule
  $L$     & 10 &  20 &  35 &   50 &  75  &  100 &   150\\
  charges & 68 & 263 & 795 & 1614 & 3619 & 6417 & 14399\\
  \bottomrule
  \end{tabular}
\end{table}

\begin{figure}
  \centering
  \resizebox{\picsize}{!}{\large \input{convergence_integration.tex}}
  \caption{Pearson correlation coefficient of the electrostatic potential $u_\mathrm{ES}$ for a variable value of $L$, displayed at the x-axis, and the value of $L = 150$. Here $L$ is the parameter controlling the number of charges per patch. Notice that the correlation is better for larger $r$ as the values of the energy are lower due to the screening. The energies were determined with the screening length $\kappa^{-1} = 0.1$ and with $K = 10$ subintervals in the angular dimensions.}
  \label{fig:convergence_integration}
\end{figure}
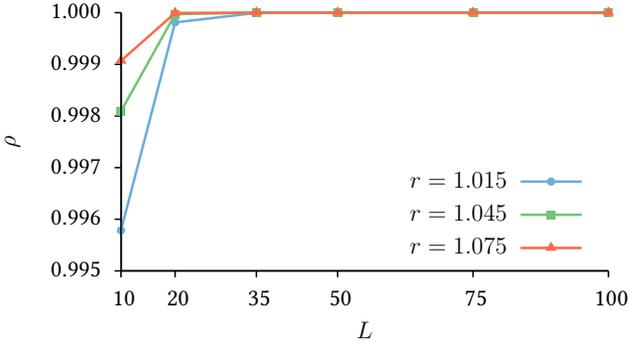

The accuracy of the resulting potential energy landscape depends on the number $K^3$ of grid points and the number $\num{0.64} L^2$ of charges per patch. To find a reasonable compromise between accuracy of the integration and computational time, we calculated the energies for $L = \num{10}, \num{20}, \num{35}, \num{50}, \num{75}, \num{100}$ and checked via the Pearson coefficient
\begin{equation}
  \rho = \frac{\sum \left(u_\mathrm{1} - \langle u_\mathrm{1}\rangle\right) \left(u_\mathrm{2} - \langle u_\mathrm{2}\rangle\right)} {\sqrt{\sum\left(u_\mathrm{1} - \langle u_\mathrm{1}\rangle\right)^2} \sqrt{\sum\left(u_\mathrm{2} - \langle u_\mathrm{2}\rangle\right)^2}}
\end{equation}
the correlation between these energies and those for $L = \num{150}$. Here the sum is over all grid points. It shows that $L = \num{35}$, \ie $\num{795}$ charges per patch, is already large enough to have a basically perfect determination of the integral (Fig.~\ref{fig:convergence_integration}). For soft Janus spheres with Coulomb interactions between the patches, a number of about $\num{500}$ point charges per patch is found as the minimum value\autocite{Hagy2012}, corroborating our findings.

\section{EKH potential}
\label{sec:EKH_potential}
In what follows, the orientation of two interacting Janus particles are characterized by the two unit vectors $\uvec{n}_{1,2}$ pointing from the south pole to the north pole. The model described by Erdmann, Kröger and Hess (EKH) makes use of spherical harmonic tensors to capture the angular dependence of the patch interaction\autocite{Hess2003}. These tensors need to fulfil the symmetry conditions
\begin{equation}
\label{eq:tensor_conditions}
\begin{split}
  u(\uvec{n}_1, \uvec{n}_2, \vec{r}) &= -u(-\uvec{n}_1, \uvec{n}_2, \vec{r})\\
  u(\uvec{n}_1, \uvec{n}_2, \vec{r}) &= -u(\uvec{n}_1, -\uvec{n}_2, \vec{r})\\
  u(\uvec{n}_1, \uvec{n}_2, \vec{r}) &=  u(\uvec{n}_2, \uvec{n}_1, -\vec{r})
\end{split}
\end{equation}
in order to reflect the Janus geometry. The first and second condition imply that the energy changes its sign if a particle is flipped, the third means that the energy keeps its sign if both particles swap their positions and orientations. As a side effect, the orientational average vanishes for the anisotropic part, \ie $\iint \, u_\mathrm{aniso} \D \uvec{n}_1 \D \uvec{n}_2 = 0$. For Janus particles, second rank tensors
\begin{equation}
\label{eq:tensors_EKH}
\begin{split}
  \psi_1 &= (\uvec{n}_1 \cdot \uvec{r})(\uvec{n}_2 \cdot \uvec{r}) - \frac{\uvec{n}_1 \cdot \uvec{n}_2}{3}\\
  \psi_2 &= \uvec{n}_1 \cdot \uvec{n}_2
\end{split}
\end{equation}
are used by Erdmann, Kröger and Hess to define the scalar anisotropy function
\begin{equation}
  \psi_\mathrm{EKH}(\uvec{n}_1, \uvec{n}_2, \vec{r}) = a_1 \psi_1 + a_2 \psi_2
\end{equation}
via linear superposition. The tensors, coming from the $S$ functions derived by Stone\autocite{Stone1978}, are restricted to those compatible with conditions \eqref{eq:tensor_conditions}. The angular dependence of $\psi_1$ corresponds to that of a dipole-dipole interaction and $\psi_2$ acts as a perturbation that determines whether the equatorial or the polar configurations from Fig.~\ref{fig:configurations} are preferred. An attractive interaction of like patches, as for hydrophilic-hydrophobic interaction, is indicated by positive signs of the coefficients $a_i$, while negative signs indicate repulsive interaction as for charged patches.

As we want to model the previously described electrostatic potential, we mapped the anisotropy function to the values of the electrostatic energies for every point in the discretised configurational space while using the hard sphere potential for the isotropic part. The distance dependence incorporated in the Yukawa potential was fitted with the function
\begin{equation}
\label{eq:distance_fit}
  a(r) = \frac{\exp{-\kappa r}}{r^3} \left(a_0 + a_1 r + a_2 r^2\right)
\end{equation}
to the prefactors of the tensors. The Levenberg-Marquardt algorithm was implemented to solve the nonlinear least-square fits.

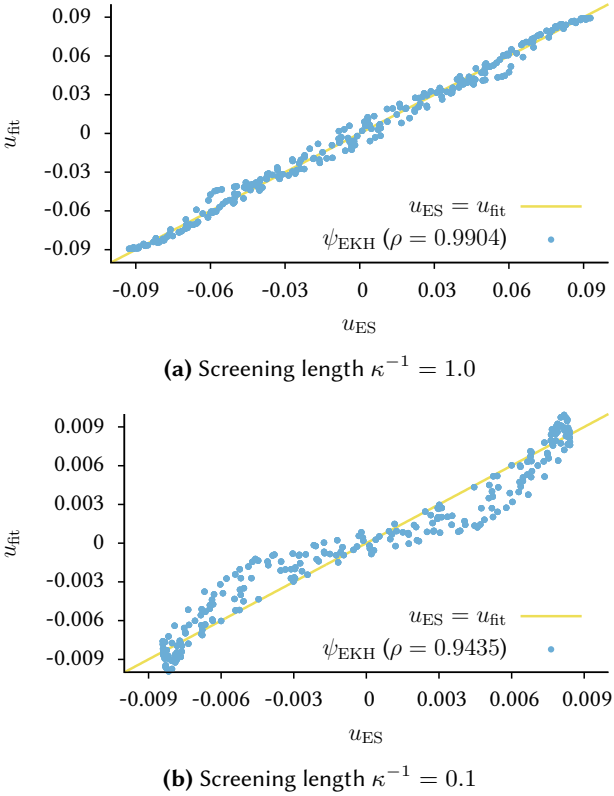
\begin{figure}[t,b]
  \centering
  \begin{subfigure}{\picsize}
    \resizebox{\textwidth}{!}{\large \input{error_lambda=10_L=50.tex}}
    \caption{Screening length $\kappa^{-1} = 1.0$}
    \label{fig:comparison_EKH_delta=10}
  \end{subfigure}
  \begin{subfigure}{\picsize}
    \resizebox{\textwidth}{!}{\large \input{error_lambda=01_L=50.tex}}
    \caption{Screening length $\kappa^{-1} = 0.1$}
    \label{fig:comparison_EKH_delta=01}
  \end{subfigure}
  \caption{Comparison of electrostatic potential $u_\mathrm{ES}$ and the fit potential $u_\mathrm{fit}$ using the EKH tensors. The energies were determined for fixed interparticle distance $r = 1.015$ and with $K = 10$ and $L = 50$ integration subintervals. The yellow line visualizes an ideal fit.}
\end{figure}

Within the limits of large screening lengths $\kappa^{-1}$ and large interparticle distances $r$, the EKH potential proves to reproduce the electrostatic potential well (Fig.~\ref{fig:comparison_EKH_delta=10}). However, with stronger screening the deviations get large especially for configurations where two patches are pointing directly towards each other (Fig.~\ref{fig:comparison_EKH_delta=01}). These configurations have the largest energy contributions and are therefore most important for the energy of Janus clusters.

\section{Mapping on a systematic tensor expansion}
\label{sec:fit_potential}
To overcome the shortcomings of the EKH model a larger set of tensors is needed. Based on the $S$ functions from Stone\autocite{Stone1978} for a dipolar interaction, we get the tensors
\begin{equation}
\begin{split}
  \psi_a &= \uvec{n}_1 \cdot \uvec{r}\\
  \psi_b &= \uvec{n}_2 \cdot \uvec{r}\\
  \psi_c &= \uvec{n}_1 \cdot \uvec{n}_2
\end{split}
\end{equation}
as a basis that is sufficient to describe the whole configurational space. The tensors $\psi_{a,b,c}$ unambiguously determine the related energy: if two configurations have identical basis tensors, they also have an identical energy. Employing the symmetry conditions \eqref{eq:tensor_conditions}, only certain combinations of $\psi_{a,b,c}$ are allowed. By flipping one of the particles the energy changes its sign, which means that in all combinations the normal vector $\uvec{n}_{k,l}$ has to occur with an odd exponent. Furthermore, by swapping the positions of the particles the sign of the energy needs to be preserved, which means that the interparticle distance $\vec{r}_{ij}$ needs to occur with an even exponent in all combinations. This yields
\begin{equation}
\label{eq:tensor_summation}
  \psi = \sum_k^B \sum_j^B \sum_i^B a_{ijk} \, (\psi_a^i \psi_b^j \psi_c^k + \psi_a^j \psi_b^i \psi_c^k)
\end{equation}
as a systematic linear combination where $j = i, i+2, i+4, \dots$ so that $i + j$ is always even. Furthermore $i + k$ has to be odd. The actual number of tensors in the final fit function is determined by the upper bound $B$ of summation, as shown in Table~\ref{tab:number_tensors}.

\begin{table}
  \centering
  \caption{Actual number of tensors used in the fit function $u_\mathrm{fit}$ corresponding to the upper bound $B$ in Eq.~\eqref{eq:tensor_summation}}
  \label{tab:number_tensors}
  \begin{tabular}{c | S[table-format=1.0] S[table-format=1.0] S[table-format=2.0] S[table-format=2.0] S[table-format=2.0] S[table-format=2.0] S[table-format=2.0] S[table-format=3.0] S[table-format=3.0]}
  \toprule
  $B$     & 1 & 2 &  3 &  4 &  5 &  6 &  7 &   8 &   9 \\
  tensors & 2 & 5 & 12 & 21 & 36 & 54 & 80 & 110 & 150 \\
  \bottomrule
  \end{tabular}
\end{table}

\begin{figure}
  \centering
  \begin{subfigure}{0.44\textwidth}
    \resizebox{\textwidth}{!}{\large \input{convergence_tensors.tex}}
    \caption{Correlation between the electrostatic potential $u_\mathrm{ES}$ and the fit function $u_\mathrm{fit}$ for upper bound $B$ that determines the actual number of fit tensors.}
    \label{fig:convergence_tensors}
  \end{subfigure}
  \begin{subfigure}{0.44\textwidth}
    \resizebox{\textwidth}{!}{\large \input{error_lambda=01_N=5.tex}}
    \caption{Comparison of electrostatic potential $u_\mathrm{ES}$ and the fit potential $u_\mathrm{fit}$ using a number of tensors corresponding to $B$. The yellow line visualises an ideal fit.}
    \label{fig:comparison_tensors}
  \end{subfigure}
  \caption{Determination of a reasonable number of tensors for the fit potential. The energies were determined with interparticle distance  $r = 1.015$, screening length $\kappa^{-1} = 0.1$, and with $K = 30$ and $L = 50$ integration subintervals.}
  \label{fig:tensors}
\end{figure}
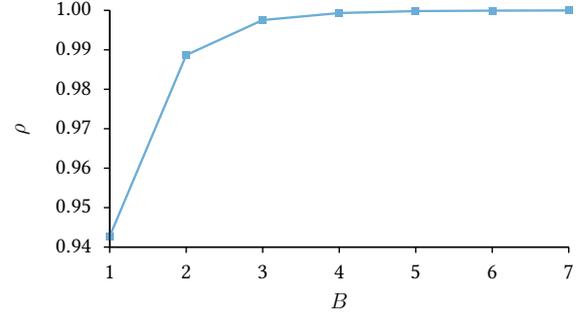
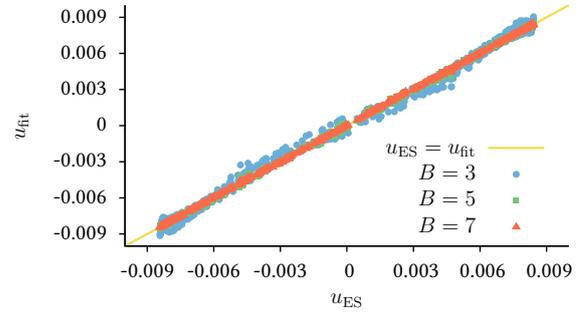

The next step is to find the smallest number of tensors that is needed to reproduce the numerically determined electrostatic potential $u_\mathrm{ES}$ with sufficient accuracy. Again we use the Pearson coefficient and check the correlation between the values of the complete electrostatic and the fitted potential for different $B$ (Fig.~\ref{fig:convergence_tensors}). It is desirable to keep the number of tensors small for two reasons. First, it enables faster evaluation of the potential, which is of importance for time-efficient simulations. Second, the more parameters are available to the least-square fit algorithm the more numerical data is needed to avoid overfitting the potential. While this holds for the angular dimensions, the fit of the distance-dependent function $a(r)$ requires only a small number of points. We found that $\num{36}$ tensors, corresponding to $B = \num{5}$, yield good agreement between fit and electrostatic potential (Fig.~\ref{fig:tensors}). It is our reference to be compared to the potential with two tensors in the following.

We want to point out that for $B = \num{1}$ the two tensors are
\begin{equation}
\begin{split}
  \psi_1 &= \psi_{110} = 2 \, (\uvec{n}_1 \cdot \uvec{r})(\uvec{n}_2 \cdot \uvec{r})\\
  \psi_2 &= \psi_{001} = 2 \, \uvec{n}_1 \cdot \uvec{n}_2
\end{split}
\end{equation}
and that these are completely equivalent to the original EKH tensors \eqref{eq:tensors_EKH} after straightforward definition of the prefactors.

\subsection*{Discussion of the potential}
\begin{figure}
  \centering
  \begin{subfigure}{\picsize}
    \resizebox{\textwidth}{!}{\large \input{landscape_EKH_L=50_K=30_10_lambda=01_N=5_r=1015_theta1=157.tex}}
    \caption{Potential energy landscape for setup shown in \textbf{(b)}.}
  \end{subfigure}
  \begin{subfigure}{\halfpicsize}
    \includegraphics[width=0.9\textwidth]{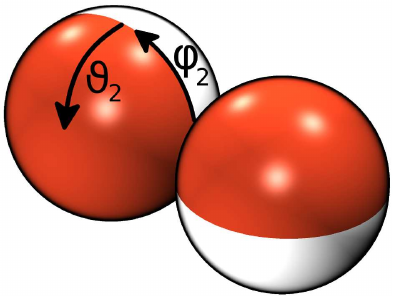}
    \label{fig:landscape_configuration}
    \caption{Setup of the two particles. The two rotation angles $\varphi_2$ and $\vartheta_2$ are indicated.}
  \end{subfigure}
  \begin{subfigure}{\halfpicsize}
    \includegraphics[width=0.9\textwidth]{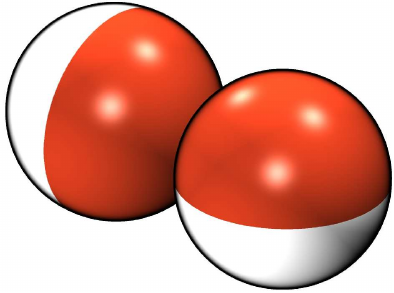}
    \label{fig:landscape_configuration_example}
    \caption{The \enquote{north pole} of one particle pointing towards the other particle.}
  \end{subfigure}
  \caption{An example for a potential energy landscape of two particles. It was determined with fixed interparticle distance $r_{ij} = 1.015$, screening length  $\kappa^{-1} = 0.1$, $L = 50$ integration subintervals and $36$ tensors corresponding to $B = 5$. The maximum of energy is found for $\vartheta_1 = 0$ where the second particle is located at one of the poles of the fixed particle. The maximum of energy is found for $\vartheta_1 = 0$ where the second particle is located at one of the poles of the fixed particle. Important are the plateaus where a rotation of the particles does not change the interaction energy. This is the case if one of the poles of the second particle points towards the fixed particle, as \eg for the configuration shown in \textbf{(c)} that is also marked in the energy landscape.}
  \label{fig:landscape}
\end{figure}

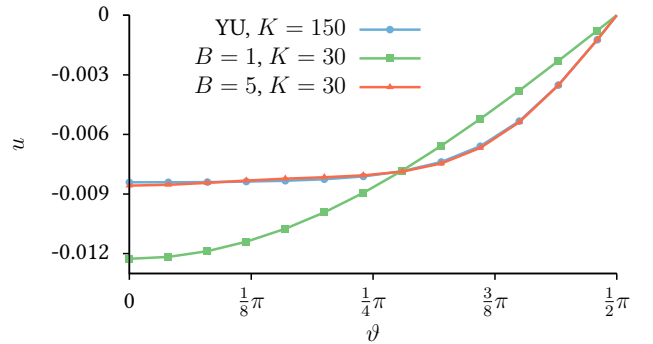
\begin{figure}
  \centering
  \resizebox{\picsize}{!}{\large \input{convergence_angular.tex}}
  \caption{Angular dependence of the interaction energy for $\kappa^{-1} = 0.1$. The angle $\vartheta$ is defined between the normal vector $\uvec{n}_k$ of the first particle and the interparticle vector $\vec{r}$, while the \enquote{north pole} of the second particle is pointing perpendicular to the surface of the first particle.}
  \label{fig:convergence_angular}
\end{figure}

For small screening lengths and small interparticle distances the interaction range is largely exceeded by the particle size. As was already mentioned, in this case only two tensors are not sufficient to reflect the subtleties of the interaction of charged Janus particles. Due to the distance dependence of the Yukawa potential, it is to be expected that nearby charges contribute more strongly to the particle interaction than charges far away from each other. For very small particle separations the interaction of charges on opposite sides of the particles is insignificant compared to neighboured sides. This effect is even more increased by the screening to the point where only the charges in a small area of the patches \enquote{feel} each other. Then, the particles are largely insensitive to rotation in the polar configurations similar to \textit{ns} and \textit{nn} shown in Fig.~\ref{fig:configurations} as it does not change the neighbourhood of the interacting areas. Only when one or both particles are crossing the equatorial regions the neighbourhood changes and hence also the interaction energy. This behaviour leads to plateaus in the potential energy landscape (Fig.~\ref{fig:landscape}) and has also been reported for square-type charge interactions if the interaction range is less than \SI{30}{\percent} of the particle diameter\autocite{Granick2006}. A fit with only two tensors is not capable to capture this feature: it distinguishes too strongly between the energies of these \enquote{polar} configurations, as clearly indicated by the angular dependence of the potential. The plot in Fig.~\ref{fig:convergence_angular} shows the potential energy of all configurations where the \enquote{north pole} of the second particles points towards the surface of the first one. Indeed, significant deviations are observed and the plateau is reproduced only for $B = 5$.

\section{DLVO potential}
\label{sec:DLVO_potential}
\begin{figure}[t,b]
  \centering
  \begin{subfigure}{\picsize}
    \resizebox{\textwidth}{!}{\large \input{DLVO_deviation.tex}}
    \caption{Ratio of the prefactors $a_{110} / a_{001}$ of the tensors for EKH with $B = 1$ and $b_1/b_2$ for the DLVO potential, determined with $K = 30$ and $L = 50$ integration subintervals.}
    \label{fig:DLVO_deviation}
  \end{subfigure}
  \begin{subfigure}{\picsize}
    \resizebox{\textwidth}{!}{\large \input{DLVO_angular.tex}}
    \caption{Angular dependence of the potentials for $\kappa^{-1} = 0.1$. The second particle points downwards, \ie parallel to the first particle, and perpendicular to the surface of the first particle, respectively.}
    \label{fig:DLVO_angular}
  \end{subfigure}
  \caption{Comparison of the fit potential for $B = 1$ and the DLVO potential.}
  \label{fig:DLVO}
\end{figure}
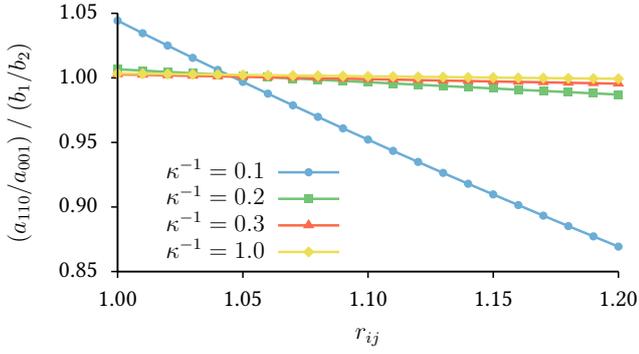
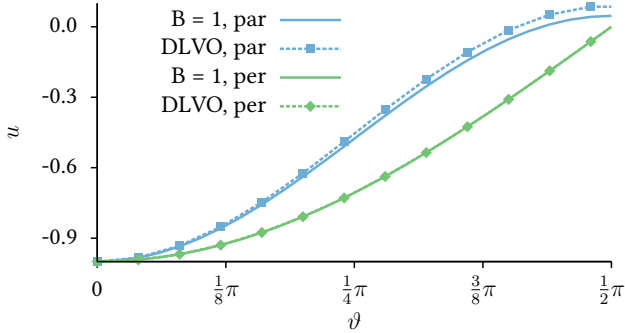

As mentioned in the introduction, the use of the Yukawa potential $u_\mathrm{Yu}$ wrongly assumes that screening effects are also present within the hard sphere particles. The DLVO theory takes the inaccessibility of the hard sphere volume correctly into account. Within lowest order of a multipole expansion it is possible to calculate the terms of the dipole-dipole interaction. These are directly related to the tensors for $B = 1$, and their prefactors\cite{Dijkstra2012}
\begin{equation}
\begin{split}
  b_1 &= -b_0 \frac{\exp{-\kappa r}}{r^3} \left(3+3 \kappa r+\left(\kappa r\right)^2\right)\\
  b_2 &= b_0 \frac{\exp{-\kappa r}}{r^3} \left(1+\kappa r\right)
\end{split}
\end{equation}
can be compared to the prefactors $a_{110}$ and $a_{001}$ of the respective tensors of the previously described potential. The ratio $\frac{a_{110} / a_{001}}{b_1 / b_2}$ between these prefactors is shown in Fig.~\ref{fig:DLVO_deviation}. We find that in the relevant range of $r-1 \le \kappa^{-1}$ the deviations are smaller than \SI{5}{\percent} for $\kappa^{-1} = 0.1$ and even smaller for larger values of $\kappa^{-1}$. These deviations occur for those configurations where the interaction happens at the \enquote{equators} of the particles, while the agreement between the potentials is very good if at least one particle is interacting via one of its \enquote{poles}. The corresponding configurations have a favourable energy and are therefore expected to be more important for the cluster formation. Furthermore, only small particle separations are relevant. We therefore expect the DLVO approximation to yield results similar to the Yukawa potential fitted with $B = 1$. Indeed, this will be shown explicitly for the formation of clusters in the subsequent section.

\section{Cluster formation}
\label{sec:cluster_formation}
We performed Monte Carlo simulations in the canonical $NVT$ ensemble in order to determine the lowest-energy states of systems with $N = 4, \dots, 13$ particles. For a faster equilibration of the system we applied the well-known parallel tempering method\autocite{Swendsen1986} with two different sets of parameters as described in the appendix. Furthermore, multiple runs were performed for each set of $B$ and $\kappa^{-1}$ to ensure that the global minima have been found. The pair interaction is normalized for all potentials so that at $r = 1.0$ the maximum of the potential, corresponding to the \textit{ns} configuration, is equal to unity.

\begin{figure}[p]
  \captionsetup[subfigure]{labelformat=empty}
  \centering
  \begin{subfigure}{\clusterpicsize}
    \includegraphics[width=0.9\textwidth]{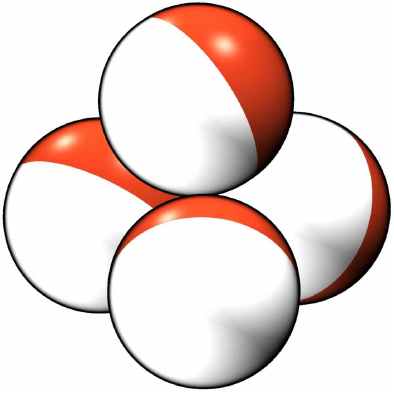}
    \caption{$N = 4$}
  \end{subfigure}
  \begin{subfigure}{\clusterpicsize}
    \includegraphics[width=0.9\textwidth]{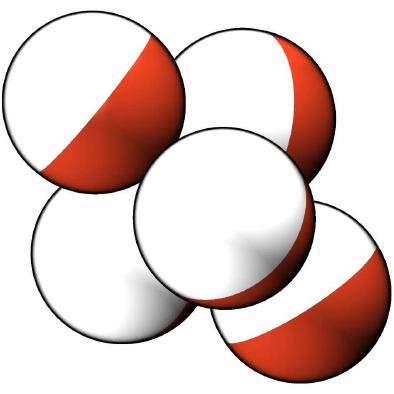}
    \caption{$N = 5$}
  \end{subfigure}
  \begin{subfigure}{\clusterpicsize}
    \includegraphics[width=0.9\textwidth]{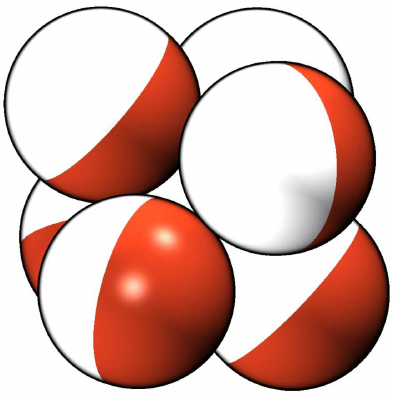}
    \caption{$N = 6$}
  \end{subfigure}
  \begin{subfigure}{\clusterpicsize}
    \includegraphics[width=0.9\textwidth]{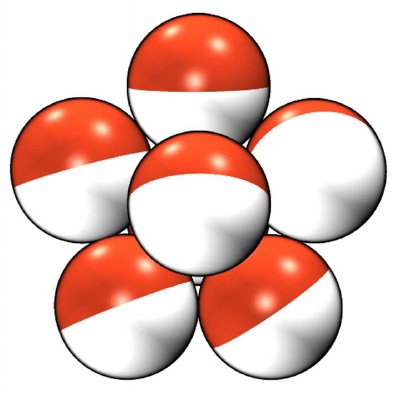}
    \caption{$N = 7$}
  \end{subfigure}
  \begin{subfigure}{\clusterpicsize}
    \includegraphics[width=0.9\textwidth]{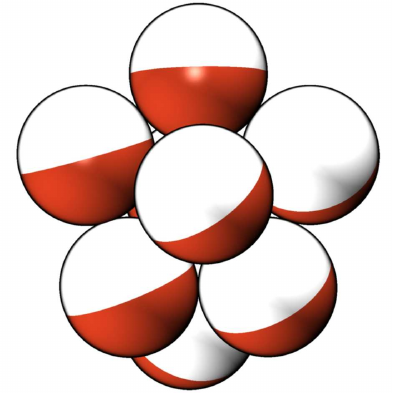}
    \caption{$N = 8$}
  \end{subfigure}
  \begin{subfigure}{\clusterpicsize}
    \includegraphics[width=0.9\textwidth]{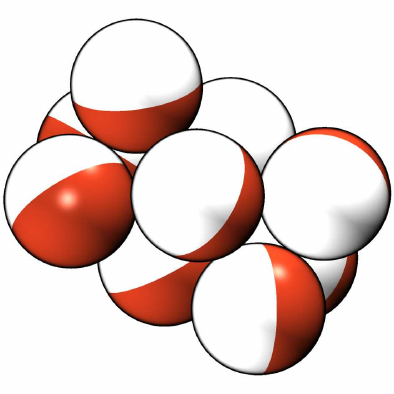}
    \caption{$N = 9$}
  \end{subfigure}
  \begin{subfigure}{\clusterpicsize}
    \includegraphics[width=0.9\textwidth]{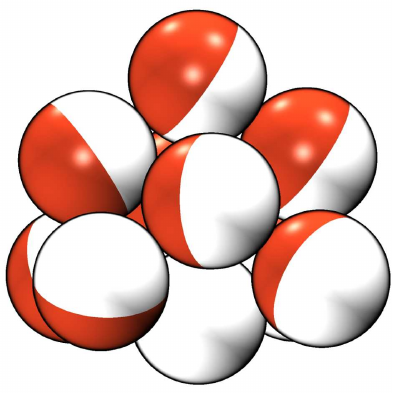}
    \caption{$N = 10$}
  \end{subfigure}
  \begin{subfigure}{\clusterpicsize}
    \includegraphics[width=0.9\textwidth]{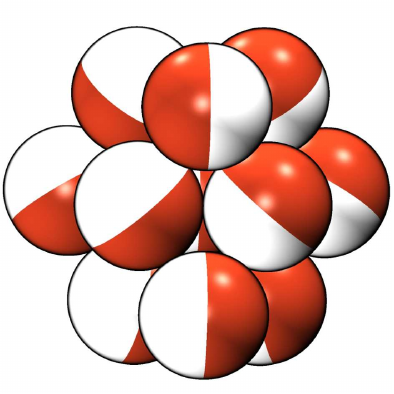}
    \caption{$N = 11$}
  \end{subfigure}
  \begin{subfigure}{\clusterpicsize}
    \includegraphics[width=0.9\textwidth]{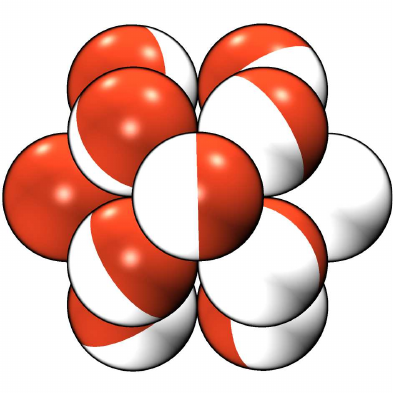}
    \caption{$N = 12$}
  \end{subfigure}
  \begin{subfigure}{\clusterpicsize}
    \includegraphics[width=0.9\textwidth]{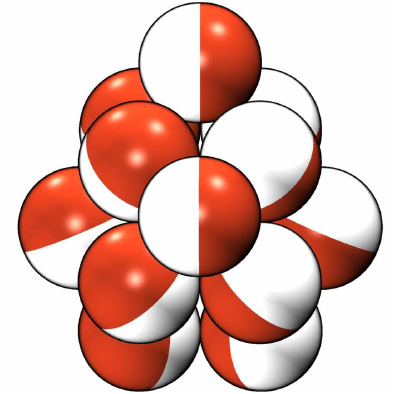}
    \caption{$N = 13$}
  \end{subfigure}
  \caption{Cluster structures with lowest energy obtained for $B = 5$ and $\kappa^{-1} = 0.1$. We notably find a tetrahedral ($N = 4$), trigonal bipyramidal ($N = 5$), octahedral ($N = 6$), pentagonal bipyramidal ($N = 7$) and tricapped trigonal prismatic ($N = 9$) geometry. Larger systems form hexagonal close-packed layers.}
  \label{fig:clusters_B=5_kappa=01}
\end{figure}

\begin{figure}[p]
  \captionsetup[subfigure]{labelformat=empty}
  \centering
  \begin{subfigure}{\clusterpicsize}
    \includegraphics[width=0.9\textwidth]{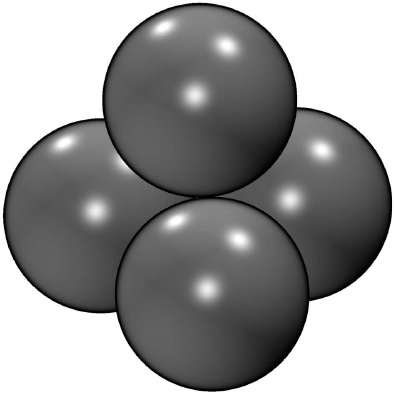}
    \caption{$N = 4$}
  \end{subfigure}
  \begin{subfigure}{\clusterpicsize}
    \includegraphics[width=0.9\textwidth]{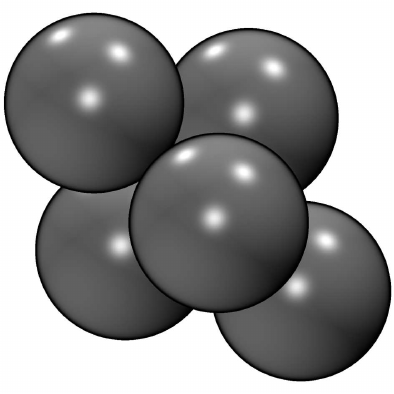}
    \caption{$N = 5$}
  \end{subfigure}
  \begin{subfigure}{\clusterpicsize}
    \includegraphics[width=0.9\textwidth]{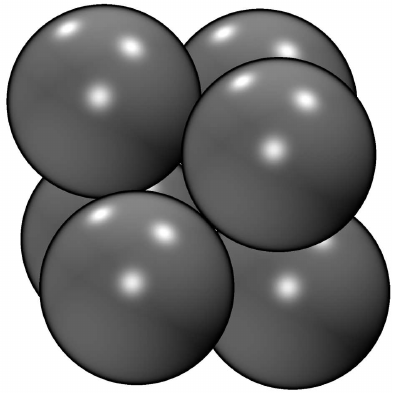}
    \caption{$N = 6$}
  \end{subfigure}
  \begin{subfigure}{\clusterpicsize}
    \includegraphics[width=0.9\textwidth]{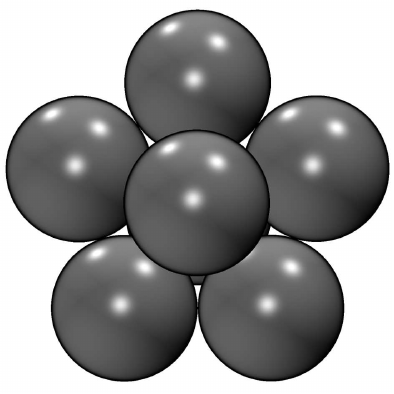}
    \caption{$N = 7$}
  \end{subfigure}
  \begin{subfigure}{\clusterpicsize}
    \includegraphics[width=0.9\textwidth]{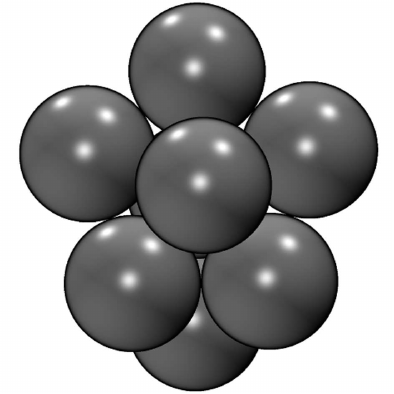}
    \caption{$N = 8$}
  \end{subfigure}
  \begin{subfigure}{\clusterpicsize}
    \includegraphics[width=0.9\textwidth]{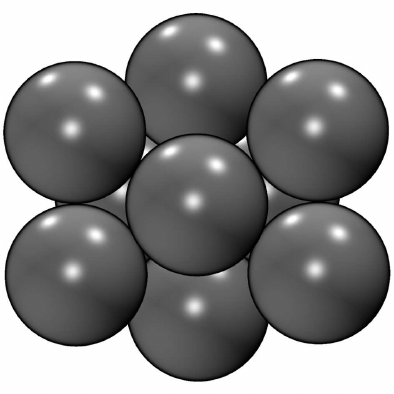}
    \caption{$N = 9$}
  \end{subfigure}
  \begin{subfigure}{\clusterpicsize}
    \includegraphics[width=0.9\textwidth]{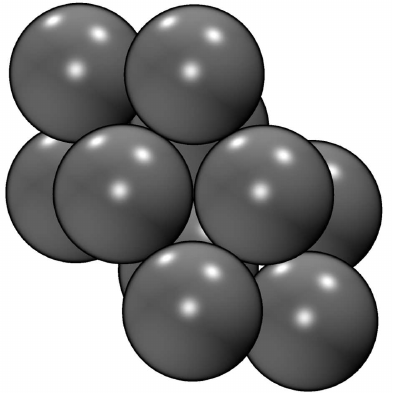}
    \caption{$N = 10$}
  \end{subfigure}
  \begin{subfigure}{\clusterpicsize}
    \includegraphics[width=0.9\textwidth]{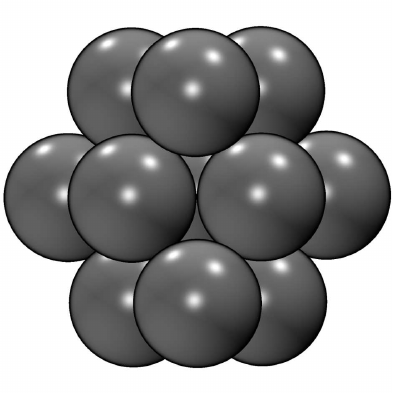}
    \caption{$N = 11$}
  \end{subfigure}
  \begin{subfigure}{\clusterpicsize}
    \includegraphics[width=0.9\textwidth]{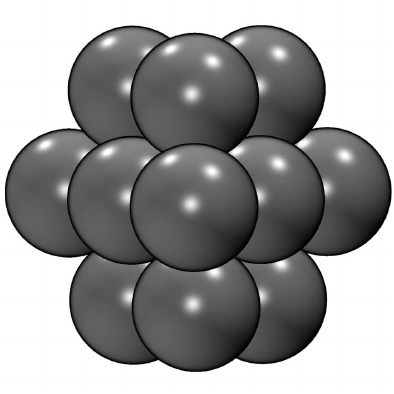}
    \caption{$N = 12$}
  \end{subfigure}
  \begin{subfigure}{\clusterpicsize}
    \includegraphics[width=0.9\textwidth]{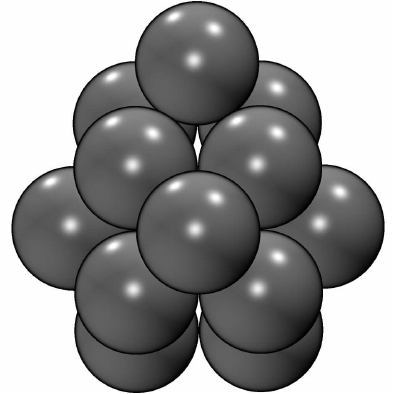}
    \caption{$N = 13$}
  \end{subfigure}
  \caption{Cluster structures of sticky hard spheres interacting via the Yukawa potential from eq.~\eqref{eq:potential_yukawa} with screening length $\kappa^{-1} = 0.1$. The arrangements are largely identical to those of Janus particles, except for $N = 9, 10$ and $12$. Again, hexagonal close-packed layers are formed for large systems.}
  \label{fig:clusters_HS}
\end{figure}

First, we discuss the cluster formation for $B = 5$ and $\kappa^{-1} = 0.1$ where the electrostatic potential is fitted with high accuracy. We obtain the structures shown in Fig.~\ref{fig:clusters_B=5_kappa=01}: a tetrahedral ($N = 4$), trigonal bipyramidal ($N = 5$), octahedral ($N = 6$), pentagonal bipyramidal ($N = 7$) and tricapped trigonal prismatic ($N = 9$) arrangement, and larger systems arrange in hexagonal close-packed layers. The particularly favourable tetrahedral, octahedral and pentagonal bipyramidal structures are reoccurring substructures of larger clusters. These structures are for $N \leq 8$ also found numerically and experimentally as reported in Ref.~\cite{Granick2006}, while no specific arrangements are described for larger particle numbers.

It is instructive to compare these structures with minimum energy structures where no anisotropic interaction is present. We use the Yukawa potential from Eq.~\eqref{eq:potential_yukawa} to obtain clusters of sticky hard spheres with only a short-ranged attractive interaction. Due to this choice the distance dependence is equal to that of the Janus particles, without having their angular dependence. As shown in Fig.~\ref{fig:clusters_HS}, at $\kappa^{-1} = 0.1$ the resulting clusters are identical for almost all particle numbers except for $N = 9, 10$ and $12$. In the latter case the clusters are both moieties of the same hexagonal close-packed structure. Only in the former two cases the patches are not able to orientate in a way that enables a favourable interaction. The anisotropy therefore does not play that an important role for the formation of clusters of Janus particles.

\begin{figure}
  \centering
  \resizebox{\picsize}{!}{\large \input{fingerprint_B=5_HS_kappa=01.tex}}
  \caption{Displacement between the particles $i$ and the particle closest to the geometric centre of a cluster. Lowest-energy clusters of size $N$ are compared: the points above the bars belong to $B=5$, $\kappa^{-1} = 0.1$, those below the bars to the sticky hard spheres. Each cluster size is shown in a different color as a guide to the eye.}
  \label{fig:fingerprint_B=5_HS_kappa=01}
\end{figure}
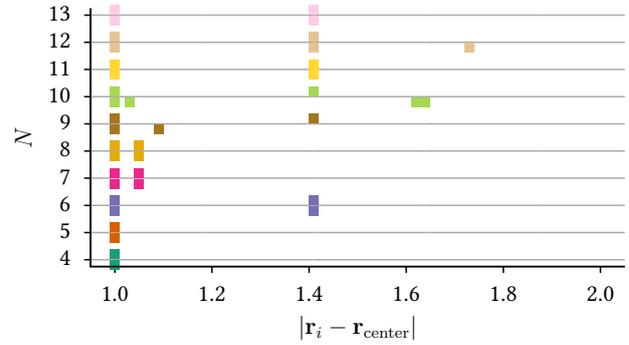

The similarity between clusters of sticky hard spheres and Janus particles can also be understood with the help of a \enquote{fingerprint} plot as shown in Fig.~\ref{fig:fingerprint_B=5_HS_kappa=01}. Here, the displacement between every particle in a cluster and the particle closest to the geometrical centre of that cluster is plotted. Identical structures can be identified by having the same pattern of displacement, and symmetric structures show only a small number of distances in the plot. It shows that deviations of the displacements do only occur for $N = 9, 10$ and $12$, while the other structures are identical.

Another instructive comparison can be made to one-patch IPCs with a hemispherical cap that are effectively identical to charged Janus particles. A system with exponential screening of the potential and screening length $\kappa^{-1} = 0.08$ has recently been found to aggregate to close-packed structures at low temperatures\autocite{Dempster2016}, similar to our findings. These aggregates are strongly polarized and keep their polarization with growth, \ie the clusters are not limited in size. Interestingly, it was also found that this behaviour is neither influenced by polydispersity in the patch size nor by the exact shape of the patch boundary.

\begin{figure}[t,b]
  \captionsetup[subfigure]{labelformat=empty}
  \centering
  \begin{subfigure}{\clusterpicsize}
    \includegraphics[width=0.9\textwidth]{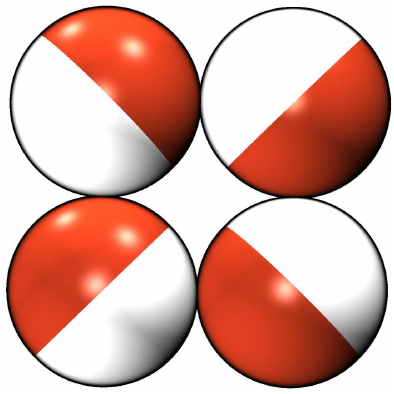}
    \caption{$N = 4$}
  \end{subfigure}
  \begin{subfigure}{\clusterpicsize}
    \includegraphics[width=0.9\textwidth]{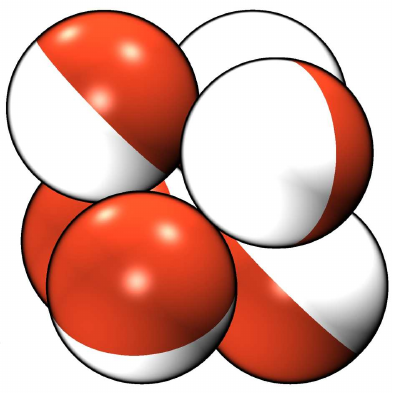}
    \caption{$N = 6$}
  \end{subfigure}
  \begin{subfigure}{\clusterpicsize}
    \includegraphics[width=0.9\textwidth]{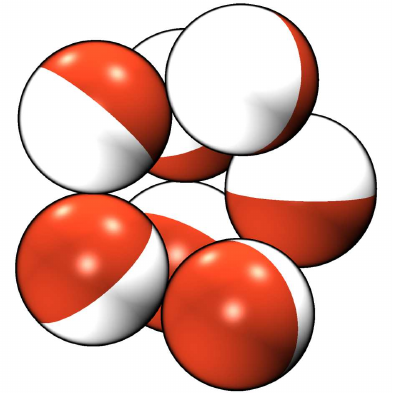}
    \caption{$N = 7$}
  \end{subfigure}
  \begin{subfigure}{\clusterpicsize}
    \includegraphics[width=0.9\textwidth]{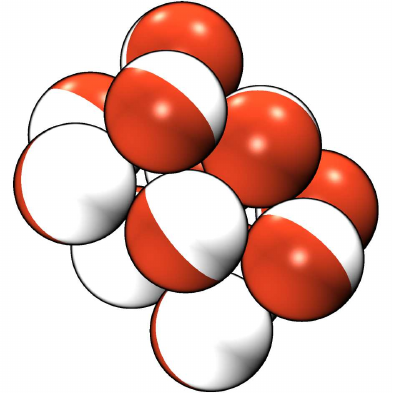}
    \caption{$N = 12$}
  \end{subfigure}
  \caption{Examples of cluster structures with lowest energy obtained for $B = 5$ with $\kappa^{-1} = 1.0$.}
  \label{fig:clusters_B=5_kappa=10}
\end{figure}

By increasing the screening length to $\kappa^{-1} = 1.0$, the distance-dependent decay of the potential is smaller and it is more important for oppositely charged patches to be far away from each other. This results in less closely packed structures compared to the smaller screening length. We observe two reoccurring patterns: first, a ring or stacked four-rings for $N = 4, 8, 12$ and incomplete stacked rings for $N = 7$ and $11$, and also the octahedral structure for $N = 6$ can be seen as bicapped four-ring. Second, for $N = 9$ we get a tricapped prism less dense than that for $\kappa^{-1} = 0.1$ allowing for larger distances between opposite patches. The structures for $N = 10$ and $13$ are rather irregular and somewhere in between stacked four-rings and prismatic. Examples are shown in Fig.~\ref{fig:clusters_B=5_kappa=10}. In total, only the octahedral and prismatic geometries are similar to the smaller screening length, beeing a clear indicator of the importance of the screening for Janus-like behaviour.

\begin{figure}[t,b]
  \captionsetup[subfigure]{labelformat=empty}
  \centering
  \begin{subfigure}{\clusterpicsize}
    \includegraphics[width=0.9\textwidth]{cluster_N=4_kappa=01_B=5.pdf}
    \caption{$\kappa^{-1} = 0.1$}
  \end{subfigure}
  \begin{subfigure}{\clusterpicsize}
    \includegraphics[width=0.9\textwidth]{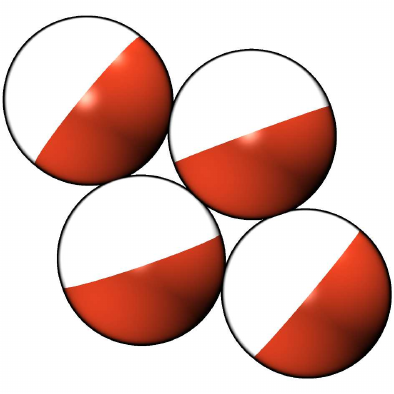}
    \caption{$\kappa^{-1} = 0.2$}
  \end{subfigure}
  \begin{subfigure}{\clusterpicsize}
    \includegraphics[width=0.9\textwidth]{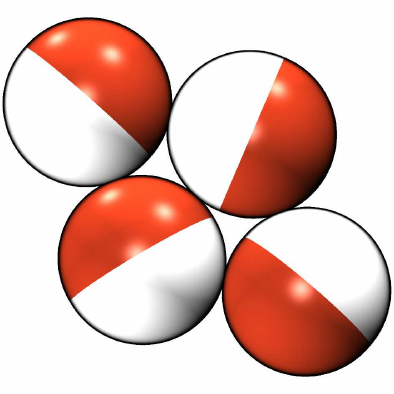}
    \caption{$\kappa^{-1} = 0.4$}
  \end{subfigure}
  \begin{subfigure}{\clusterpicsize}
    \includegraphics[width=0.9\textwidth]{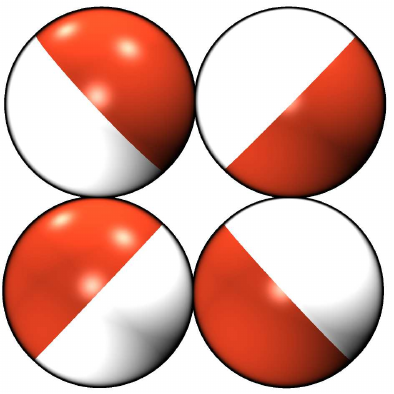}
    \caption{$\kappa^{-1} = 0.5$}
  \end{subfigure}
  \caption{Influence of the screening length $\kappa^{-1}$ on the structure formation for $N = 4$.}
  \label{fig:clusters_screening}
\end{figure}

We performed explicit simulations for $N = 4$ at intermediate screening lengths. The resulting structures in Fig.~\ref{fig:clusters_screening} shown that there is a transition from tetrahedral to ring geometry. Already for $\kappa^{-1} = 0.2$ we observe a planar rhombic structure with a mean absolute dipole moment $|\mu|/N = 0.96$. Between $\kappa^{-1} = 0.3$ and $0.4$ a rearrangement of the particles occurs, resulting in a lower dipole moment $|\mu|/N = 0.15$. A further increase of the screening length to $\kappa^{-1} = 0.5$ leads to the ring structures also obtained for $\kappa^{-1} = 1.0$ with a dipole moment close to zero.

\begin{figure}[t,b]
  \captionsetup[subfigure]{labelformat=empty}
  \centering
  \begin{subfigure}{\clusterpicsize}
    \includegraphics[width=0.9\textwidth]{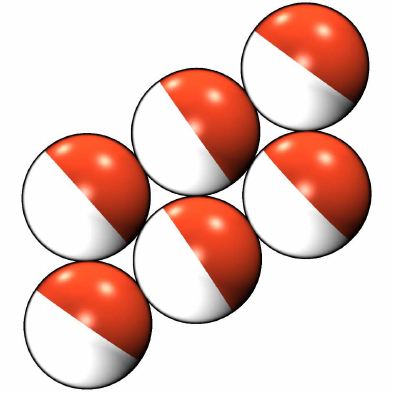}
    \caption{$N = 6$, $B = 1$}
  \end{subfigure}
  \begin{subfigure}{\clusterpicsize}
    \includegraphics[width=0.9\textwidth]{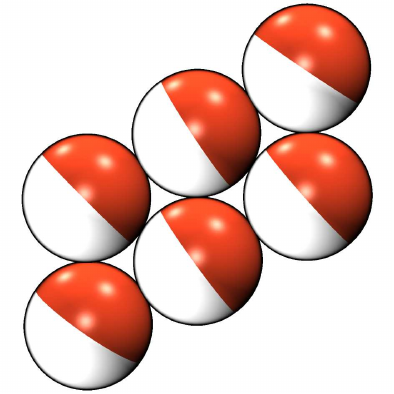}
    \caption{$N = 6$, DLVO}
  \end{subfigure}
  \begin{subfigure}{\clusterpicsize}
    \includegraphics[width=0.9\textwidth]{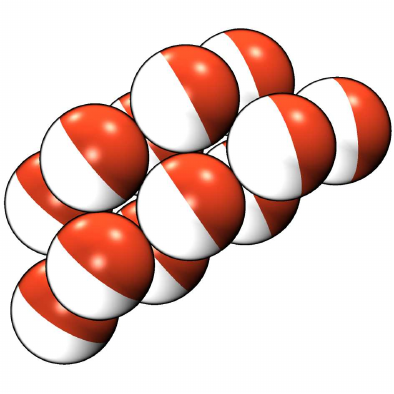}
    \caption{$N = 12$, $B = 1$}
  \end{subfigure}
  \begin{subfigure}{\clusterpicsize}
    \includegraphics[width=0.9\textwidth]{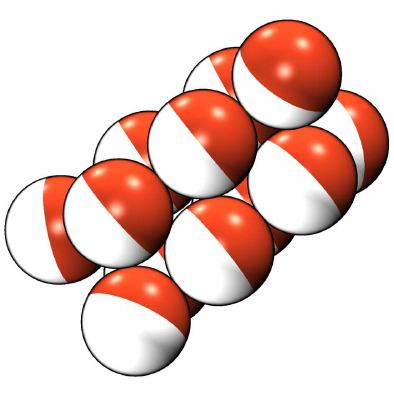}
    \caption{$N = 12$, DLVO}
  \end{subfigure}
  \caption{Comparison of cluster structures with lowest energy obtained for $B = 1$ and DLVO with $\kappa^{-1} = 0.1$.}
  \label{fig:clusters_B=1_DLVO}
\end{figure}

Finally, by chosing $B = 1$ and setting the number of tensors to $2$ we get a dipole-like potential. The minimum of the particle-particle interaction then is only achieved for configurations corresponding to the \textit{ns} configuration where the poles of the particles are pointing towards each other. This leads to the formation of chain-like structures for $\kappa^{-1} = 0.1$, and to oval structures or rings for $\kappa^{-1} = 1.0$ as here the repulsive interaction of the \enquote{antipodal} patches again is relevant. This aggregation behaviour is known from dipoles\autocite{Douglas2006, Moghani2013} but not from Janus particles. Both chain and ring structures are similarly found for the DLVO potential, even though not identical in all cases. Two examples of the interconnected chain structures for $\kappa^{-1} = 0.1$ are compared in Fig.~\ref{fig:clusters_B=1_DLVO}.

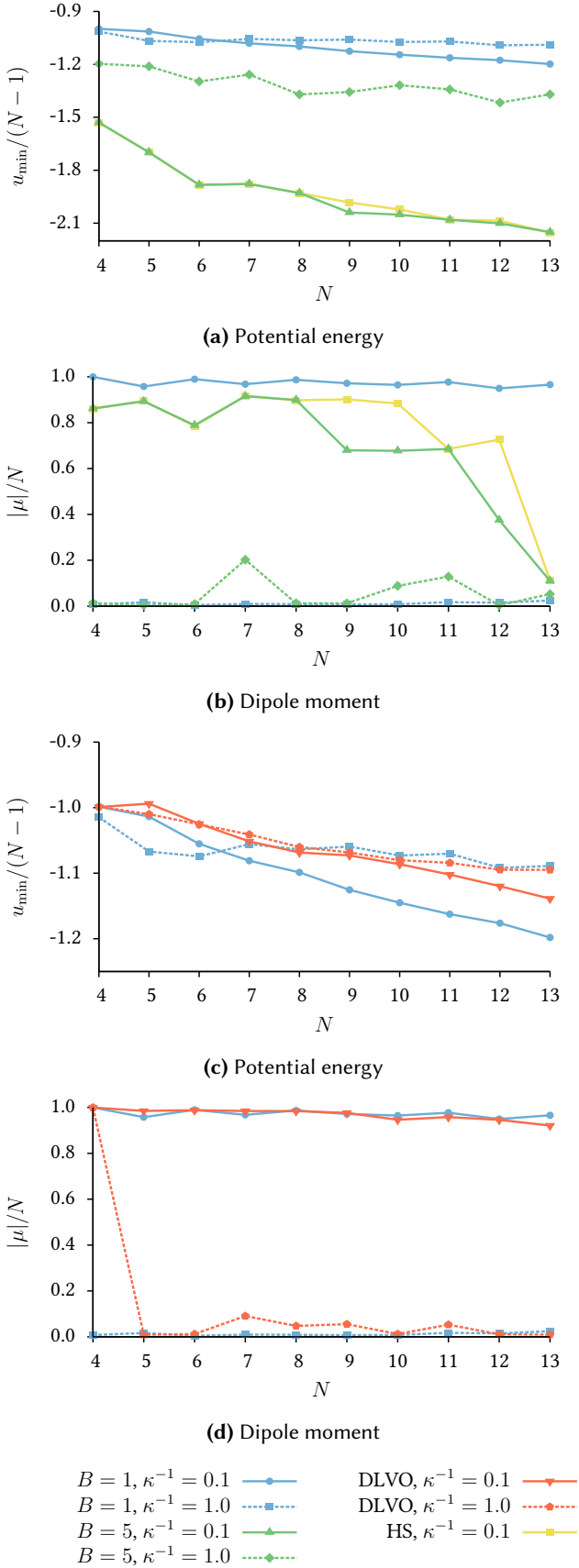
\begin{figure}[t,b]
  \centering
  \begin{subfigure}{0.465\textwidth}
    \resizebox{\textwidth}{!}{\large \input{particle_energy_1.tex}}
    \caption{Potential energy}
    \label{fig:particle_energy_1}
  \end{subfigure}
  \begin{subfigure}{0.465\textwidth}
    \resizebox{\textwidth}{!}{\large \input{particle_dipole_1.tex}}
    \caption{Dipole moment}
    \label{fig:particle_dipole_1}
  \end{subfigure}
  \begin{subfigure}{0.465\textwidth}
    \resizebox{\textwidth}{!}{\large \input{particle_energy_2.tex}}
    \caption{Potential energy}
    \label{fig:particle_energy_2}
  \end{subfigure}
  \begin{subfigure}{0.465\textwidth}
    \resizebox{\textwidth}{!}{\large \input{particle_dipole_2.tex}}
    \caption{Dipole moment}
    \label{fig:particle_dipole_2}
  \end{subfigure}
  \vskip1em
  \begin{subfigure}{0.465\textwidth}
    \resizebox{\textwidth}{!}{\large \input{particle_key.tex}}
  \end{subfigure}
  \caption{Comparison of the absolute value $|\mu|$ of the mean dipole moment and the potential energy $u$ of clusters obtained for $B = 1$ and $5$ with $\kappa^{-1} = 0.1$ and $1.0$, respectively, as well as for DLVO theory with $\kappa^{-1} = 0.1$. Furthermore, HS denotes Janus particles with $B = 5$ and $\kappa^{-1} = 0.1$ that are restricted in their coordinates to the arrangement of the sticky hard sphere clusters from Fig.~\ref{fig:clusters_HS}.}
  \label{fig:particle_comparison}
\end{figure}

A quantitative comparison of the potentials is possible via the potential energy of the clusters. The energy per particle is plotted against the cluster size in Fig.~\ref{fig:particle_energy_1} for $B = 1$ and $5$. The lowest energies are obtained for $B = 5$ and $\kappa^{-1} = 0.1$ where the particles can rotate to some degree without change in energy, as oppositely charged patches cannot interact due to the small screening length. This favours the formation of close packed clusters with a high number of connections, and the energy per particle even increases with the cluster size. In order to evaluate the clusters of the sticky hard spheres from Fig.~\ref{fig:clusters_HS}, we put Janus particles on the corresponding coordinates and allowed them to optimize their orientations, showing differences again only for $N = 9$ and $10$. For the other potentials, the particles favour linear or ring structures as the degree of rotational freedom is smaller than for $B = 5$ and $\kappa^{-1} = 0.1$, which results in a comparatively smaller energy gain per particle number. The plot of the absolute mean dipole moment of the structures, shown in Fig.~\ref{fig:particle_dipole_1}, indicates that the orientations in the ring structures for $B = 1$ and $B = 5$ with $\kappa^{-1} = 1.0$ cancel out each other. In contrast, the linear configurations for $B = 1$ and $\kappa^{-1} = 0.1$ have a dipole moment constantly close to unity, as additional particles are added to one end of a chain. The particles in the clusters for $B = 5$ and $\kappa^{-1} = 0.1$ are orientated such that unequal patches are facing each other, leading to a significant dipole moment of the clusters and similarly enabling the addition of particles to a cluster.

The close relation of the dipolar fit for $B = 1$ and the dipolar DLVO approximation is illustrated by the plot of the potential energy in Fig.~\ref{fig:particle_energy_2}. For $\kappa^{-1} = 0.1$ interconnected chain structures are formed, leading to an energy per particle slightly increasing with the system size. However, the energy gain is smaller than for $B = 5$ due to the lower connectivity and the stronger angular dependence. Also the effect of the weak DLVO potential for \enquote{equatorial} configurations, as previously discussed, is reflected in the plot and explains why even identical structures like those shown in Fig.~\ref{fig:clusters_B=1_DLVO} for $N = 6$ do not have the same energy. This effect does not occur for $N = 4$ as well as for the ring structures obtained for $\kappa^{-1} = 1.0$ where \enquote{polar} configurations are predominant. Here, the energies of DLVO and fit potential are equal. The plot of the dipole moments in Fig.~\ref{fig:particle_dipole_2} also shows that chain structures for $\kappa^{-1} = 0.1$ and ring structures for $\kappa^{-1} = 1.0$ are comparable.

\section{Conclusion}
\label{sec:conclusion}
In this paper we studied the structure formation of charged Janus particles with respect to the underlying potential. As a reference we set up an electrostatic potential that includes screening effects via an empirically motivated Yukawa term. For small screening lengths the resulting potential energy landscape features plateaus for certain configurations. We systematically constructed a fit potential by linear superposition of basis tensors that were mapped to the electrostatic potential. The parameter $B$ controls the number of tensors used for the fit. For sufficiently large $B$ it captures both the angle and distance dependence of the reference even for small screening lengths. For $B = 1$ the fit is identical to a dipole-like interaction\autocite{Hess2003}, and comparable to the dipole approximation in the DLVO limit\autocite{Dijkstra2012}.

We obtained minimum potential energy structures via parallel tempering Monte Carlo simulations for the three mentioned potentials. The resulting structures are for DLVO and dipolar approximation similar to aggregated dipoles\autocite{Douglas2006, Moghani2013}: mostly chain and ring formation was observed. This is also reflected in the linear scaling of the energy per particle with the cluster size. In contrast, for $B = 5$ and a better fit of the electrostatic potential we observed an increasing energy per particle and cluster formation that is similar to experimental results from Hong \etal\autocite{Granick2006}. Interestingly, the dipolar DLVO approximation cannot reproduce these results and leads to structurally different clustering behaviour, even though it correctly treats the hard core of the particles. This suggest that, in order to capture the properties of clusters formed by Janus particles, the consideration of higher-order non-dipolar-like terms is at least as important as the correct treatment of the impact of hard core effects on the shielding properties.

\section*{Acknowledgments}
  We gratefully acknowledge the support by the DFG (TRR 61, B12) and helpful discussions with B.-J. Ravoo.

\appendix
\section{Parallel tempering routine}
With increasing particle number the energy landscapes become more and more rugged. Therefore we make use of the parallel tempering method: multiple replicas of a system are simulated in parallel at different temperatures, and configurations of neighbouring replicas are exchanged with a probability depending on the energy difference between the replicas. Like this, the system can escape local minima at higher temperatures and may explore the low energy regions at lower temperatures.

For an efficient sampling of a landscape, the energy histograms of the replicas should partially overlap to enable a frequent exchange of configurations. This is given for small temperature differences $\Delta T$ between replicas $i$ and $j$. The exchange probability is
\begin{equation}
  p_\mathrm{ex} = \mathrm{min}\left(1, \exp{\left(u_{i} - u_j\right) \left(\frac{1}{T_i} - \frac{1}{T_j}\right)}\right)
\end{equation}
according to the Metropolis-Hastings criterion. We performed each simulation with 16 replicas at temperatures
\begin{equation}
  T_i = T_0 \ \exp{\alpha i}
\end{equation}
starting with $T_0 = 0.001$. The highest temperature $T_\mathrm{max} = 0.12$ is chosen such that it prevents the formation of clusters, compensating for the interaction potential of the particles.

The number $n_\mathrm{sim}$ of simulation steps is in the range from \num{1.65} to \num{6.5} million steps and is increased with the number of particles in the system. Translational and rotational moves are attempted in every step. For parallel tempering an exchange of replicas is attempted on average in every third step. The step sizes of the trial moves are adapted in certain intervals between two MC steps throughout the simulation. This is to achieve an acceptance rate of roughly $\num{0.33}$ which is benefiting for a fast equilibration of the system.

In practice it turns out that the system tends to stick in spatially elongated structures with relatively high energy. To avoid this problem we add during the first fifth of the simulation an additional energy penalty which is proportional to the sum of distances between all particles. This term helps to find the more compact low-energy states. Afterwards this term is switched off.

We also use a second set of parameters for the parallel tempering routine with $T_0 = 0.001$, $T_\mathrm{max} = 0.3$, up to \num{10} million simulations steps and an additional energy penalty in the first half of the simulation. In general, this showed to be less efficient, but for $B=5$ and $\kappa^{-1} = 0.1$ it helped to obtain the low-energy states specifically for $N = 11,12,13$.

\section{Parameters of the fitted potential with B = 5}
The quality of the fitted potential is mainly determined by the number of tensors, but also by the number $K$ of grid points where the electrostatic potential is evaluated. For a large number of tensors also $K$ needs to be large in order to generate enough data to avoid overfitting of the potential described by Eq.~\eqref{eq:tensor_summation}. The prefactors $a_{ijk}$ of the tensors are again determined by Eq.~\ref{eq:distance_fit} and every tensor is given by three parameters $a_{0,1,2}$. An exemplary set of parameters obtained for $B = 5$ and $\kappa^{-1} = 0.1$ with $K = 50$ and $L = 50$ is given in Table~\ref{tab:parameters_fit}.

\begin{table}
  \caption{Parameters determining the three prefactors $a_{0, 1, 2}$ of the $36$ tensors for $B = 5$ and $\kappa^{-1} = 0.1$.}
  \small
  \label{tab:parameters_fit}
  \begin{tabular}{c | S[table-format=1.6e-1] S[table-format=1.6e-1] S[table-format=1.6e-1]}
    \toprule
    tensor & {$a_0$}      & {$a_1$}      & {$a_2$}\\
    1      & 2.495724E+02 & 5.567518E+00 & 3.566689E+02\\
    2      & 1.177006E+03 & 4.428369E+02 & 5.939162E+02\\
    3      & 7.576554E+02 & 3.545202E+02 & 2.427296E+02\\
    4      & 6.148736E+02 & 8.977681E+01 & 3.457705E+01\\
    5      & 2.675007E+02 & 1.044046E+03 & 5.550898E+02\\
    6      & 9.870923E+02 & 1.429238E+03 & 5.958074E+02\\
    7      & 3.705798E+01 & 2.495840E-01 & 1.067393E+01\\
    8      & 8.150465E+01 & 2.669996E+02 & 1.039591E+02\\
    9      & 4.946397E+01 & 3.717428E+02 & 1.695175E+02\\
    10     & 4.788286E+02 & 1.119554E+03 & 4.971669E+02\\
    11     & 2.412448E+03 & 2.916522E+03 & 1.134817E+03\\
    12     & 3.099282E+03 & 2.342143E+03 & 6.617654E+02\\
    13     & 1.926071E+02 & 6.367596E+01 & 2.712171E+00\\
    14     & 3.756181E+03 & 4.725410E+03 & 1.868790E+03\\
    15     & 3.320513E+03 & 3.355056E+03 & 1.174777E+03\\
    16     & 1.328071E+04 & 1.751503E+04 & 7.222651E+03\\
    17     & 2.380848E+04 & 2.569263E+04 & 9.735365E+03\\
    18     & 4.176866E+03 & 3.206092E+03 & 6.678584E+02\\
    19     & 1.003277E+01 & 7.434166E+00 & 7.538136E+00\\
    20     & 1.985735E+03 & 2.909333E+03 & 1.205150E+03\\
    21     & 1.686291E+03 & 2.187038E+03 & 8.558277E+02\\
    22     & 6.597944E+03 & 7.658536E+03 & 2.939687E+03\\
    23     & 4.501884E+03 & 3.568398E+03 & 3.011543E+03\\
    24     & 1.761495E+04 & 3.090680E+04 & 1.454623E+04\\
    25     & 9.031020E+02 & 1.333005E+03 & 5.251129E+02\\
    26     & 3.832313E+03 & 5.578704E+03 & 2.079896E+03\\
    27     & 1.203150E+04 & 2.135641E+04 & 9.358615E+03\\
    28     & 1.269046E+03 & 1.008296E+03 & 1.372714E+02\\
    29     & 3.659269E+04 & 6.157140E+04 & 2.626564E+04\\
    30     & 1.392928E+04 & 2.172823E+04 & 8.926257E+03\\
    31     & 5.123095E+01 & 9.265441E+01 & 4.061949E+01\\
    32     & 5.167195E+02 & 8.395294E+02 & 3.890510E+02\\
    33     & 2.773006E+03 & 4.719220E+03 & 2.047147E+03\\
    34     & 1.045070E+03 & 1.180300E+03 & 6.688967E+02\\
    35     & 1.341567E+04 & 2.072763E+04 & 8.553302E+03\\
    36     & 6.470530E+03 & 8.909971E+03 & 3.442863E+03\\
    \bottomrule
  \end{tabular}
\end{table}

\balance
\printbibliography
\end{document}

%% file: convergence_integration.tex
% GNUPLOT: LaTeX picture with Postscript
\begingroup
  \makeatletter
  \providecommand\color[2][]{%
    \GenericError{(gnuplot) \space\space\space\@spaces}{%
      Package color not loaded in conjunction with
      terminal option `colourtext'%
    }{See the gnuplot documentation for explanation.%
    }{Either use 'blacktext' in gnuplot or load the package
      color.sty in LaTeX.}%
    \renewcommand\color[2][]{}%
  }%
  \providecommand\includegraphics[2][]{%
    \GenericError{(gnuplot) \space\space\space\@spaces}{%
      Package graphicx or graphics not loaded%
    }{See the gnuplot documentation for explanation.%
    }{The gnuplot epslatex terminal needs graphicx.sty or graphics.sty.}%
    \renewcommand\includegraphics[2][]{}%
  }%
  \providecommand\rotatebox[2]{#2}%
  \@ifundefined{ifGPcolor}{%
    \newif\ifGPcolor
    \GPcolortrue
  }{}%
  \@ifundefined{ifGPblacktext}{%
    \newif\ifGPblacktext
    \GPblacktextfalse
  }{}%
  % define a \g@addto@macro without @ in the name:
  \let\gplgaddtomacro\g@addto@macro
  % define empty templates for all commands taking text:
  \gdef\gplbacktext{}%
  \gdef\gplfronttext{}%
  \makeatother
  \ifGPblacktext
    % no textcolor at all
    \def\colorrgb#1{}%
    \def\colorgray#1{}%
  \else
    % gray or color?
    \ifGPcolor
      \def\colorrgb#1{\color[rgb]{#1}}%
      \def\colorgray#1{\color[gray]{#1}}%
      \expandafter\def\csname LTw\endcsname{\color{white}}%
      \expandafter\def\csname LTb\endcsname{\color{black}}%
      \expandafter\def\csname LTa\endcsname{\color{black}}%
      \expandafter\def\csname LT0\endcsname{\color[rgb]{1,0,0}}%
      \expandafter\def\csname LT1\endcsname{\color[rgb]{0,1,0}}%
      \expandafter\def\csname LT2\endcsname{\color[rgb]{0,0,1}}%
      \expandafter\def\csname LT3\endcsname{\color[rgb]{1,0,1}}%
      \expandafter\def\csname LT4\endcsname{\color[rgb]{0,1,1}}%
      \expandafter\def\csname LT5\endcsname{\color[rgb]{1,1,0}}%
      \expandafter\def\csname LT6\endcsname{\color[rgb]{0,0,0}}%
      \expandafter\def\csname LT7\endcsname{\color[rgb]{1,0.3,0}}%
      \expandafter\def\csname LT8\endcsname{\color[rgb]{0.5,0.5,0.5}}%
    \else
      % gray
      \def\colorrgb#1{\color{black}}%
      \def\colorgray#1{\color[gray]{#1}}%
      \expandafter\def\csname LTw\endcsname{\color{white}}%
      \expandafter\def\csname LTb\endcsname{\color{black}}%
      \expandafter\def\csname LTa\endcsname{\color{black}}%
      \expandafter\def\csname LT0\endcsname{\color{black}}%
      \expandafter\def\csname LT1\endcsname{\color{black}}%
      \expandafter\def\csname LT2\endcsname{\color{black}}%
      \expandafter\def\csname LT3\endcsname{\color{black}}%
      \expandafter\def\csname LT4\endcsname{\color{black}}%
      \expandafter\def\csname LT5\endcsname{\color{black}}%
      \expandafter\def\csname LT6\endcsname{\color{black}}%
      \expandafter\def\csname LT7\endcsname{\color{black}}%
      \expandafter\def\csname LT8\endcsname{\color{black}}%
    \fi
  \fi
  \setlength{\unitlength}{0.0500bp}%
  \begin{picture}(6480.00,3600.00)%
    \gplgaddtomacro\gplbacktext{%
      \colorrgb{0.00,0.00,0.00}%
      \put(1078,767){\makebox(0,0)[r]{\strut{}0.995}}%
      \colorrgb{0.00,0.00,0.00}%
      \put(1078,1281){\makebox(0,0)[r]{\strut{}0.996}}%
      \colorrgb{0.00,0.00,0.00}%
      \put(1078,1794){\makebox(0,0)[r]{\strut{}0.997}}%
      \colorrgb{0.00,0.00,0.00}%
      \put(1078,2308){\makebox(0,0)[r]{\strut{}0.998}}%
      \colorrgb{0.00,0.00,0.00}%
      \put(1078,2821){\makebox(0,0)[r]{\strut{}0.999}}%
      \colorrgb{0.00,0.00,0.00}%
      \put(1078,3335){\makebox(0,0)[r]{\strut{}1.000}}%
      \colorrgb{0.00,0.00,0.00}%
      \put(1273,484){\makebox(0,0){\strut{} 10}}%
      \colorrgb{0.00,0.00,0.00}%
      \put(1807,484){\makebox(0,0){\strut{} 20}}%
      \colorrgb{0.00,0.00,0.00}%
      \put(2609,484){\makebox(0,0){\strut{} 35}}%
      \colorrgb{0.00,0.00,0.00}%
      \put(3411,484){\makebox(0,0){\strut{} 50}}%
      \colorrgb{0.00,0.00,0.00}%
      \put(4747,484){\makebox(0,0){\strut{} 75}}%
      \colorrgb{0.00,0.00,0.00}%
      \put(6083,484){\makebox(0,0){\strut{} 100}}%
      \csname LTb\endcsname%
      \put(176,2051){\rotatebox{-270}{\makebox(0,0){\strut{}$\rho$}}}%
      \put(3678,154){\makebox(0,0){\strut{}$L$}}%
    }%
    \gplgaddtomacro\gplfronttext{%
      \csname LTb\endcsname%
      \put(5084,1655){\makebox(0,0)[r]{\strut{}$r = \num{1.015}$}}%
      \csname LTb\endcsname%
      \put(5084,1325){\makebox(0,0)[r]{\strut{}$r = \num{1.045}$}}%
      \csname LTb\endcsname%
      \put(5084,995){\makebox(0,0)[r]{\strut{}$r = \num{1.075}$}}%
    }%
    \gplbacktext
    \put(0,0){\includegraphics{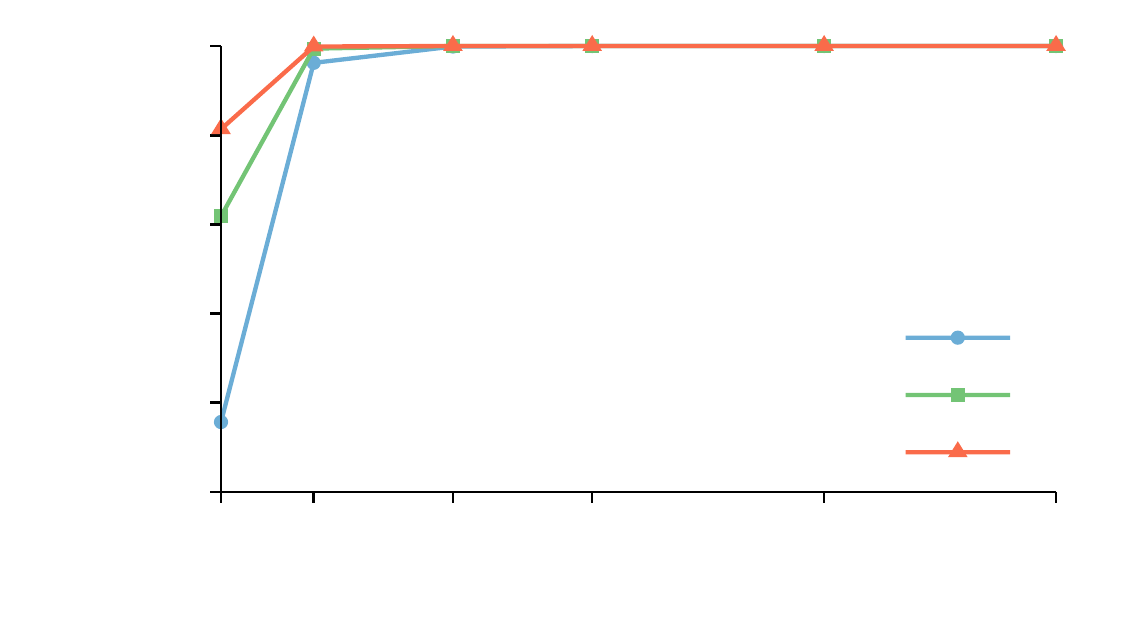}}%
    \gplfronttext
  \end{picture}%
\endgroup

%% file: error_lambda=10_L=50.tex
% GNUPLOT: LaTeX picture with Postscript
\begingroup
  \makeatletter
  \providecommand\color[2][]{%
    \GenericError{(gnuplot) \space\space\space\@spaces}{%
      Package color not loaded in conjunction with
      terminal option `colourtext'%
    }{See the gnuplot documentation for explanation.%
    }{Either use 'blacktext' in gnuplot or load the package
      color.sty in LaTeX.}%
    \renewcommand\color[2][]{}%
  }%
  \providecommand\includegraphics[2][]{%
    \GenericError{(gnuplot) \space\space\space\@spaces}{%
      Package graphicx or graphics not loaded%
    }{See the gnuplot documentation for explanation.%
    }{The gnuplot epslatex terminal needs graphicx.sty or graphics.sty.}%
    \renewcommand\includegraphics[2][]{}%
  }%
  \providecommand\rotatebox[2]{#2}%
  \@ifundefined{ifGPcolor}{%
    \newif\ifGPcolor
    \GPcolortrue
  }{}%
  \@ifundefined{ifGPblacktext}{%
    \newif\ifGPblacktext
    \GPblacktextfalse
  }{}%
  % define a \g@addto@macro without @ in the name:
  \let\gplgaddtomacro\g@addto@macro
  % define empty templates for all commands taking text:
  \gdef\gplbacktext{}%
  \gdef\gplfronttext{}%
  \makeatother
  \ifGPblacktext
    % no textcolor at all
    \def\colorrgb#1{}%
    \def\colorgray#1{}%
  \else
    % gray or color?
    \ifGPcolor
      \def\colorrgb#1{\color[rgb]{#1}}%
      \def\colorgray#1{\color[gray]{#1}}%
      \expandafter\def\csname LTw\endcsname{\color{white}}%
      \expandafter\def\csname LTb\endcsname{\color{black}}%
      \expandafter\def\csname LTa\endcsname{\color{black}}%
      \expandafter\def\csname LT0\endcsname{\color[rgb]{1,0,0}}%
      \expandafter\def\csname LT1\endcsname{\color[rgb]{0,1,0}}%
      \expandafter\def\csname LT2\endcsname{\color[rgb]{0,0,1}}%
      \expandafter\def\csname LT3\endcsname{\color[rgb]{1,0,1}}%
      \expandafter\def\csname LT4\endcsname{\color[rgb]{0,1,1}}%
      \expandafter\def\csname LT5\endcsname{\color[rgb]{1,1,0}}%
      \expandafter\def\csname LT6\endcsname{\color[rgb]{0,0,0}}%
      \expandafter\def\csname LT7\endcsname{\color[rgb]{1,0.3,0}}%
      \expandafter\def\csname LT8\endcsname{\color[rgb]{0.5,0.5,0.5}}%
    \else
      % gray
      \def\colorrgb#1{\color{black}}%
      \def\colorgray#1{\color[gray]{#1}}%
      \expandafter\def\csname LTw\endcsname{\color{white}}%
      \expandafter\def\csname LTb\endcsname{\color{black}}%
      \expandafter\def\csname LTa\endcsname{\color{black}}%
      \expandafter\def\csname LT0\endcsname{\color{black}}%
      \expandafter\def\csname LT1\endcsname{\color{black}}%
      \expandafter\def\csname LT2\endcsname{\color{black}}%
      \expandafter\def\csname LT3\endcsname{\color{black}}%
      \expandafter\def\csname LT4\endcsname{\color{black}}%
      \expandafter\def\csname LT5\endcsname{\color{black}}%
      \expandafter\def\csname LT6\endcsname{\color{black}}%
      \expandafter\def\csname LT7\endcsname{\color{black}}%
      \expandafter\def\csname LT8\endcsname{\color{black}}%
    \fi
  \fi
  \setlength{\unitlength}{0.0500bp}%
  \begin{picture}(6480.00,3600.00)%
    \gplgaddtomacro\gplbacktext{%
      \colorrgb{0.00,0.00,0.00}%
      \put(979,895){\makebox(0,0)[r]{\strut{}-0.09}}%
      \colorrgb{0.00,0.00,0.00}%
      \put(979,1281){\makebox(0,0)[r]{\strut{}-0.06}}%
      \colorrgb{0.00,0.00,0.00}%
      \put(979,1666){\makebox(0,0)[r]{\strut{}-0.03}}%
      \colorrgb{0.00,0.00,0.00}%
      \put(979,2051){\makebox(0,0)[r]{\strut{} 0}}%
      \colorrgb{0.00,0.00,0.00}%
      \put(979,2436){\makebox(0,0)[r]{\strut{} 0.03}}%
      \colorrgb{0.00,0.00,0.00}%
      \put(979,2821){\makebox(0,0)[r]{\strut{} 0.06}}%
      \colorrgb{0.00,0.00,0.00}%
      \put(979,3207){\makebox(0,0)[r]{\strut{} 0.09}}%
      \colorrgb{0.00,0.00,0.00}%
      \put(1419,484){\makebox(0,0){\strut{}-0.09}}%
      \colorrgb{0.00,0.00,0.00}%
      \put(2156,484){\makebox(0,0){\strut{}-0.06}}%
      \colorrgb{0.00,0.00,0.00}%
      \put(2892,484){\makebox(0,0){\strut{}-0.03}}%
      \colorrgb{0.00,0.00,0.00}%
      \put(3629,484){\makebox(0,0){\strut{} 0}}%
      \colorrgb{0.00,0.00,0.00}%
      \put(4365,484){\makebox(0,0){\strut{} 0.03}}%
      \colorrgb{0.00,0.00,0.00}%
      \put(5101,484){\makebox(0,0){\strut{} 0.06}}%
      \colorrgb{0.00,0.00,0.00}%
      \put(5838,484){\makebox(0,0){\strut{} 0.09}}%
      \csname LTb\endcsname%
      \put(176,2051){\rotatebox{-270}{\makebox(0,0){\strut{}$u_\mathrm{fit}$}}}%
      \put(3628,154){\makebox(0,0){\strut{}$u_\mathrm{ES}$}}%
    }%
    \gplgaddtomacro\gplfronttext{%
      \csname LTb\endcsname%
      \put(5084,1325){\makebox(0,0)[r]{\strut{}$u_\mathrm{ES} = u_\mathrm{fit}$}}%
      \csname LTb\endcsname%
      \put(5084,995){\makebox(0,0)[r]{\strut{}$\psi_\mathrm{EKH}$ ($\rho = \num{0.9904}$)}}%
    }%
    \gplbacktext
    \put(0,0){\includegraphics{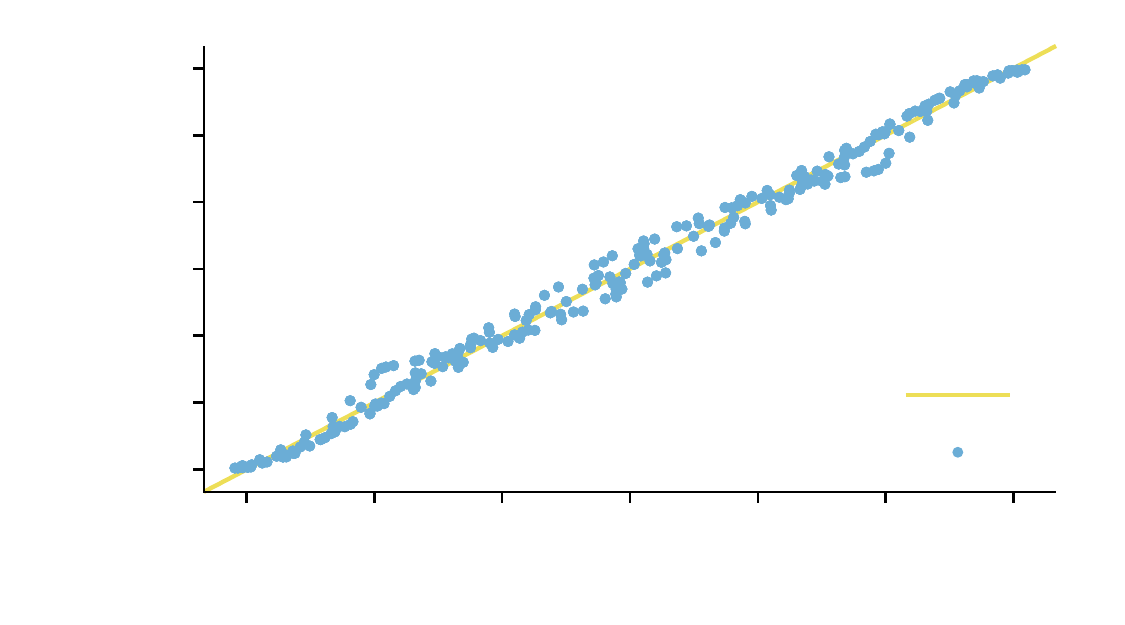}}%
    \gplfronttext
  \end{picture}%
\endgroup

%% file: error_lambda=01_L=50.tex
% GNUPLOT: LaTeX picture with Postscript
\begingroup
  \makeatletter
  \providecommand\color[2][]{%
    \GenericError{(gnuplot) \space\space\space\@spaces}{%
      Package color not loaded in conjunction with
      terminal option `colourtext'%
    }{See the gnuplot documentation for explanation.%
    }{Either use 'blacktext' in gnuplot or load the package
      color.sty in LaTeX.}%
    \renewcommand\color[2][]{}%
  }%
  \providecommand\includegraphics[2][]{%
    \GenericError{(gnuplot) \space\space\space\@spaces}{%
      Package graphicx or graphics not loaded%
    }{See the gnuplot documentation for explanation.%
    }{The gnuplot epslatex terminal needs graphicx.sty or graphics.sty.}%
    \renewcommand\includegraphics[2][]{}%
  }%
  \providecommand\rotatebox[2]{#2}%
  \@ifundefined{ifGPcolor}{%
    \newif\ifGPcolor
    \GPcolortrue
  }{}%
  \@ifundefined{ifGPblacktext}{%
    \newif\ifGPblacktext
    \GPblacktextfalse
  }{}%
  % define a \g@addto@macro without @ in the name:
  \let\gplgaddtomacro\g@addto@macro
  % define empty templates for all commands taking text:
  \gdef\gplbacktext{}%
  \gdef\gplfronttext{}%
  \makeatother
  \ifGPblacktext
    % no textcolor at all
    \def\colorrgb#1{}%
    \def\colorgray#1{}%
  \else
    % gray or color?
    \ifGPcolor
      \def\colorrgb#1{\color[rgb]{#1}}%
      \def\colorgray#1{\color[gray]{#1}}%
      \expandafter\def\csname LTw\endcsname{\color{white}}%
      \expandafter\def\csname LTb\endcsname{\color{black}}%
      \expandafter\def\csname LTa\endcsname{\color{black}}%
      \expandafter\def\csname LT0\endcsname{\color[rgb]{1,0,0}}%
      \expandafter\def\csname LT1\endcsname{\color[rgb]{0,1,0}}%
      \expandafter\def\csname LT2\endcsname{\color[rgb]{0,0,1}}%
      \expandafter\def\csname LT3\endcsname{\color[rgb]{1,0,1}}%
      \expandafter\def\csname LT4\endcsname{\color[rgb]{0,1,1}}%
      \expandafter\def\csname LT5\endcsname{\color[rgb]{1,1,0}}%
      \expandafter\def\csname LT6\endcsname{\color[rgb]{0,0,0}}%
      \expandafter\def\csname LT7\endcsname{\color[rgb]{1,0.3,0}}%
      \expandafter\def\csname LT8\endcsname{\color[rgb]{0.5,0.5,0.5}}%
    \else
      % gray
      \def\colorrgb#1{\color{black}}%
      \def\colorgray#1{\color[gray]{#1}}%
      \expandafter\def\csname LTw\endcsname{\color{white}}%
      \expandafter\def\csname LTb\endcsname{\color{black}}%
      \expandafter\def\csname LTa\endcsname{\color{black}}%
      \expandafter\def\csname LT0\endcsname{\color{black}}%
      \expandafter\def\csname LT1\endcsname{\color{black}}%
      \expandafter\def\csname LT2\endcsname{\color{black}}%
      \expandafter\def\csname LT3\endcsname{\color{black}}%
      \expandafter\def\csname LT4\endcsname{\color{black}}%
      \expandafter\def\csname LT5\endcsname{\color{black}}%
      \expandafter\def\csname LT6\endcsname{\color{black}}%
      \expandafter\def\csname LT7\endcsname{\color{black}}%
      \expandafter\def\csname LT8\endcsname{\color{black}}%
    \fi
  \fi
  \setlength{\unitlength}{0.0500bp}%
  \begin{picture}(6480.00,3600.00)%
    \gplgaddtomacro\gplbacktext{%
      \colorrgb{0.00,0.00,0.00}%
      \put(1111,895){\makebox(0,0)[r]{\strut{}-0.009}}%
      \colorrgb{0.00,0.00,0.00}%
      \put(1111,1281){\makebox(0,0)[r]{\strut{}-0.006}}%
      \colorrgb{0.00,0.00,0.00}%
      \put(1111,1666){\makebox(0,0)[r]{\strut{}-0.003}}%
      \colorrgb{0.00,0.00,0.00}%
      \put(1111,2051){\makebox(0,0)[r]{\strut{} 0}}%
      \colorrgb{0.00,0.00,0.00}%
      \put(1111,2436){\makebox(0,0)[r]{\strut{} 0.003}}%
      \colorrgb{0.00,0.00,0.00}%
      \put(1111,2821){\makebox(0,0)[r]{\strut{} 0.006}}%
      \colorrgb{0.00,0.00,0.00}%
      \put(1111,3207){\makebox(0,0)[r]{\strut{} 0.009}}%
      \colorrgb{0.00,0.00,0.00}%
      \put(1545,484){\makebox(0,0){\strut{}-0.009}}%
      \colorrgb{0.00,0.00,0.00}%
      \put(2261,484){\makebox(0,0){\strut{}-0.006}}%
      \colorrgb{0.00,0.00,0.00}%
      \put(2978,484){\makebox(0,0){\strut{}-0.003}}%
      \colorrgb{0.00,0.00,0.00}%
      \put(3694,484){\makebox(0,0){\strut{} 0}}%
      \colorrgb{0.00,0.00,0.00}%
      \put(4411,484){\makebox(0,0){\strut{} 0.003}}%
      \colorrgb{0.00,0.00,0.00}%
      \put(5128,484){\makebox(0,0){\strut{} 0.006}}%
      \colorrgb{0.00,0.00,0.00}%
      \put(5844,484){\makebox(0,0){\strut{} 0.009}}%
      \csname LTb\endcsname%
      \put(176,2051){\rotatebox{-270}{\makebox(0,0){\strut{}$u_\mathrm{fit}$}}}%
      \put(3694,154){\makebox(0,0){\strut{}$u_\mathrm{ES}$}}%
    }%
    \gplgaddtomacro\gplfronttext{%
      \csname LTb\endcsname%
      \put(5084,1325){\makebox(0,0)[r]{\strut{}$u_\mathrm{ES} = u_\mathrm{fit}$}}%
      \csname LTb\endcsname%
      \put(5084,995){\makebox(0,0)[r]{\strut{}$\psi_\mathrm{EKH}$ ($\rho = \num{0.9435}$)}}%
    }%
    \gplbacktext
    \put(0,0){\includegraphics{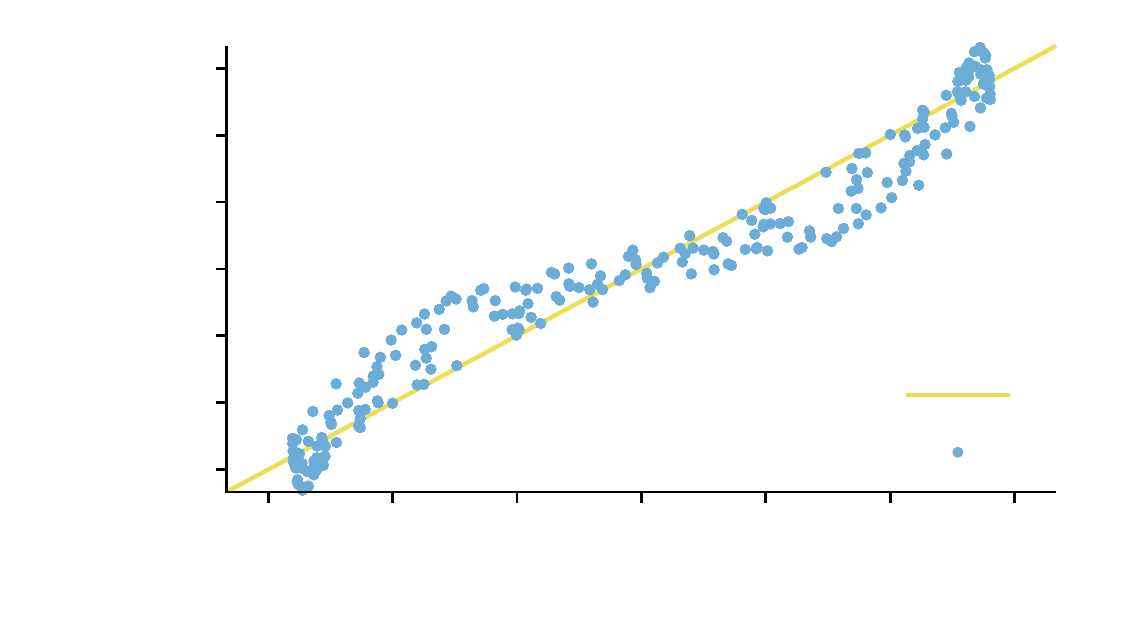}}%
    \gplfronttext
  \end{picture}%
\endgroup

%% file: convergence_tensors.tex
% GNUPLOT: LaTeX picture with Postscript
\begingroup
  \makeatletter
  \providecommand\color[2][]{%
    \GenericError{(gnuplot) \space\space\space\@spaces}{%
      Package color not loaded in conjunction with
      terminal option `colourtext'%
    }{See the gnuplot documentation for explanation.%
    }{Either use 'blacktext' in gnuplot or load the package
      color.sty in LaTeX.}%
    \renewcommand\color[2][]{}%
  }%
  \providecommand\includegraphics[2][]{%
    \GenericError{(gnuplot) \space\space\space\@spaces}{%
      Package graphicx or graphics not loaded%
    }{See the gnuplot documentation for explanation.%
    }{The gnuplot epslatex terminal needs graphicx.sty or graphics.sty.}%
    \renewcommand\includegraphics[2][]{}%
  }%
  \providecommand\rotatebox[2]{#2}%
  \@ifundefined{ifGPcolor}{%
    \newif\ifGPcolor
    \GPcolortrue
  }{}%
  \@ifundefined{ifGPblacktext}{%
    \newif\ifGPblacktext
    \GPblacktextfalse
  }{}%
  % define a \g@addto@macro without @ in the name:
  \let\gplgaddtomacro\g@addto@macro
  % define empty templates for all commands taking text:
  \gdef\gplbacktext{}%
  \gdef\gplfronttext{}%
  \makeatother
  \ifGPblacktext
    % no textcolor at all
    \def\colorrgb#1{}%
    \def\colorgray#1{}%
  \else
    % gray or color?
    \ifGPcolor
      \def\colorrgb#1{\color[rgb]{#1}}%
      \def\colorgray#1{\color[gray]{#1}}%
      \expandafter\def\csname LTw\endcsname{\color{white}}%
      \expandafter\def\csname LTb\endcsname{\color{black}}%
      \expandafter\def\csname LTa\endcsname{\color{black}}%
      \expandafter\def\csname LT0\endcsname{\color[rgb]{1,0,0}}%
      \expandafter\def\csname LT1\endcsname{\color[rgb]{0,1,0}}%
      \expandafter\def\csname LT2\endcsname{\color[rgb]{0,0,1}}%
      \expandafter\def\csname LT3\endcsname{\color[rgb]{1,0,1}}%
      \expandafter\def\csname LT4\endcsname{\color[rgb]{0,1,1}}%
      \expandafter\def\csname LT5\endcsname{\color[rgb]{1,1,0}}%
      \expandafter\def\csname LT6\endcsname{\color[rgb]{0,0,0}}%
      \expandafter\def\csname LT7\endcsname{\color[rgb]{1,0.3,0}}%
      \expandafter\def\csname LT8\endcsname{\color[rgb]{0.5,0.5,0.5}}%
    \else
      % gray
      \def\colorrgb#1{\color{black}}%
      \def\colorgray#1{\color[gray]{#1}}%
      \expandafter\def\csname LTw\endcsname{\color{white}}%
      \expandafter\def\csname LTb\endcsname{\color{black}}%
      \expandafter\def\csname LTa\endcsname{\color{black}}%
      \expandafter\def\csname LT0\endcsname{\color{black}}%
      \expandafter\def\csname LT1\endcsname{\color{black}}%
      \expandafter\def\csname LT2\endcsname{\color{black}}%
      \expandafter\def\csname LT3\endcsname{\color{black}}%
      \expandafter\def\csname LT4\endcsname{\color{black}}%
      \expandafter\def\csname LT5\endcsname{\color{black}}%
      \expandafter\def\csname LT6\endcsname{\color{black}}%
      \expandafter\def\csname LT7\endcsname{\color{black}}%
      \expandafter\def\csname LT8\endcsname{\color{black}}%
    \fi
  \fi
  \setlength{\unitlength}{0.0500bp}%
  \begin{picture}(6480.00,3600.00)%
    \gplgaddtomacro\gplbacktext{%
      \colorrgb{0.00,0.00,0.00}%
      \put(946,767){\makebox(0,0)[r]{\strut{}0.94}}%
      \colorrgb{0.00,0.00,0.00}%
      \put(946,1195){\makebox(0,0)[r]{\strut{}0.95}}%
      \colorrgb{0.00,0.00,0.00}%
      \put(946,1623){\makebox(0,0)[r]{\strut{}0.96}}%
      \colorrgb{0.00,0.00,0.00}%
      \put(946,2051){\makebox(0,0)[r]{\strut{}0.97}}%
      \colorrgb{0.00,0.00,0.00}%
      \put(946,2479){\makebox(0,0)[r]{\strut{}0.98}}%
      \colorrgb{0.00,0.00,0.00}%
      \put(946,2907){\makebox(0,0)[r]{\strut{}0.99}}%
      \colorrgb{0.00,0.00,0.00}%
      \put(946,3335){\makebox(0,0)[r]{\strut{}1.00}}%
      \colorrgb{0.00,0.00,0.00}%
      \put(1141,484){\makebox(0,0){\strut{}1}}%
      \colorrgb{0.00,0.00,0.00}%
      \put(1965,484){\makebox(0,0){\strut{}2}}%
      \colorrgb{0.00,0.00,0.00}%
      \put(2788,484){\makebox(0,0){\strut{}3}}%
      \colorrgb{0.00,0.00,0.00}%
      \put(3612,484){\makebox(0,0){\strut{}4}}%
      \colorrgb{0.00,0.00,0.00}%
      \put(4436,484){\makebox(0,0){\strut{}5}}%
      \colorrgb{0.00,0.00,0.00}%
      \put(5259,484){\makebox(0,0){\strut{}6}}%
      \colorrgb{0.00,0.00,0.00}%
      \put(6083,484){\makebox(0,0){\strut{}7}}%
      \csname LTb\endcsname%
      \put(176,2051){\rotatebox{-270}{\makebox(0,0){\strut{}$\rho$}}}%
      \put(3612,154){\makebox(0,0){\strut{}$B$}}%
    }%
    \gplgaddtomacro\gplfronttext{%
    }%
    \gplbacktext
    \put(0,0){\includegraphics{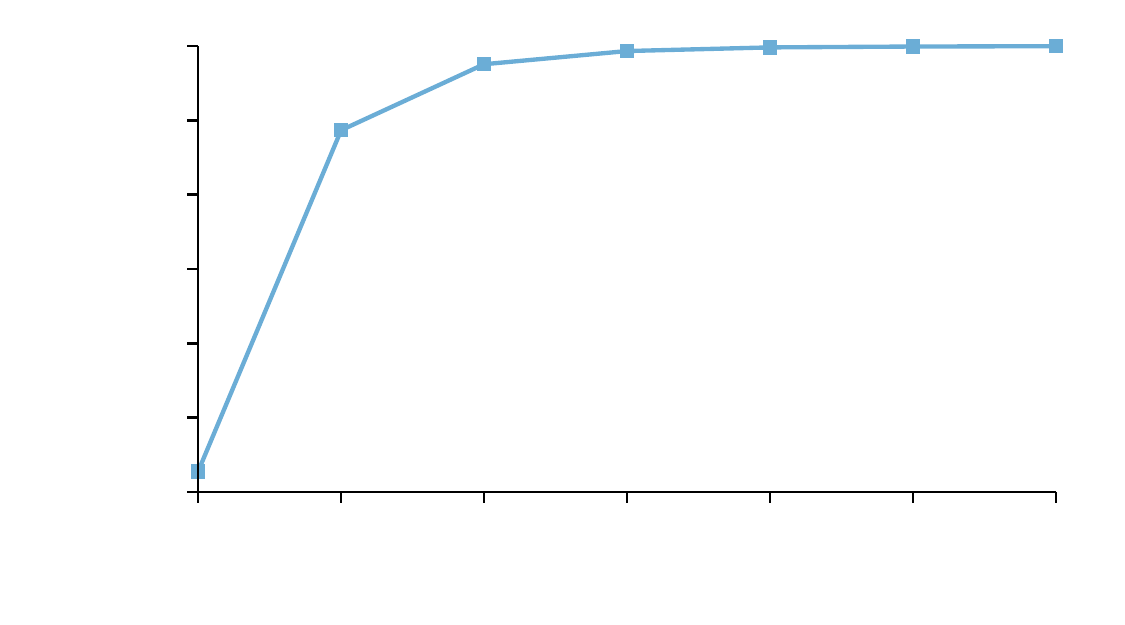}}%
    \gplfronttext
  \end{picture}%
\endgroup

%% file: error_lambda=01_N=5.tex
% GNUPLOT: LaTeX picture with Postscript
\begingroup
  \makeatletter
  \providecommand\color[2][]{%
    \GenericError{(gnuplot) \space\space\space\@spaces}{%
      Package color not loaded in conjunction with
      terminal option `colourtext'%
    }{See the gnuplot documentation for explanation.%
    }{Either use 'blacktext' in gnuplot or load the package
      color.sty in LaTeX.}%
    \renewcommand\color[2][]{}%
  }%
  \providecommand\includegraphics[2][]{%
    \GenericError{(gnuplot) \space\space\space\@spaces}{%
      Package graphicx or graphics not loaded%
    }{See the gnuplot documentation for explanation.%
    }{The gnuplot epslatex terminal needs graphicx.sty or graphics.sty.}%
    \renewcommand\includegraphics[2][]{}%
  }%
  \providecommand\rotatebox[2]{#2}%
  \@ifundefined{ifGPcolor}{%
    \newif\ifGPcolor
    \GPcolortrue
  }{}%
  \@ifundefined{ifGPblacktext}{%
    \newif\ifGPblacktext
    \GPblacktextfalse
  }{}%
  % define a \g@addto@macro without @ in the name:
  \let\gplgaddtomacro\g@addto@macro
  % define empty templates for all commands taking text:
  \gdef\gplbacktext{}%
  \gdef\gplfronttext{}%
  \makeatother
  \ifGPblacktext
    % no textcolor at all
    \def\colorrgb#1{}%
    \def\colorgray#1{}%
  \else
    % gray or color?
    \ifGPcolor
      \def\colorrgb#1{\color[rgb]{#1}}%
      \def\colorgray#1{\color[gray]{#1}}%
      \expandafter\def\csname LTw\endcsname{\color{white}}%
      \expandafter\def\csname LTb\endcsname{\color{black}}%
      \expandafter\def\csname LTa\endcsname{\color{black}}%
      \expandafter\def\csname LT0\endcsname{\color[rgb]{1,0,0}}%
      \expandafter\def\csname LT1\endcsname{\color[rgb]{0,1,0}}%
      \expandafter\def\csname LT2\endcsname{\color[rgb]{0,0,1}}%
      \expandafter\def\csname LT3\endcsname{\color[rgb]{1,0,1}}%
      \expandafter\def\csname LT4\endcsname{\color[rgb]{0,1,1}}%
      \expandafter\def\csname LT5\endcsname{\color[rgb]{1,1,0}}%
      \expandafter\def\csname LT6\endcsname{\color[rgb]{0,0,0}}%
      \expandafter\def\csname LT7\endcsname{\color[rgb]{1,0.3,0}}%
      \expandafter\def\csname LT8\endcsname{\color[rgb]{0.5,0.5,0.5}}%
    \else
      % gray
      \def\colorrgb#1{\color{black}}%
      \def\colorgray#1{\color[gray]{#1}}%
      \expandafter\def\csname LTw\endcsname{\color{white}}%
      \expandafter\def\csname LTb\endcsname{\color{black}}%
      \expandafter\def\csname LTa\endcsname{\color{black}}%
      \expandafter\def\csname LT0\endcsname{\color{black}}%
      \expandafter\def\csname LT1\endcsname{\color{black}}%
      \expandafter\def\csname LT2\endcsname{\color{black}}%
      \expandafter\def\csname LT3\endcsname{\color{black}}%
      \expandafter\def\csname LT4\endcsname{\color{black}}%
      \expandafter\def\csname LT5\endcsname{\color{black}}%
      \expandafter\def\csname LT6\endcsname{\color{black}}%
      \expandafter\def\csname LT7\endcsname{\color{black}}%
      \expandafter\def\csname LT8\endcsname{\color{black}}%
    \fi
  \fi
  \setlength{\unitlength}{0.0500bp}%
  \begin{picture}(6480.00,3600.00)%
    \gplgaddtomacro\gplbacktext{%
      \colorrgb{0.00,0.00,0.00}%
      \put(1111,854){\makebox(0,0)[r]{\strut{}-0.009}}%
      \colorrgb{0.00,0.00,0.00}%
      \put(1111,1245){\makebox(0,0)[r]{\strut{}-0.006}}%
      \colorrgb{0.00,0.00,0.00}%
      \put(1111,1637){\makebox(0,0)[r]{\strut{}-0.003}}%
      \colorrgb{0.00,0.00,0.00}%
      \put(1111,2029){\makebox(0,0)[r]{\strut{} 0}}%
      \colorrgb{0.00,0.00,0.00}%
      \put(1111,2421){\makebox(0,0)[r]{\strut{} 0.003}}%
      \colorrgb{0.00,0.00,0.00}%
      \put(1111,2813){\makebox(0,0)[r]{\strut{} 0.006}}%
      \colorrgb{0.00,0.00,0.00}%
      \put(1111,3204){\makebox(0,0)[r]{\strut{} 0.009}}%
      \colorrgb{0.00,0.00,0.00}%
      \put(1545,440){\makebox(0,0){\strut{}-0.009}}%
      \colorrgb{0.00,0.00,0.00}%
      \put(2261,440){\makebox(0,0){\strut{}-0.006}}%
      \colorrgb{0.00,0.00,0.00}%
      \put(2978,440){\makebox(0,0){\strut{}-0.003}}%
      \colorrgb{0.00,0.00,0.00}%
      \put(3694,440){\makebox(0,0){\strut{} 0}}%
      \colorrgb{0.00,0.00,0.00}%
      \put(4411,440){\makebox(0,0){\strut{} 0.003}}%
      \colorrgb{0.00,0.00,0.00}%
      \put(5128,440){\makebox(0,0){\strut{} 0.006}}%
      \colorrgb{0.00,0.00,0.00}%
      \put(5844,440){\makebox(0,0){\strut{} 0.009}}%
      \csname LTb\endcsname%
      \put(176,2029){\rotatebox{-270}{\makebox(0,0){\strut{}$u_\mathrm{fit}$}}}%
      \put(3694,154){\makebox(0,0){\strut{}$u_\mathrm{ES}$}}%
    }%
    \gplgaddtomacro\gplfronttext{%
      \csname LTb\endcsname%
      \put(5084,1787){\makebox(0,0)[r]{\strut{}$u_\mathrm{ES} = u_\mathrm{fit}$}}%
      \csname LTb\endcsname%
      \put(5084,1501){\makebox(0,0)[r]{\strut{}$B = \num{3}$}}%
      \csname LTb\endcsname%
      \put(5084,1215){\makebox(0,0)[r]{\strut{}$B = \num{5}$}}%
      \csname LTb\endcsname%
      \put(5084,929){\makebox(0,0)[r]{\strut{}$B = \num{7}$}}%
    }%
    \gplbacktext
    \put(0,0){\includegraphics{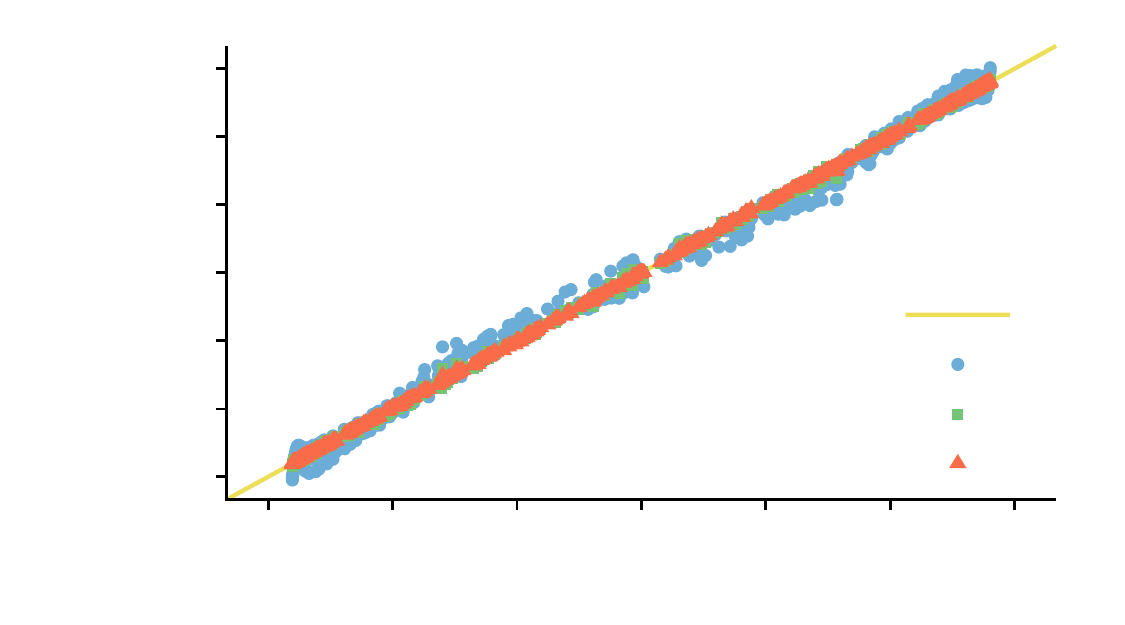}}%
    \gplfronttext
  \end{picture}%
\endgroup

%% file: landscape_EKH_L=50_K=30_10_lambda=01_N=5_r=1015_theta1=157.tex
% GNUPLOT: LaTeX picture with Postscript
\begingroup
  \makeatletter
  \providecommand\color[2][]{%
    \GenericError{(gnuplot) \space\space\space\@spaces}{%
      Package color not loaded in conjunction with
      terminal option `colourtext'%
    }{See the gnuplot documentation for explanation.%
    }{Either use 'blacktext' in gnuplot or load the package
      color.sty in LaTeX.}%
    \renewcommand\color[2][]{}%
  }%
  \providecommand\includegraphics[2][]{%
    \GenericError{(gnuplot) \space\space\space\@spaces}{%
      Package graphicx or graphics not loaded%
    }{See the gnuplot documentation for explanation.%
    }{The gnuplot epslatex terminal needs graphicx.sty or graphics.sty.}%
    \renewcommand\includegraphics[2][]{}%
  }%
  \providecommand\rotatebox[2]{#2}%
  \@ifundefined{ifGPcolor}{%
    \newif\ifGPcolor
    \GPcolortrue
  }{}%
  \@ifundefined{ifGPblacktext}{%
    \newif\ifGPblacktext
    \GPblacktextfalse
  }{}%
  % define a \g@addto@macro without @ in the name:
  \let\gplgaddtomacro\g@addto@macro
  % define empty templates for all commands taking text:
  \gdef\gplbacktext{}%
  \gdef\gplfronttext{}%
  \makeatother
  \ifGPblacktext
    % no textcolor at all
    \def\colorrgb#1{}%
    \def\colorgray#1{}%
  \else
    % gray or color?
    \ifGPcolor
      \def\colorrgb#1{\color[rgb]{#1}}%
      \def\colorgray#1{\color[gray]{#1}}%
      \expandafter\def\csname LTw\endcsname{\color{white}}%
      \expandafter\def\csname LTb\endcsname{\color{black}}%
      \expandafter\def\csname LTa\endcsname{\color{black}}%
      \expandafter\def\csname LT0\endcsname{\color[rgb]{1,0,0}}%
      \expandafter\def\csname LT1\endcsname{\color[rgb]{0,1,0}}%
      \expandafter\def\csname LT2\endcsname{\color[rgb]{0,0,1}}%
      \expandafter\def\csname LT3\endcsname{\color[rgb]{1,0,1}}%
      \expandafter\def\csname LT4\endcsname{\color[rgb]{0,1,1}}%
      \expandafter\def\csname LT5\endcsname{\color[rgb]{1,1,0}}%
      \expandafter\def\csname LT6\endcsname{\color[rgb]{0,0,0}}%
      \expandafter\def\csname LT7\endcsname{\color[rgb]{1,0.3,0}}%
      \expandafter\def\csname LT8\endcsname{\color[rgb]{0.5,0.5,0.5}}%
    \else
      % gray
      \def\colorrgb#1{\color{black}}%
      \def\colorgray#1{\color[gray]{#1}}%
      \expandafter\def\csname LTw\endcsname{\color{white}}%
      \expandafter\def\csname LTb\endcsname{\color{black}}%
      \expandafter\def\csname LTa\endcsname{\color{black}}%
      \expandafter\def\csname LT0\endcsname{\color{black}}%
      \expandafter\def\csname LT1\endcsname{\color{black}}%
      \expandafter\def\csname LT2\endcsname{\color{black}}%
      \expandafter\def\csname LT3\endcsname{\color{black}}%
      \expandafter\def\csname LT4\endcsname{\color{black}}%
      \expandafter\def\csname LT5\endcsname{\color{black}}%
      \expandafter\def\csname LT6\endcsname{\color{black}}%
      \expandafter\def\csname LT7\endcsname{\color{black}}%
      \expandafter\def\csname LT8\endcsname{\color{black}}%
    \fi
  \fi
  \setlength{\unitlength}{0.0500bp}%
  \begin{picture}(6480.00,5760.00)%
    \gplgaddtomacro\gplbacktext{%
      \colorrgb{0.00,0.00,0.00}%
      \put(770,1955){\makebox(0,0)[r]{\strut{}0}}%
      \colorrgb{0.00,0.00,0.00}%
      \put(1305,1578){\makebox(0,0)[r]{\strut{}$\frac{1}{2}\pi$}}%
      \colorrgb{0.00,0.00,0.00}%
      \put(1840,1200){\makebox(0,0)[r]{\strut{}$\pi$}}%
      \colorrgb{0.00,0.00,0.00}%
      \put(2375,823){\makebox(0,0)[r]{\strut{}$\frac{3}{2}\pi$}}%
      \colorrgb{0.00,0.00,0.00}%
      \put(2910,445){\makebox(0,0)[r]{\strut{}$2\pi$}}%
      \colorrgb{0.00,0.00,0.00}%
      \put(3240,446){\makebox(0,0){\strut{}0}}%
      \colorrgb{0.00,0.00,0.00}%
      \put(3876,762){\makebox(0,0){\strut{}$\frac{1}{2}\pi$}}%
      \colorrgb{0.00,0.00,0.00}%
      \put(4514,1079){\makebox(0,0){\strut{}$\pi$}}%
      \colorrgb{0.00,0.00,0.00}%
      \put(5151,1396){\makebox(0,0){\strut{}$\frac{3}{2}\pi$}}%
      \colorrgb{0.00,0.00,0.00}%
      \put(5789,1712){\makebox(0,0){\strut{}$2\pi$}}%
      \colorrgb{0.00,0.00,0.00}%
      \put(706,2149){\makebox(0,0)[r]{\strut{}-0.6}}%
      \colorrgb{0.00,0.00,0.00}%
      \put(706,2641){\makebox(0,0)[r]{\strut{}-0.3}}%
      \colorrgb{0.00,0.00,0.00}%
      \put(706,3133){\makebox(0,0)[r]{\strut{} 0}}%
      \colorrgb{0.00,0.00,0.00}%
      \put(706,3626){\makebox(0,0)[r]{\strut{} 0.3}}%
      \colorrgb{0.00,0.00,0.00}%
      \put(706,4118){\makebox(0,0)[r]{\strut{} 0.6}}%
      \csname LTb\endcsname%
      \put(-3,3111){\rotatebox{90}{\makebox(0,0){\strut{}$u/u_{max}$}}}%
    }%
    \gplgaddtomacro\gplfronttext{%
      \csname LTb\endcsname%
      \put(1296,1026){\makebox(0,0){\strut{}$\varphi_2$}}%
      \put(4991,844){\makebox(0,0){\strut{}$\vartheta_2$}}%
      \put(-3,3111){\rotatebox{90}{\makebox(0,0){\strut{}$u/u_{max}$}}}%
      \colorrgb{0.00,0.00,0.00}%
      \put(5961,3477){\makebox(0,0)[l]{\strut{} 0}}%
      \colorrgb{0.00,0.00,0.00}%
      \put(5961,2580){\makebox(0,0)[l]{\strut{}-0.6}}%
      \colorrgb{0.00,0.00,0.00}%
      \put(5961,2879){\makebox(0,0)[l]{\strut{}-0.4}}%
      \colorrgb{0.00,0.00,0.00}%
      \put(5961,3178){\makebox(0,0)[l]{\strut{}-0.2}}%
      \colorrgb{0.00,0.00,0.00}%
      \put(5961,3776){\makebox(0,0)[l]{\strut{} 0.2}}%
      \colorrgb{0.00,0.00,0.00}%
      \put(5961,4075){\makebox(0,0)[l]{\strut{} 0.4}}%
      \colorrgb{0.00,0.00,0.00}%
      \put(5961,4375){\makebox(0,0)[l]{\strut{} 0.6}}%
    }%
    \gplbacktext
    \put(0,0){\includegraphics{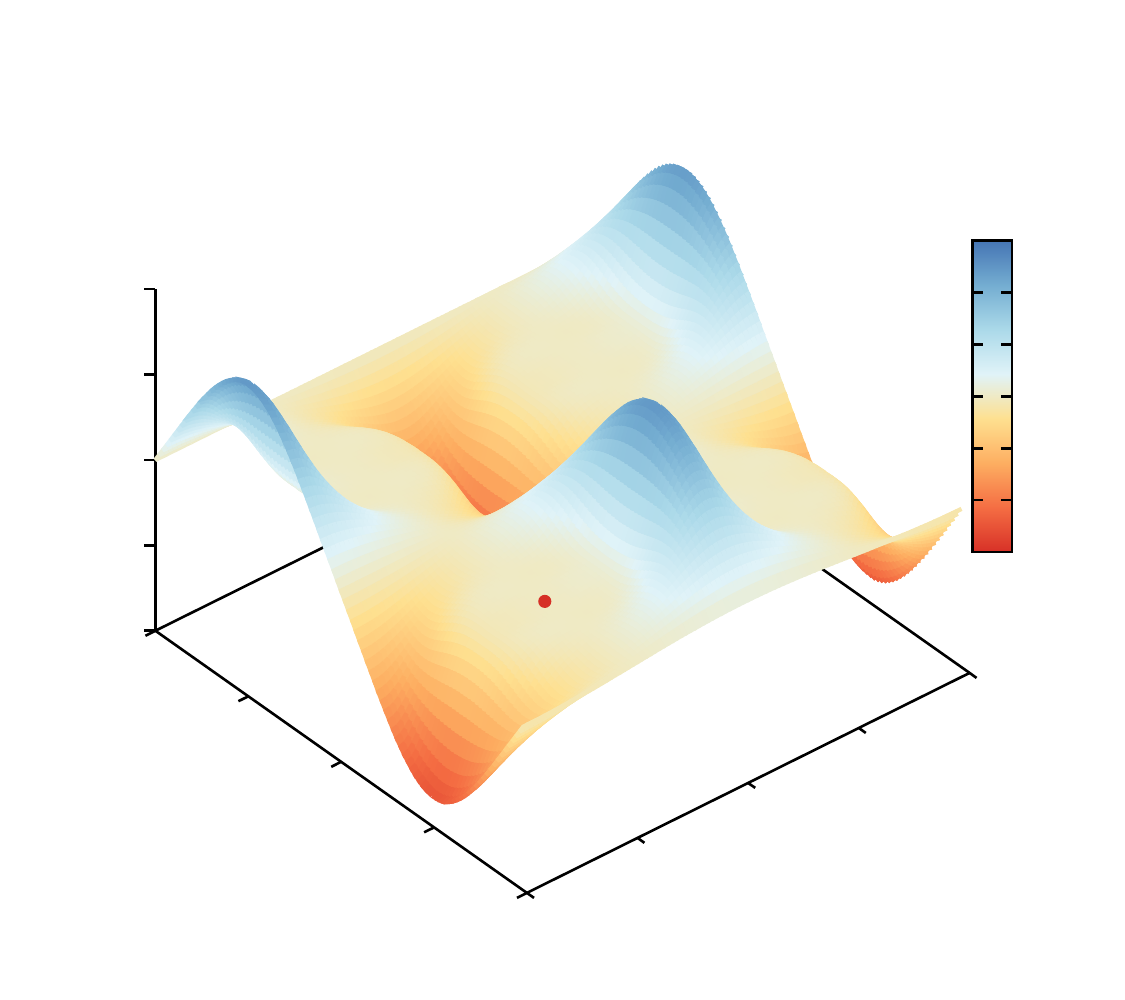}}%
    \gplfronttext
  \end{picture}%
\endgroup

%% file: convergence_angular.tex
% GNUPLOT: LaTeX picture with Postscript
\begingroup
  \makeatletter
  \providecommand\color[2][]{%
    \GenericError{(gnuplot) \space\space\space\@spaces}{%
      Package color not loaded in conjunction with
      terminal option `colourtext'%
    }{See the gnuplot documentation for explanation.%
    }{Either use 'blacktext' in gnuplot or load the package
      color.sty in LaTeX.}%
    \renewcommand\color[2][]{}%
  }%
  \providecommand\includegraphics[2][]{%
    \GenericError{(gnuplot) \space\space\space\@spaces}{%
      Package graphicx or graphics not loaded%
    }{See the gnuplot documentation for explanation.%
    }{The gnuplot epslatex terminal needs graphicx.sty or graphics.sty.}%
    \renewcommand\includegraphics[2][]{}%
  }%
  \providecommand\rotatebox[2]{#2}%
  \@ifundefined{ifGPcolor}{%
    \newif\ifGPcolor
    \GPcolortrue
  }{}%
  \@ifundefined{ifGPblacktext}{%
    \newif\ifGPblacktext
    \GPblacktextfalse
  }{}%
  % define a \g@addto@macro without @ in the name:
  \let\gplgaddtomacro\g@addto@macro
  % define empty templates for all commands taking text:
  \gdef\gplbacktext{}%
  \gdef\gplfronttext{}%
  \makeatother
  \ifGPblacktext
    % no textcolor at all
    \def\colorrgb#1{}%
    \def\colorgray#1{}%
  \else
    % gray or color?
    \ifGPcolor
      \def\colorrgb#1{\color[rgb]{#1}}%
      \def\colorgray#1{\color[gray]{#1}}%
      \expandafter\def\csname LTw\endcsname{\color{white}}%
      \expandafter\def\csname LTb\endcsname{\color{black}}%
      \expandafter\def\csname LTa\endcsname{\color{black}}%
      \expandafter\def\csname LT0\endcsname{\color[rgb]{1,0,0}}%
      \expandafter\def\csname LT1\endcsname{\color[rgb]{0,1,0}}%
      \expandafter\def\csname LT2\endcsname{\color[rgb]{0,0,1}}%
      \expandafter\def\csname LT3\endcsname{\color[rgb]{1,0,1}}%
      \expandafter\def\csname LT4\endcsname{\color[rgb]{0,1,1}}%
      \expandafter\def\csname LT5\endcsname{\color[rgb]{1,1,0}}%
      \expandafter\def\csname LT6\endcsname{\color[rgb]{0,0,0}}%
      \expandafter\def\csname LT7\endcsname{\color[rgb]{1,0.3,0}}%
      \expandafter\def\csname LT8\endcsname{\color[rgb]{0.5,0.5,0.5}}%
    \else
      % gray
      \def\colorrgb#1{\color{black}}%
      \def\colorgray#1{\color[gray]{#1}}%
      \expandafter\def\csname LTw\endcsname{\color{white}}%
      \expandafter\def\csname LTb\endcsname{\color{black}}%
      \expandafter\def\csname LTa\endcsname{\color{black}}%
      \expandafter\def\csname LT0\endcsname{\color{black}}%
      \expandafter\def\csname LT1\endcsname{\color{black}}%
      \expandafter\def\csname LT2\endcsname{\color{black}}%
      \expandafter\def\csname LT3\endcsname{\color{black}}%
      \expandafter\def\csname LT4\endcsname{\color{black}}%
      \expandafter\def\csname LT5\endcsname{\color{black}}%
      \expandafter\def\csname LT6\endcsname{\color{black}}%
      \expandafter\def\csname LT7\endcsname{\color{black}}%
      \expandafter\def\csname LT8\endcsname{\color{black}}%
    \fi
  \fi
  \setlength{\unitlength}{0.0500bp}%
  \begin{picture}(6480.00,3600.00)%
    \gplgaddtomacro\gplbacktext{%
      \colorrgb{0.00,0.00,0.00}%
      \put(1078,3335){\makebox(0,0)[r]{\strut{}0}}%
      \colorrgb{0.00,0.00,0.00}%
      \put(1078,965){\makebox(0,0)[r]{\strut{}-0.012}}%
      \colorrgb{0.00,0.00,0.00}%
      \put(1078,1557){\makebox(0,0)[r]{\strut{}-0.009}}%
      \colorrgb{0.00,0.00,0.00}%
      \put(1078,2150){\makebox(0,0)[r]{\strut{}-0.006}}%
      \colorrgb{0.00,0.00,0.00}%
      \put(1078,2742){\makebox(0,0)[r]{\strut{}-0.003}}%
      \colorrgb{0.00,0.00,0.00}%
      \put(1273,484){\makebox(0,0){\strut{}0}}%
      \colorrgb{0.00,0.00,0.00}%
      \put(2476,484){\makebox(0,0){\strut{}$\frac{1}{8}\pi$}}%
      \colorrgb{0.00,0.00,0.00}%
      \put(3678,484){\makebox(0,0){\strut{}$\frac{1}{4}\pi$}}%
      \colorrgb{0.00,0.00,0.00}%
      \put(4881,484){\makebox(0,0){\strut{}$\frac{3}{8}\pi$}}%
      \colorrgb{0.00,0.00,0.00}%
      \put(6083,484){\makebox(0,0){\strut{}$\frac{1}{2}\pi$}}%
      \csname LTb\endcsname%
      \put(176,2051){\rotatebox{-270}{\makebox(0,0){\strut{}$u$}}}%
      \put(3678,154){\makebox(0,0){\strut{}$\vartheta$}}%
    }%
    \gplgaddtomacro\gplfronttext{%
      \csname LTb\endcsname%
      \put(3412,3192){\makebox(0,0)[r]{\strut{}YU, $K = 150$}}%
      \csname LTb\endcsname%
      \put(3412,2906){\makebox(0,0)[r]{\strut{}$B = 1$, $K = 30$}}%
      \csname LTb\endcsname%
      \put(3412,2620){\makebox(0,0)[r]{\strut{}$B = 5$, $K = 30$}}%
    }%
    \gplbacktext
    \put(0,0){\includegraphics{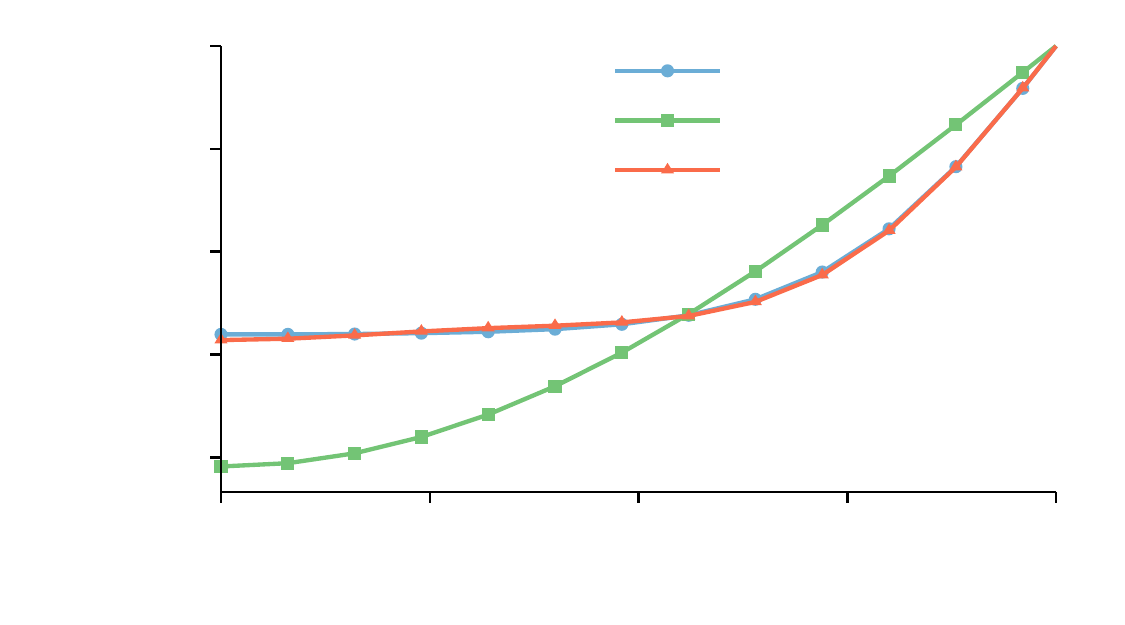}}%
    \gplfronttext
  \end{picture}%
\endgroup

%% file: DLVO_deviation.tex
% GNUPLOT: LaTeX picture with Postscript
\begingroup
  \makeatletter
  \providecommand\color[2][]{%
    \GenericError{(gnuplot) \space\space\space\@spaces}{%
      Package color not loaded in conjunction with
      terminal option `colourtext'%
    }{See the gnuplot documentation for explanation.%
    }{Either use 'blacktext' in gnuplot or load the package
      color.sty in LaTeX.}%
    \renewcommand\color[2][]{}%
  }%
  \providecommand\includegraphics[2][]{%
    \GenericError{(gnuplot) \space\space\space\@spaces}{%
      Package graphicx or graphics not loaded%
    }{See the gnuplot documentation for explanation.%
    }{The gnuplot epslatex terminal needs graphicx.sty or graphics.sty.}%
    \renewcommand\includegraphics[2][]{}%
  }%
  \providecommand\rotatebox[2]{#2}%
  \@ifundefined{ifGPcolor}{%
    \newif\ifGPcolor
    \GPcolortrue
  }{}%
  \@ifundefined{ifGPblacktext}{%
    \newif\ifGPblacktext
    \GPblacktextfalse
  }{}%
  % define a \g@addto@macro without @ in the name:
  \let\gplgaddtomacro\g@addto@macro
  % define empty templates for all commands taking text:
  \gdef\gplbacktext{}%
  \gdef\gplfronttext{}%
  \makeatother
  \ifGPblacktext
    % no textcolor at all
    \def\colorrgb#1{}%
    \def\colorgray#1{}%
  \else
    % gray or color?
    \ifGPcolor
      \def\colorrgb#1{\color[rgb]{#1}}%
      \def\colorgray#1{\color[gray]{#1}}%
      \expandafter\def\csname LTw\endcsname{\color{white}}%
      \expandafter\def\csname LTb\endcsname{\color{black}}%
      \expandafter\def\csname LTa\endcsname{\color{black}}%
      \expandafter\def\csname LT0\endcsname{\color[rgb]{1,0,0}}%
      \expandafter\def\csname LT1\endcsname{\color[rgb]{0,1,0}}%
      \expandafter\def\csname LT2\endcsname{\color[rgb]{0,0,1}}%
      \expandafter\def\csname LT3\endcsname{\color[rgb]{1,0,1}}%
      \expandafter\def\csname LT4\endcsname{\color[rgb]{0,1,1}}%
      \expandafter\def\csname LT5\endcsname{\color[rgb]{1,1,0}}%
      \expandafter\def\csname LT6\endcsname{\color[rgb]{0,0,0}}%
      \expandafter\def\csname LT7\endcsname{\color[rgb]{1,0.3,0}}%
      \expandafter\def\csname LT8\endcsname{\color[rgb]{0.5,0.5,0.5}}%
    \else
      % gray
      \def\colorrgb#1{\color{black}}%
      \def\colorgray#1{\color[gray]{#1}}%
      \expandafter\def\csname LTw\endcsname{\color{white}}%
      \expandafter\def\csname LTb\endcsname{\color{black}}%
      \expandafter\def\csname LTa\endcsname{\color{black}}%
      \expandafter\def\csname LT0\endcsname{\color{black}}%
      \expandafter\def\csname LT1\endcsname{\color{black}}%
      \expandafter\def\csname LT2\endcsname{\color{black}}%
      \expandafter\def\csname LT3\endcsname{\color{black}}%
      \expandafter\def\csname LT4\endcsname{\color{black}}%
      \expandafter\def\csname LT5\endcsname{\color{black}}%
      \expandafter\def\csname LT6\endcsname{\color{black}}%
      \expandafter\def\csname LT7\endcsname{\color{black}}%
      \expandafter\def\csname LT8\endcsname{\color{black}}%
    \fi
  \fi
  \setlength{\unitlength}{0.0500bp}%
  \begin{picture}(6480.00,3600.00)%
    \gplgaddtomacro\gplbacktext{%
      \colorrgb{0.00,0.00,0.00}%
      \put(946,767){\makebox(0,0)[r]{\strut{}0.85}}%
      \colorrgb{0.00,0.00,0.00}%
      \put(946,1409){\makebox(0,0)[r]{\strut{}0.90}}%
      \colorrgb{0.00,0.00,0.00}%
      \put(946,2051){\makebox(0,0)[r]{\strut{}0.95}}%
      \colorrgb{0.00,0.00,0.00}%
      \put(946,2693){\makebox(0,0)[r]{\strut{}1.00}}%
      \colorrgb{0.00,0.00,0.00}%
      \put(946,3335){\makebox(0,0)[r]{\strut{}1.05}}%
      \colorrgb{0.00,0.00,0.00}%
      \put(1141,484){\makebox(0,0){\strut{}1.00}}%
      \colorrgb{0.00,0.00,0.00}%
      \put(2377,484){\makebox(0,0){\strut{}1.05}}%
      \colorrgb{0.00,0.00,0.00}%
      \put(3612,484){\makebox(0,0){\strut{}1.10}}%
      \colorrgb{0.00,0.00,0.00}%
      \put(4848,484){\makebox(0,0){\strut{}1.15}}%
      \colorrgb{0.00,0.00,0.00}%
      \put(6083,484){\makebox(0,0){\strut{}1.20}}%
      \csname LTb\endcsname%
      \put(176,2051){\rotatebox{-270}{\makebox(0,0){\strut{}$(a_{110} / a_{001})\ /\ (b_1 / b_2)$}}}%
      \put(3612,154){\makebox(0,0){\strut{}$r_{ij}$}}%
    }%
    \gplgaddtomacro\gplfronttext{%
      \csname LTb\endcsname%
      \put(2593,1754){\makebox(0,0)[r]{\strut{}$\kappa^{-1} = \num{0.1}$}}%
      \csname LTb\endcsname%
      \put(2593,1490){\makebox(0,0)[r]{\strut{}$\kappa^{-1} = \num{0.2}$}}%
      \csname LTb\endcsname%
      \put(2593,1226){\makebox(0,0)[r]{\strut{}$\kappa^{-1} = \num{0.3}$}}%
      \csname LTb\endcsname%
      \put(2593,962){\makebox(0,0)[r]{\strut{}$\kappa^{-1} = \num{1.0}$}}%
    }%
    \gplbacktext
    \put(0,0){\includegraphics{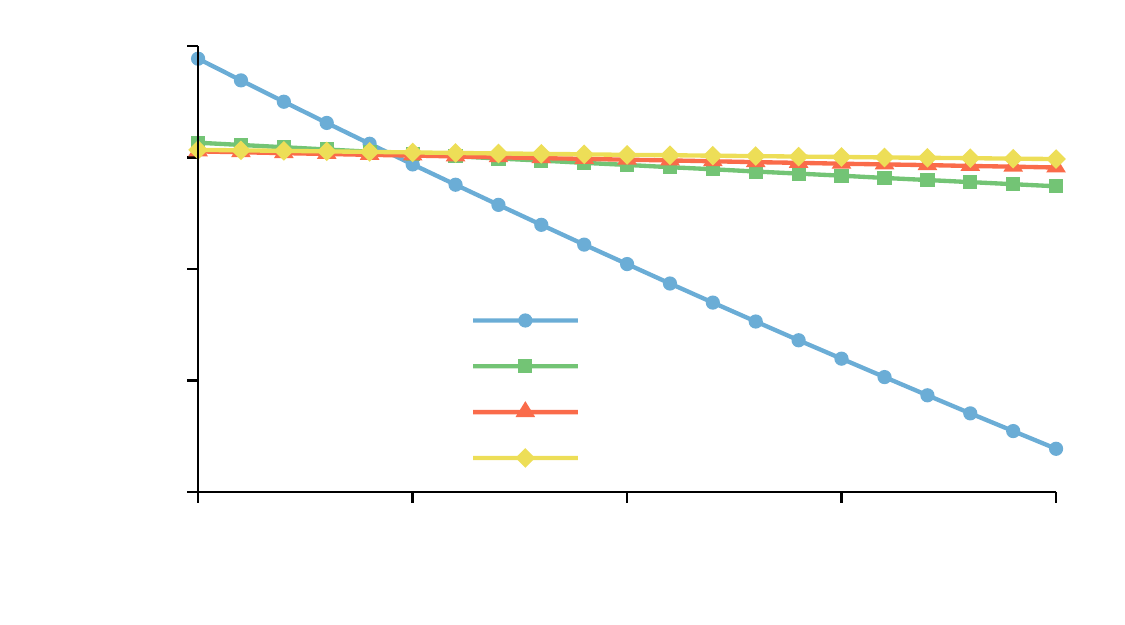}}%
    \gplfronttext
  \end{picture}%
\endgroup

%% file: DLVO_angular.tex
% GNUPLOT: LaTeX picture with Postscript
\begingroup
  \makeatletter
  \providecommand\color[2][]{%
    \GenericError{(gnuplot) \space\space\space\@spaces}{%
      Package color not loaded in conjunction with
      terminal option `colourtext'%
    }{See the gnuplot documentation for explanation.%
    }{Either use 'blacktext' in gnuplot or load the package
      color.sty in LaTeX.}%
    \renewcommand\color[2][]{}%
  }%
  \providecommand\includegraphics[2][]{%
    \GenericError{(gnuplot) \space\space\space\@spaces}{%
      Package graphicx or graphics not loaded%
    }{See the gnuplot documentation for explanation.%
    }{The gnuplot epslatex terminal needs graphicx.sty or graphics.sty.}%
    \renewcommand\includegraphics[2][]{}%
  }%
  \providecommand\rotatebox[2]{#2}%
  \@ifundefined{ifGPcolor}{%
    \newif\ifGPcolor
    \GPcolortrue
  }{}%
  \@ifundefined{ifGPblacktext}{%
    \newif\ifGPblacktext
    \GPblacktextfalse
  }{}%
  % define a \g@addto@macro without @ in the name:
  \let\gplgaddtomacro\g@addto@macro
  % define empty templates for all commands taking text:
  \gdef\gplbacktext{}%
  \gdef\gplfronttext{}%
  \makeatother
  \ifGPblacktext
    % no textcolor at all
    \def\colorrgb#1{}%
    \def\colorgray#1{}%
  \else
    % gray or color?
    \ifGPcolor
      \def\colorrgb#1{\color[rgb]{#1}}%
      \def\colorgray#1{\color[gray]{#1}}%
      \expandafter\def\csname LTw\endcsname{\color{white}}%
      \expandafter\def\csname LTb\endcsname{\color{black}}%
      \expandafter\def\csname LTa\endcsname{\color{black}}%
      \expandafter\def\csname LT0\endcsname{\color[rgb]{1,0,0}}%
      \expandafter\def\csname LT1\endcsname{\color[rgb]{0,1,0}}%
      \expandafter\def\csname LT2\endcsname{\color[rgb]{0,0,1}}%
      \expandafter\def\csname LT3\endcsname{\color[rgb]{1,0,1}}%
      \expandafter\def\csname LT4\endcsname{\color[rgb]{0,1,1}}%
      \expandafter\def\csname LT5\endcsname{\color[rgb]{1,1,0}}%
      \expandafter\def\csname LT6\endcsname{\color[rgb]{0,0,0}}%
      \expandafter\def\csname LT7\endcsname{\color[rgb]{1,0.3,0}}%
      \expandafter\def\csname LT8\endcsname{\color[rgb]{0.5,0.5,0.5}}%
    \else
      % gray
      \def\colorrgb#1{\color{black}}%
      \def\colorgray#1{\color[gray]{#1}}%
      \expandafter\def\csname LTw\endcsname{\color{white}}%
      \expandafter\def\csname LTb\endcsname{\color{black}}%
      \expandafter\def\csname LTa\endcsname{\color{black}}%
      \expandafter\def\csname LT0\endcsname{\color{black}}%
      \expandafter\def\csname LT1\endcsname{\color{black}}%
      \expandafter\def\csname LT2\endcsname{\color{black}}%
      \expandafter\def\csname LT3\endcsname{\color{black}}%
      \expandafter\def\csname LT4\endcsname{\color{black}}%
      \expandafter\def\csname LT5\endcsname{\color{black}}%
      \expandafter\def\csname LT6\endcsname{\color{black}}%
      \expandafter\def\csname LT7\endcsname{\color{black}}%
      \expandafter\def\csname LT8\endcsname{\color{black}}%
    \fi
  \fi
  \setlength{\unitlength}{0.0500bp}%
  \begin{picture}(6480.00,3600.00)%
    \gplgaddtomacro\gplbacktext{%
      \colorrgb{0.00,0.00,0.00}%
      \put(814,1000){\makebox(0,0)[r]{\strut{}-0.9}}%
      \colorrgb{0.00,0.00,0.00}%
      \put(814,1701){\makebox(0,0)[r]{\strut{}-0.6}}%
      \colorrgb{0.00,0.00,0.00}%
      \put(814,2401){\makebox(0,0)[r]{\strut{}-0.3}}%
      \colorrgb{0.00,0.00,0.00}%
      \put(814,3102){\makebox(0,0)[r]{\strut{}0.0}}%
      \colorrgb{0.00,0.00,0.00}%
      \put(1009,484){\makebox(0,0){\strut{}0}}%
      \colorrgb{0.00,0.00,0.00}%
      \put(2278,484){\makebox(0,0){\strut{}$\frac{1}{8}\pi$}}%
      \colorrgb{0.00,0.00,0.00}%
      \put(3546,484){\makebox(0,0){\strut{}$\frac{1}{4}\pi$}}%
      \colorrgb{0.00,0.00,0.00}%
      \put(4815,484){\makebox(0,0){\strut{}$\frac{3}{8}\pi$}}%
      \colorrgb{0.00,0.00,0.00}%
      \put(6083,484){\makebox(0,0){\strut{}$\frac{1}{2}\pi$}}%
      \csname LTb\endcsname%
      \put(176,2051){\rotatebox{-270}{\makebox(0,0){\strut{}$u$}}}%
      \put(3546,154){\makebox(0,0){\strut{}$\vartheta$}}%
    }%
    \gplgaddtomacro\gplfronttext{%
      \csname LTb\endcsname%
      \put(2679,3169){\makebox(0,0)[r]{\strut{}B = 1, par}}%
      \csname LTb\endcsname%
      \put(2679,2883){\makebox(0,0)[r]{\strut{} DLVO, par}}%
      \csname LTb\endcsname%
      \put(2679,2597){\makebox(0,0)[r]{\strut{}B = 1, per}}%
      \csname LTb\endcsname%
      \put(2679,2311){\makebox(0,0)[r]{\strut{}DLVO, per}}%
    }%
    \gplbacktext
    \put(0,0){\includegraphics{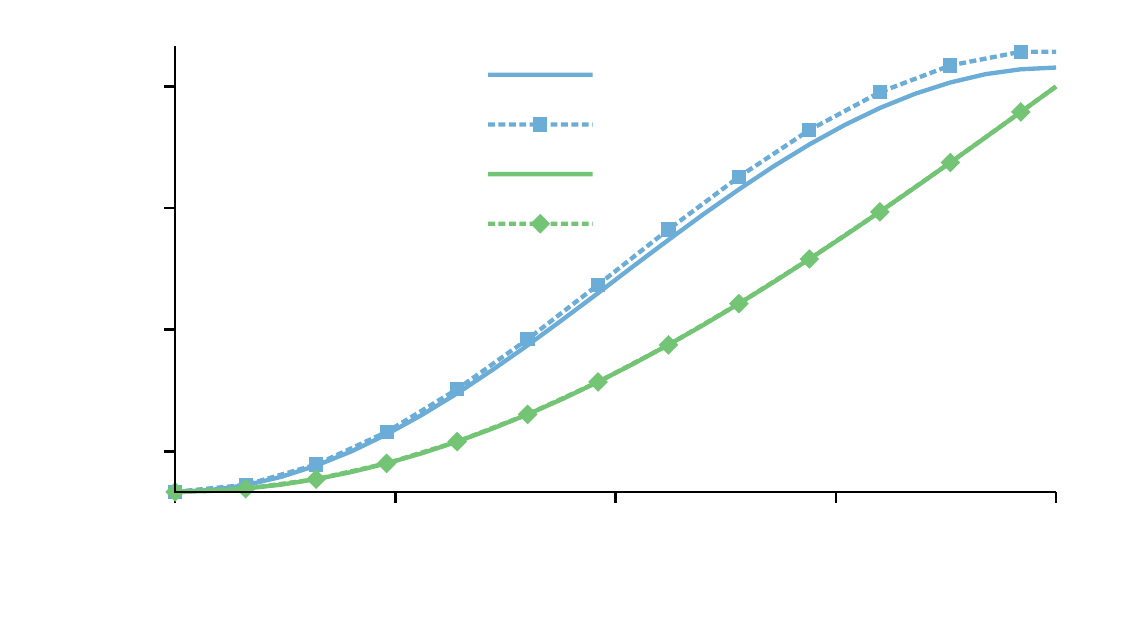}}%
    \gplfronttext
  \end{picture}%
\endgroup

%% file: fingerprint_B=5_HS_kappa=01.tex
% GNUPLOT: LaTeX picture with Postscript
\begingroup
  \makeatletter
  \providecommand\color[2][]{%
    \GenericError{(gnuplot) \space\space\space\@spaces}{%
      Package color not loaded in conjunction with
      terminal option `colourtext'%
    }{See the gnuplot documentation for explanation.%
    }{Either use 'blacktext' in gnuplot or load the package
      color.sty in LaTeX.}%
    \renewcommand\color[2][]{}%
  }%
  \providecommand\includegraphics[2][]{%
    \GenericError{(gnuplot) \space\space\space\@spaces}{%
      Package graphicx or graphics not loaded%
    }{See the gnuplot documentation for explanation.%
    }{The gnuplot epslatex terminal needs graphicx.sty or graphics.sty.}%
    \renewcommand\includegraphics[2][]{}%
  }%
  \providecommand\rotatebox[2]{#2}%
  \@ifundefined{ifGPcolor}{%
    \newif\ifGPcolor
    \GPcolortrue
  }{}%
  \@ifundefined{ifGPblacktext}{%
    \newif\ifGPblacktext
    \GPblacktextfalse
  }{}%
  % define a \g@addto@macro without @ in the name:
  \let\gplgaddtomacro\g@addto@macro
  % define empty templates for all commands taking text:
  \gdef\gplbacktext{}%
  \gdef\gplfronttext{}%
  \makeatother
  \ifGPblacktext
    % no textcolor at all
    \def\colorrgb#1{}%
    \def\colorgray#1{}%
  \else
    % gray or color?
    \ifGPcolor
      \def\colorrgb#1{\color[rgb]{#1}}%
      \def\colorgray#1{\color[gray]{#1}}%
      \expandafter\def\csname LTw\endcsname{\color{white}}%
      \expandafter\def\csname LTb\endcsname{\color{black}}%
      \expandafter\def\csname LTa\endcsname{\color{black}}%
      \expandafter\def\csname LT0\endcsname{\color[rgb]{1,0,0}}%
      \expandafter\def\csname LT1\endcsname{\color[rgb]{0,1,0}}%
      \expandafter\def\csname LT2\endcsname{\color[rgb]{0,0,1}}%
      \expandafter\def\csname LT3\endcsname{\color[rgb]{1,0,1}}%
      \expandafter\def\csname LT4\endcsname{\color[rgb]{0,1,1}}%
      \expandafter\def\csname LT5\endcsname{\color[rgb]{1,1,0}}%
      \expandafter\def\csname LT6\endcsname{\color[rgb]{0,0,0}}%
      \expandafter\def\csname LT7\endcsname{\color[rgb]{1,0.3,0}}%
      \expandafter\def\csname LT8\endcsname{\color[rgb]{0.5,0.5,0.5}}%
    \else
      % gray
      \def\colorrgb#1{\color{black}}%
      \def\colorgray#1{\color[gray]{#1}}%
      \expandafter\def\csname LTw\endcsname{\color{white}}%
      \expandafter\def\csname LTb\endcsname{\color{black}}%
      \expandafter\def\csname LTa\endcsname{\color{black}}%
      \expandafter\def\csname LT0\endcsname{\color{black}}%
      \expandafter\def\csname LT1\endcsname{\color{black}}%
      \expandafter\def\csname LT2\endcsname{\color{black}}%
      \expandafter\def\csname LT3\endcsname{\color{black}}%
      \expandafter\def\csname LT4\endcsname{\color{black}}%
      \expandafter\def\csname LT5\endcsname{\color{black}}%
      \expandafter\def\csname LT6\endcsname{\color{black}}%
      \expandafter\def\csname LT7\endcsname{\color{black}}%
      \expandafter\def\csname LT8\endcsname{\color{black}}%
    \fi
  \fi
    \setlength{\unitlength}{0.0500bp}%
    \ifx\gptboxheight\undefined%
      \newlength{\gptboxheight}%
      \newlength{\gptboxwidth}%
      \newsavebox{\gptboxtext}%
    \fi%
    \setlength{\fboxrule}{0.5pt}%
    \setlength{\fboxsep}{1pt}%
\begin{picture}(6480.00,3600.00)%
    \gplgaddtomacro\gplbacktext{%
      \colorrgb{0.00,0.00,0.00}%
      \put(616,835){\makebox(0,0)[r]{\strut{}4}}%
      \colorrgb{0.00,0.00,0.00}%
      \put(616,1105){\makebox(0,0)[r]{\strut{}5}}%
      \colorrgb{0.00,0.00,0.00}%
      \put(616,1375){\makebox(0,0)[r]{\strut{}6}}%
      \colorrgb{0.00,0.00,0.00}%
      \put(616,1646){\makebox(0,0)[r]{\strut{}7}}%
      \colorrgb{0.00,0.00,0.00}%
      \put(616,1916){\makebox(0,0)[r]{\strut{}8}}%
      \colorrgb{0.00,0.00,0.00}%
      \put(616,2186){\makebox(0,0)[r]{\strut{}9}}%
      \colorrgb{0.00,0.00,0.00}%
      \put(616,2456){\makebox(0,0)[r]{\strut{}10}}%
      \colorrgb{0.00,0.00,0.00}%
      \put(616,2727){\makebox(0,0)[r]{\strut{}11}}%
      \colorrgb{0.00,0.00,0.00}%
      \put(616,2997){\makebox(0,0)[r]{\strut{}12}}%
      \colorrgb{0.00,0.00,0.00}%
      \put(616,3267){\makebox(0,0)[r]{\strut{}13}}%
      \colorrgb{0.00,0.00,0.00}%
      \put(1051,484){\makebox(0,0){\strut{}1.0}}%
      \colorrgb{0.00,0.00,0.00}%
      \put(2009,484){\makebox(0,0){\strut{}1.2}}%
      \colorrgb{0.00,0.00,0.00}%
      \put(2968,484){\makebox(0,0){\strut{}1.4}}%
      \colorrgb{0.00,0.00,0.00}%
      \put(3926,484){\makebox(0,0){\strut{}1.6}}%
      \colorrgb{0.00,0.00,0.00}%
      \put(4885,484){\makebox(0,0){\strut{}1.8}}%
      \colorrgb{0.00,0.00,0.00}%
      \put(5843,484){\makebox(0,0){\strut{}2.0}}%
    }%
    \gplgaddtomacro\gplfronttext{%
      \csname LTb\endcsname%
      \put(176,2051){\rotatebox{-270}{\makebox(0,0){\strut{}$N$}}}%
      \put(3447,154){\makebox(0,0){\strut{}$|\vec{r}_i - \vec{r}_\mathrm{center}|$}}%
    }%
    \gplbacktext
    \put(0,0){\includegraphics{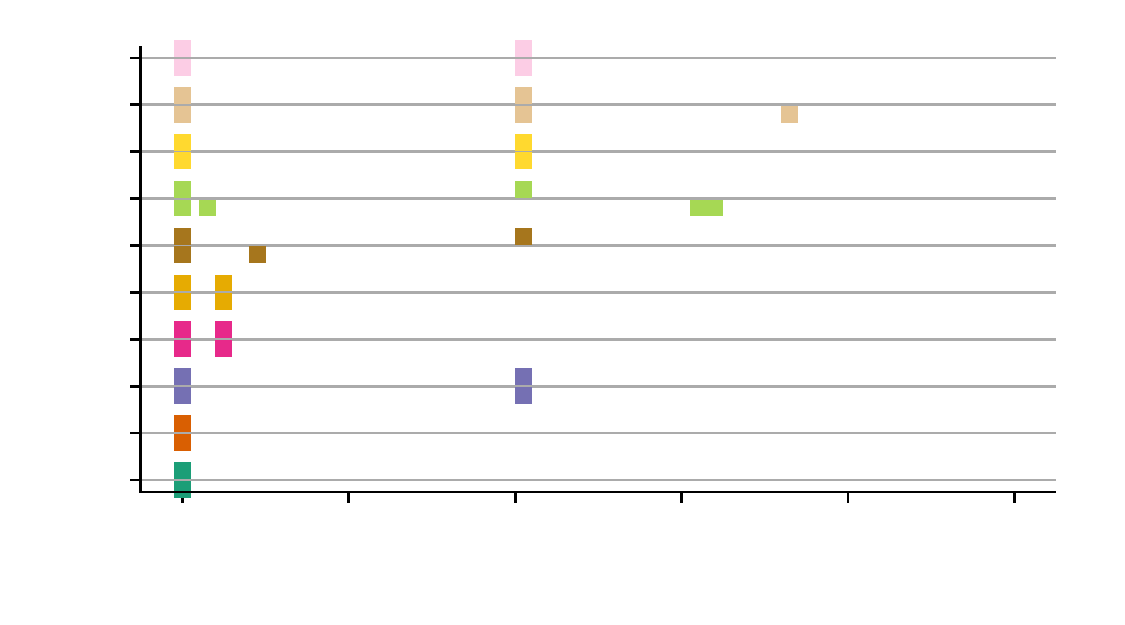}}%
    \gplfronttext
  \end{picture}%
\endgroup

%% file: particle_energy_1.tex
% GNUPLOT: LaTeX picture with Postscript
\begingroup
  \makeatletter
  \providecommand\color[2][]{%
    \GenericError{(gnuplot) \space\space\space\@spaces}{%
      Package color not loaded in conjunction with
      terminal option `colourtext'%
    }{See the gnuplot documentation for explanation.%
    }{Either use 'blacktext' in gnuplot or load the package
      color.sty in LaTeX.}%
    \renewcommand\color[2][]{}%
  }%
  \providecommand\includegraphics[2][]{%
    \GenericError{(gnuplot) \space\space\space\@spaces}{%
      Package graphicx or graphics not loaded%
    }{See the gnuplot documentation for explanation.%
    }{The gnuplot epslatex terminal needs graphicx.sty or graphics.sty.}%
    \renewcommand\includegraphics[2][]{}%
  }%
  \providecommand\rotatebox[2]{#2}%
  \@ifundefined{ifGPcolor}{%
    \newif\ifGPcolor
    \GPcolortrue
  }{}%
  \@ifundefined{ifGPblacktext}{%
    \newif\ifGPblacktext
    \GPblacktextfalse
  }{}%
  % define a \g@addto@macro without @ in the name:
  \let\gplgaddtomacro\g@addto@macro
  % define empty templates for all commands taking text:
  \gdef\gplbacktext{}%
  \gdef\gplfronttext{}%
  \makeatother
  \ifGPblacktext
    % no textcolor at all
    \def\colorrgb#1{}%
    \def\colorgray#1{}%
  \else
    % gray or color?
    \ifGPcolor
      \def\colorrgb#1{\color[rgb]{#1}}%
      \def\colorgray#1{\color[gray]{#1}}%
      \expandafter\def\csname LTw\endcsname{\color{white}}%
      \expandafter\def\csname LTb\endcsname{\color{black}}%
      \expandafter\def\csname LTa\endcsname{\color{black}}%
      \expandafter\def\csname LT0\endcsname{\color[rgb]{1,0,0}}%
      \expandafter\def\csname LT1\endcsname{\color[rgb]{0,1,0}}%
      \expandafter\def\csname LT2\endcsname{\color[rgb]{0,0,1}}%
      \expandafter\def\csname LT3\endcsname{\color[rgb]{1,0,1}}%
      \expandafter\def\csname LT4\endcsname{\color[rgb]{0,1,1}}%
      \expandafter\def\csname LT5\endcsname{\color[rgb]{1,1,0}}%
      \expandafter\def\csname LT6\endcsname{\color[rgb]{0,0,0}}%
      \expandafter\def\csname LT7\endcsname{\color[rgb]{1,0.3,0}}%
      \expandafter\def\csname LT8\endcsname{\color[rgb]{0.5,0.5,0.5}}%
    \else
      % gray
      \def\colorrgb#1{\color{black}}%
      \def\colorgray#1{\color[gray]{#1}}%
      \expandafter\def\csname LTw\endcsname{\color{white}}%
      \expandafter\def\csname LTb\endcsname{\color{black}}%
      \expandafter\def\csname LTa\endcsname{\color{black}}%
      \expandafter\def\csname LT0\endcsname{\color{black}}%
      \expandafter\def\csname LT1\endcsname{\color{black}}%
      \expandafter\def\csname LT2\endcsname{\color{black}}%
      \expandafter\def\csname LT3\endcsname{\color{black}}%
      \expandafter\def\csname LT4\endcsname{\color{black}}%
      \expandafter\def\csname LT5\endcsname{\color{black}}%
      \expandafter\def\csname LT6\endcsname{\color{black}}%
      \expandafter\def\csname LT7\endcsname{\color{black}}%
      \expandafter\def\csname LT8\endcsname{\color{black}}%
    \fi
  \fi
  \setlength{\unitlength}{0.0500bp}%
  \begin{picture}(6480.00,3600.00)%
    \gplgaddtomacro\gplbacktext{%
      \colorrgb{0.00,0.00,0.00}%
      \put(880,965){\makebox(0,0)[r]{\strut{}-2.1}}%
      \colorrgb{0.00,0.00,0.00}%
      \put(880,1557){\makebox(0,0)[r]{\strut{}-1.8}}%
      \colorrgb{0.00,0.00,0.00}%
      \put(880,2150){\makebox(0,0)[r]{\strut{}-1.5}}%
      \colorrgb{0.00,0.00,0.00}%
      \put(880,2742){\makebox(0,0)[r]{\strut{}-1.2}}%
      \colorrgb{0.00,0.00,0.00}%
      \put(880,3335){\makebox(0,0)[r]{\strut{}-0.9}}%
      \colorrgb{0.00,0.00,0.00}%
      \put(1075,484){\makebox(0,0){\strut{} 4}}%
      \colorrgb{0.00,0.00,0.00}%
      \put(1631,484){\makebox(0,0){\strut{} 5}}%
      \colorrgb{0.00,0.00,0.00}%
      \put(2188,484){\makebox(0,0){\strut{} 6}}%
      \colorrgb{0.00,0.00,0.00}%
      \put(2744,484){\makebox(0,0){\strut{} 7}}%
      \colorrgb{0.00,0.00,0.00}%
      \put(3301,484){\makebox(0,0){\strut{} 8}}%
      \colorrgb{0.00,0.00,0.00}%
      \put(3857,484){\makebox(0,0){\strut{} 9}}%
      \colorrgb{0.00,0.00,0.00}%
      \put(4414,484){\makebox(0,0){\strut{}10}}%
      \colorrgb{0.00,0.00,0.00}%
      \put(4970,484){\makebox(0,0){\strut{}11}}%
      \colorrgb{0.00,0.00,0.00}%
      \put(5527,484){\makebox(0,0){\strut{}12}}%
      \colorrgb{0.00,0.00,0.00}%
      \put(6083,484){\makebox(0,0){\strut{}13}}%
      \csname LTb\endcsname%
      \put(176,2051){\rotatebox{-270}{\makebox(0,0){\strut{}$u_\mathrm{min} / (N - 1)$}}}%
      \put(3579,154){\makebox(0,0){\strut{}$N$}}%
    }%
    \gplgaddtomacro\gplfronttext{%
    }%
    \gplbacktext
    \put(0,0){\includegraphics{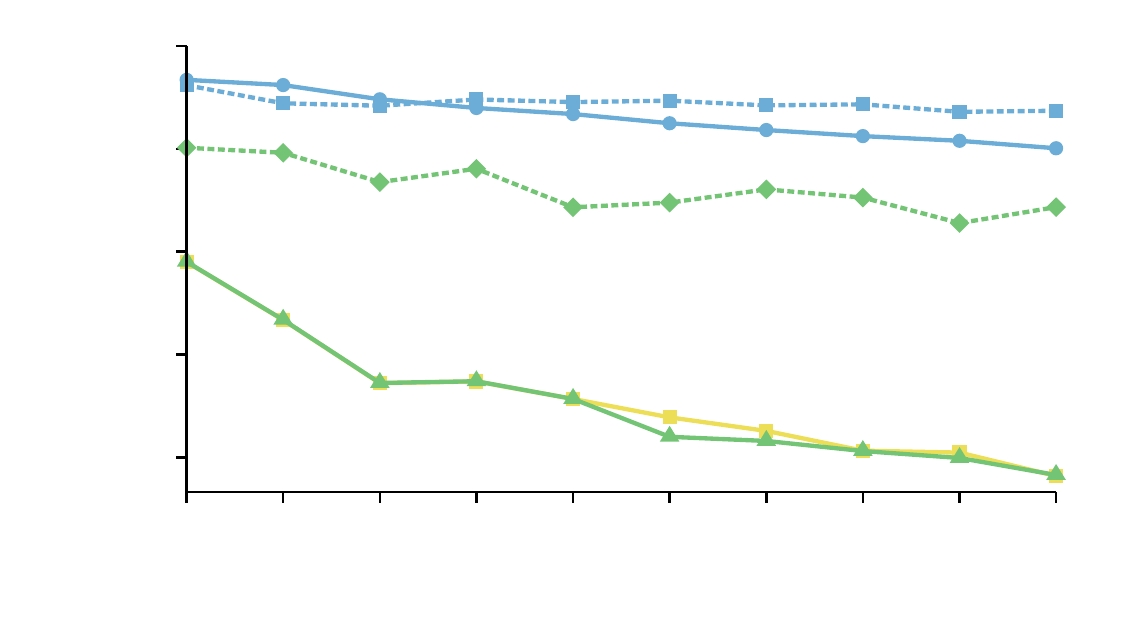}}%
    \gplfronttext
  \end{picture}%
\endgroup

%% file: particle_dipole_1.tex
% GNUPLOT: LaTeX picture with Postscript
\begingroup
  \makeatletter
  \providecommand\color[2][]{%
    \GenericError{(gnuplot) \space\space\space\@spaces}{%
      Package color not loaded in conjunction with
      terminal option `colourtext'%
    }{See the gnuplot documentation for explanation.%
    }{Either use 'blacktext' in gnuplot or load the package
      color.sty in LaTeX.}%
    \renewcommand\color[2][]{}%
  }%
  \providecommand\includegraphics[2][]{%
    \GenericError{(gnuplot) \space\space\space\@spaces}{%
      Package graphicx or graphics not loaded%
    }{See the gnuplot documentation for explanation.%
    }{The gnuplot epslatex terminal needs graphicx.sty or graphics.sty.}%
    \renewcommand\includegraphics[2][]{}%
  }%
  \providecommand\rotatebox[2]{#2}%
  \@ifundefined{ifGPcolor}{%
    \newif\ifGPcolor
    \GPcolortrue
  }{}%
  \@ifundefined{ifGPblacktext}{%
    \newif\ifGPblacktext
    \GPblacktextfalse
  }{}%
  % define a \g@addto@macro without @ in the name:
  \let\gplgaddtomacro\g@addto@macro
  % define empty templates for all commands taking text:
  \gdef\gplbacktext{}%
  \gdef\gplfronttext{}%
  \makeatother
  \ifGPblacktext
    % no textcolor at all
    \def\colorrgb#1{}%
    \def\colorgray#1{}%
  \else
    % gray or color?
    \ifGPcolor
      \def\colorrgb#1{\color[rgb]{#1}}%
      \def\colorgray#1{\color[gray]{#1}}%
      \expandafter\def\csname LTw\endcsname{\color{white}}%
      \expandafter\def\csname LTb\endcsname{\color{black}}%
      \expandafter\def\csname LTa\endcsname{\color{black}}%
      \expandafter\def\csname LT0\endcsname{\color[rgb]{1,0,0}}%
      \expandafter\def\csname LT1\endcsname{\color[rgb]{0,1,0}}%
      \expandafter\def\csname LT2\endcsname{\color[rgb]{0,0,1}}%
      \expandafter\def\csname LT3\endcsname{\color[rgb]{1,0,1}}%
      \expandafter\def\csname LT4\endcsname{\color[rgb]{0,1,1}}%
      \expandafter\def\csname LT5\endcsname{\color[rgb]{1,1,0}}%
      \expandafter\def\csname LT6\endcsname{\color[rgb]{0,0,0}}%
      \expandafter\def\csname LT7\endcsname{\color[rgb]{1,0.3,0}}%
      \expandafter\def\csname LT8\endcsname{\color[rgb]{0.5,0.5,0.5}}%
    \else
      % gray
      \def\colorrgb#1{\color{black}}%
      \def\colorgray#1{\color[gray]{#1}}%
      \expandafter\def\csname LTw\endcsname{\color{white}}%
      \expandafter\def\csname LTb\endcsname{\color{black}}%
      \expandafter\def\csname LTa\endcsname{\color{black}}%
      \expandafter\def\csname LT0\endcsname{\color{black}}%
      \expandafter\def\csname LT1\endcsname{\color{black}}%
      \expandafter\def\csname LT2\endcsname{\color{black}}%
      \expandafter\def\csname LT3\endcsname{\color{black}}%
      \expandafter\def\csname LT4\endcsname{\color{black}}%
      \expandafter\def\csname LT5\endcsname{\color{black}}%
      \expandafter\def\csname LT6\endcsname{\color{black}}%
      \expandafter\def\csname LT7\endcsname{\color{black}}%
      \expandafter\def\csname LT8\endcsname{\color{black}}%
    \fi
  \fi
  \setlength{\unitlength}{0.0500bp}%
  \begin{picture}(6480.00,3600.00)%
    \gplgaddtomacro\gplbacktext{%
      \colorrgb{0.00,0.00,0.00}%
      \put(814,767){\makebox(0,0)[r]{\strut{}0.0}}%
      \colorrgb{0.00,0.00,0.00}%
      \put(814,1281){\makebox(0,0)[r]{\strut{}0.2}}%
      \colorrgb{0.00,0.00,0.00}%
      \put(814,1794){\makebox(0,0)[r]{\strut{}0.4}}%
      \colorrgb{0.00,0.00,0.00}%
      \put(814,2308){\makebox(0,0)[r]{\strut{}0.6}}%
      \colorrgb{0.00,0.00,0.00}%
      \put(814,2821){\makebox(0,0)[r]{\strut{}0.8}}%
      \colorrgb{0.00,0.00,0.00}%
      \put(814,3335){\makebox(0,0)[r]{\strut{}1.0}}%
      \colorrgb{0.00,0.00,0.00}%
      \put(1009,484){\makebox(0,0){\strut{} 4}}%
      \colorrgb{0.00,0.00,0.00}%
      \put(1573,484){\makebox(0,0){\strut{} 5}}%
      \colorrgb{0.00,0.00,0.00}%
      \put(2137,484){\makebox(0,0){\strut{} 6}}%
      \colorrgb{0.00,0.00,0.00}%
      \put(2700,484){\makebox(0,0){\strut{} 7}}%
      \colorrgb{0.00,0.00,0.00}%
      \put(3264,484){\makebox(0,0){\strut{} 8}}%
      \colorrgb{0.00,0.00,0.00}%
      \put(3828,484){\makebox(0,0){\strut{} 9}}%
      \colorrgb{0.00,0.00,0.00}%
      \put(4392,484){\makebox(0,0){\strut{}10}}%
      \colorrgb{0.00,0.00,0.00}%
      \put(4955,484){\makebox(0,0){\strut{}11}}%
      \colorrgb{0.00,0.00,0.00}%
      \put(5519,484){\makebox(0,0){\strut{}12}}%
      \colorrgb{0.00,0.00,0.00}%
      \put(6083,484){\makebox(0,0){\strut{}13}}%
      \csname LTb\endcsname%
      \put(176,2051){\rotatebox{-270}{\makebox(0,0){\strut{}$|\mu| / N$}}}%
      \put(3546,154){\makebox(0,0){\strut{}$N$}}%
    }%
    \gplgaddtomacro\gplfronttext{%
    }%
    \gplbacktext
    \put(0,0){\includegraphics{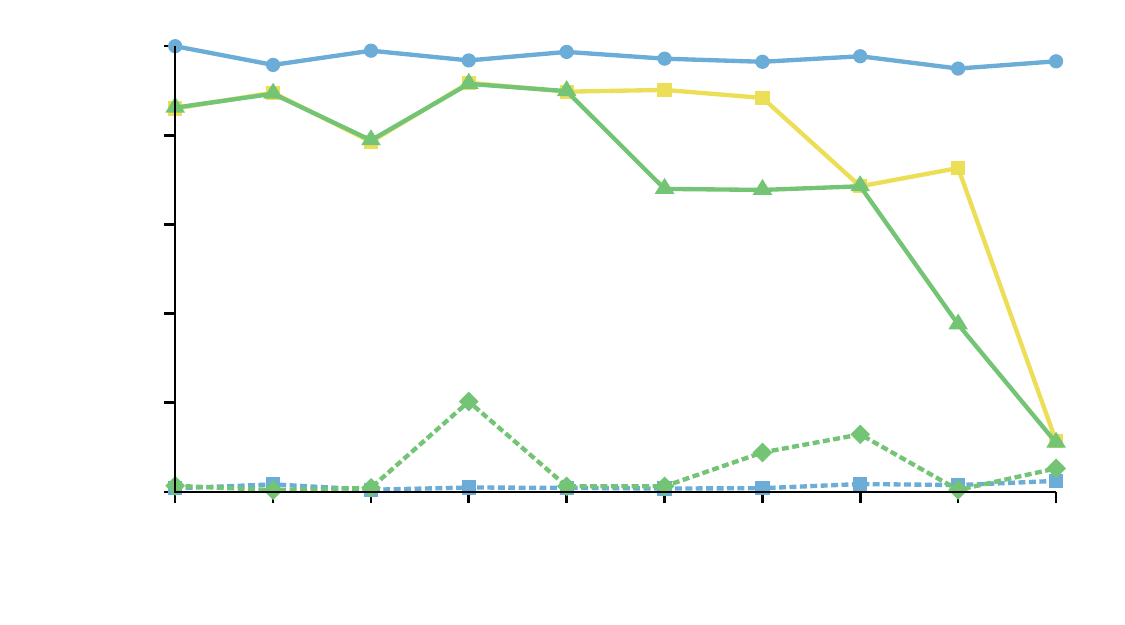}}%
    \gplfronttext
  \end{picture}%
\endgroup

%% file: particle_energy_2.tex
% GNUPLOT: LaTeX picture with Postscript
\begingroup
  \makeatletter
  \providecommand\color[2][]{%
    \GenericError{(gnuplot) \space\space\space\@spaces}{%
      Package color not loaded in conjunction with
      terminal option `colourtext'%
    }{See the gnuplot documentation for explanation.%
    }{Either use 'blacktext' in gnuplot or load the package
      color.sty in LaTeX.}%
    \renewcommand\color[2][]{}%
  }%
  \providecommand\includegraphics[2][]{%
    \GenericError{(gnuplot) \space\space\space\@spaces}{%
      Package graphicx or graphics not loaded%
    }{See the gnuplot documentation for explanation.%
    }{The gnuplot epslatex terminal needs graphicx.sty or graphics.sty.}%
    \renewcommand\includegraphics[2][]{}%
  }%
  \providecommand\rotatebox[2]{#2}%
  \@ifundefined{ifGPcolor}{%
    \newif\ifGPcolor
    \GPcolortrue
  }{}%
  \@ifundefined{ifGPblacktext}{%
    \newif\ifGPblacktext
    \GPblacktextfalse
  }{}%
  % define a \g@addto@macro without @ in the name:
  \let\gplgaddtomacro\g@addto@macro
  % define empty templates for all commands taking text:
  \gdef\gplbacktext{}%
  \gdef\gplfronttext{}%
  \makeatother
  \ifGPblacktext
    % no textcolor at all
    \def\colorrgb#1{}%
    \def\colorgray#1{}%
  \else
    % gray or color?
    \ifGPcolor
      \def\colorrgb#1{\color[rgb]{#1}}%
      \def\colorgray#1{\color[gray]{#1}}%
      \expandafter\def\csname LTw\endcsname{\color{white}}%
      \expandafter\def\csname LTb\endcsname{\color{black}}%
      \expandafter\def\csname LTa\endcsname{\color{black}}%
      \expandafter\def\csname LT0\endcsname{\color[rgb]{1,0,0}}%
      \expandafter\def\csname LT1\endcsname{\color[rgb]{0,1,0}}%
      \expandafter\def\csname LT2\endcsname{\color[rgb]{0,0,1}}%
      \expandafter\def\csname LT3\endcsname{\color[rgb]{1,0,1}}%
      \expandafter\def\csname LT4\endcsname{\color[rgb]{0,1,1}}%
      \expandafter\def\csname LT5\endcsname{\color[rgb]{1,1,0}}%
      \expandafter\def\csname LT6\endcsname{\color[rgb]{0,0,0}}%
      \expandafter\def\csname LT7\endcsname{\color[rgb]{1,0.3,0}}%
      \expandafter\def\csname LT8\endcsname{\color[rgb]{0.5,0.5,0.5}}%
    \else
      % gray
      \def\colorrgb#1{\color{black}}%
      \def\colorgray#1{\color[gray]{#1}}%
      \expandafter\def\csname LTw\endcsname{\color{white}}%
      \expandafter\def\csname LTb\endcsname{\color{black}}%
      \expandafter\def\csname LTa\endcsname{\color{black}}%
      \expandafter\def\csname LT0\endcsname{\color{black}}%
      \expandafter\def\csname LT1\endcsname{\color{black}}%
      \expandafter\def\csname LT2\endcsname{\color{black}}%
      \expandafter\def\csname LT3\endcsname{\color{black}}%
      \expandafter\def\csname LT4\endcsname{\color{black}}%
      \expandafter\def\csname LT5\endcsname{\color{black}}%
      \expandafter\def\csname LT6\endcsname{\color{black}}%
      \expandafter\def\csname LT7\endcsname{\color{black}}%
      \expandafter\def\csname LT8\endcsname{\color{black}}%
    \fi
  \fi
  \setlength{\unitlength}{0.0500bp}%
  \begin{picture}(6480.00,3600.00)%
    \gplgaddtomacro\gplbacktext{%
      \colorrgb{0.00,0.00,0.00}%
      \put(880,1134){\makebox(0,0)[r]{\strut{}-1.2}}%
      \colorrgb{0.00,0.00,0.00}%
      \put(880,1868){\makebox(0,0)[r]{\strut{}-1.1}}%
      \colorrgb{0.00,0.00,0.00}%
      \put(880,2601){\makebox(0,0)[r]{\strut{}-1.0}}%
      \colorrgb{0.00,0.00,0.00}%
      \put(880,3335){\makebox(0,0)[r]{\strut{}-0.9}}%
      \colorrgb{0.00,0.00,0.00}%
      \put(1075,484){\makebox(0,0){\strut{} 4}}%
      \colorrgb{0.00,0.00,0.00}%
      \put(1631,484){\makebox(0,0){\strut{} 5}}%
      \colorrgb{0.00,0.00,0.00}%
      \put(2188,484){\makebox(0,0){\strut{} 6}}%
      \colorrgb{0.00,0.00,0.00}%
      \put(2744,484){\makebox(0,0){\strut{} 7}}%
      \colorrgb{0.00,0.00,0.00}%
      \put(3301,484){\makebox(0,0){\strut{} 8}}%
      \colorrgb{0.00,0.00,0.00}%
      \put(3857,484){\makebox(0,0){\strut{} 9}}%
      \colorrgb{0.00,0.00,0.00}%
      \put(4414,484){\makebox(0,0){\strut{}10}}%
      \colorrgb{0.00,0.00,0.00}%
      \put(4970,484){\makebox(0,0){\strut{}11}}%
      \colorrgb{0.00,0.00,0.00}%
      \put(5527,484){\makebox(0,0){\strut{}12}}%
      \colorrgb{0.00,0.00,0.00}%
      \put(6083,484){\makebox(0,0){\strut{}13}}%
      \csname LTb\endcsname%
      \put(176,2051){\rotatebox{-270}{\makebox(0,0){\strut{}$u_\mathrm{min} / (N - 1)$}}}%
      \put(3579,154){\makebox(0,0){\strut{}$N$}}%
    }%
    \gplgaddtomacro\gplfronttext{%
    }%
    \gplbacktext
    \put(0,0){\includegraphics{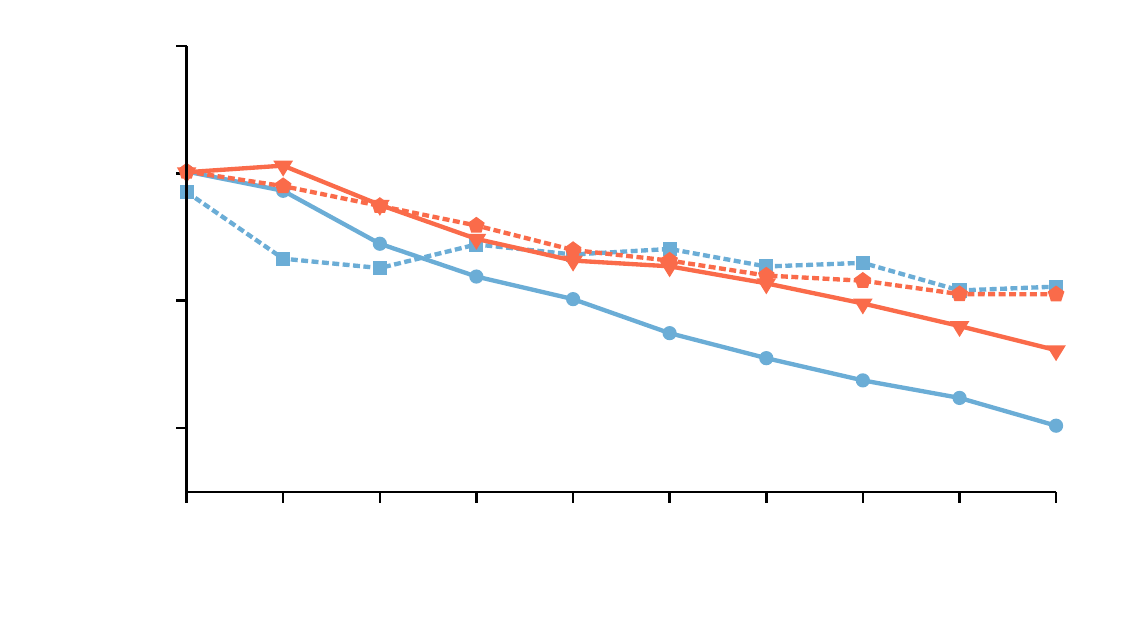}}%
    \gplfronttext
  \end{picture}%
\endgroup

%% file: particle_dipole_2.tex
% GNUPLOT: LaTeX picture with Postscript
\begingroup
  \makeatletter
  \providecommand\color[2][]{%
    \GenericError{(gnuplot) \space\space\space\@spaces}{%
      Package color not loaded in conjunction with
      terminal option `colourtext'%
    }{See the gnuplot documentation for explanation.%
    }{Either use 'blacktext' in gnuplot or load the package
      color.sty in LaTeX.}%
    \renewcommand\color[2][]{}%
  }%
  \providecommand\includegraphics[2][]{%
    \GenericError{(gnuplot) \space\space\space\@spaces}{%
      Package graphicx or graphics not loaded%
    }{See the gnuplot documentation for explanation.%
    }{The gnuplot epslatex terminal needs graphicx.sty or graphics.sty.}%
    \renewcommand\includegraphics[2][]{}%
  }%
  \providecommand\rotatebox[2]{#2}%
  \@ifundefined{ifGPcolor}{%
    \newif\ifGPcolor
    \GPcolortrue
  }{}%
  \@ifundefined{ifGPblacktext}{%
    \newif\ifGPblacktext
    \GPblacktextfalse
  }{}%
  % define a \g@addto@macro without @ in the name:
  \let\gplgaddtomacro\g@addto@macro
  % define empty templates for all commands taking text:
  \gdef\gplbacktext{}%
  \gdef\gplfronttext{}%
  \makeatother
  \ifGPblacktext
    % no textcolor at all
    \def\colorrgb#1{}%
    \def\colorgray#1{}%
  \else
    % gray or color?
    \ifGPcolor
      \def\colorrgb#1{\color[rgb]{#1}}%
      \def\colorgray#1{\color[gray]{#1}}%
      \expandafter\def\csname LTw\endcsname{\color{white}}%
      \expandafter\def\csname LTb\endcsname{\color{black}}%
      \expandafter\def\csname LTa\endcsname{\color{black}}%
      \expandafter\def\csname LT0\endcsname{\color[rgb]{1,0,0}}%
      \expandafter\def\csname LT1\endcsname{\color[rgb]{0,1,0}}%
      \expandafter\def\csname LT2\endcsname{\color[rgb]{0,0,1}}%
      \expandafter\def\csname LT3\endcsname{\color[rgb]{1,0,1}}%
      \expandafter\def\csname LT4\endcsname{\color[rgb]{0,1,1}}%
      \expandafter\def\csname LT5\endcsname{\color[rgb]{1,1,0}}%
      \expandafter\def\csname LT6\endcsname{\color[rgb]{0,0,0}}%
      \expandafter\def\csname LT7\endcsname{\color[rgb]{1,0.3,0}}%
      \expandafter\def\csname LT8\endcsname{\color[rgb]{0.5,0.5,0.5}}%
    \else
      % gray
      \def\colorrgb#1{\color{black}}%
      \def\colorgray#1{\color[gray]{#1}}%
      \expandafter\def\csname LTw\endcsname{\color{white}}%
      \expandafter\def\csname LTb\endcsname{\color{black}}%
      \expandafter\def\csname LTa\endcsname{\color{black}}%
      \expandafter\def\csname LT0\endcsname{\color{black}}%
      \expandafter\def\csname LT1\endcsname{\color{black}}%
      \expandafter\def\csname LT2\endcsname{\color{black}}%
      \expandafter\def\csname LT3\endcsname{\color{black}}%
      \expandafter\def\csname LT4\endcsname{\color{black}}%
      \expandafter\def\csname LT5\endcsname{\color{black}}%
      \expandafter\def\csname LT6\endcsname{\color{black}}%
      \expandafter\def\csname LT7\endcsname{\color{black}}%
      \expandafter\def\csname LT8\endcsname{\color{black}}%
    \fi
  \fi
  \setlength{\unitlength}{0.0500bp}%
  \begin{picture}(6480.00,3600.00)%
    \gplgaddtomacro\gplbacktext{%
      \colorrgb{0.00,0.00,0.00}%
      \put(814,767){\makebox(0,0)[r]{\strut{}0.0}}%
      \colorrgb{0.00,0.00,0.00}%
      \put(814,1281){\makebox(0,0)[r]{\strut{}0.2}}%
      \colorrgb{0.00,0.00,0.00}%
      \put(814,1794){\makebox(0,0)[r]{\strut{}0.4}}%
      \colorrgb{0.00,0.00,0.00}%
      \put(814,2308){\makebox(0,0)[r]{\strut{}0.6}}%
      \colorrgb{0.00,0.00,0.00}%
      \put(814,2821){\makebox(0,0)[r]{\strut{}0.8}}%
      \colorrgb{0.00,0.00,0.00}%
      \put(814,3335){\makebox(0,0)[r]{\strut{}1.0}}%
      \colorrgb{0.00,0.00,0.00}%
      \put(1009,484){\makebox(0,0){\strut{} 4}}%
      \colorrgb{0.00,0.00,0.00}%
      \put(1573,484){\makebox(0,0){\strut{} 5}}%
      \colorrgb{0.00,0.00,0.00}%
      \put(2137,484){\makebox(0,0){\strut{} 6}}%
      \colorrgb{0.00,0.00,0.00}%
      \put(2700,484){\makebox(0,0){\strut{} 7}}%
      \colorrgb{0.00,0.00,0.00}%
      \put(3264,484){\makebox(0,0){\strut{} 8}}%
      \colorrgb{0.00,0.00,0.00}%
      \put(3828,484){\makebox(0,0){\strut{} 9}}%
      \colorrgb{0.00,0.00,0.00}%
      \put(4392,484){\makebox(0,0){\strut{}10}}%
      \colorrgb{0.00,0.00,0.00}%
      \put(4955,484){\makebox(0,0){\strut{}11}}%
      \colorrgb{0.00,0.00,0.00}%
      \put(5519,484){\makebox(0,0){\strut{}12}}%
      \colorrgb{0.00,0.00,0.00}%
      \put(6083,484){\makebox(0,0){\strut{}13}}%
      \csname LTb\endcsname%
      \put(176,2051){\rotatebox{-270}{\makebox(0,0){\strut{}$|\mu| / N$}}}%
      \put(3546,154){\makebox(0,0){\strut{}$N$}}%
    }%
    \gplgaddtomacro\gplfronttext{%
    }%
    \gplbacktext
    \put(0,0){\includegraphics{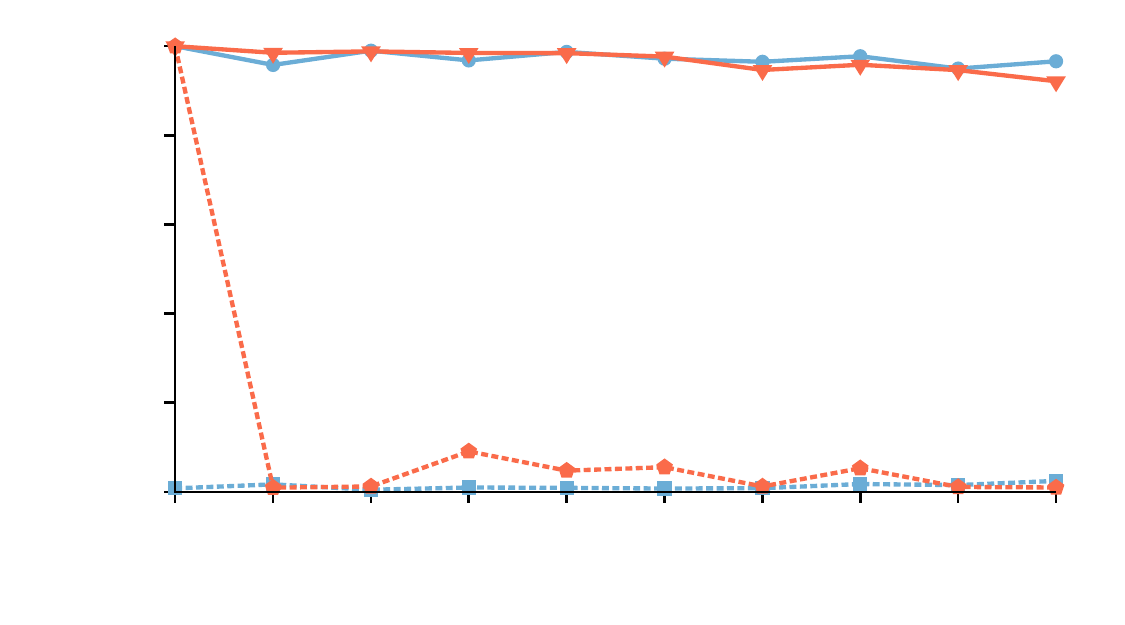}}%
    \gplfronttext
  \end{picture}%
\endgroup

%% file: particle_key.tex
% GNUPLOT: LaTeX picture with Postscript
\begingroup
  \makeatletter
  \providecommand\color[2][]{%
    \GenericError{(gnuplot) \space\space\space\@spaces}{%
      Package color not loaded in conjunction with
      terminal option `colourtext'%
    }{See the gnuplot documentation for explanation.%
    }{Either use 'blacktext' in gnuplot or load the package
      color.sty in LaTeX.}%
    \renewcommand\color[2][]{}%
  }%
  \providecommand\includegraphics[2][]{%
    \GenericError{(gnuplot) \space\space\space\@spaces}{%
      Package graphicx or graphics not loaded%
    }{See the gnuplot documentation for explanation.%
    }{The gnuplot epslatex terminal needs graphicx.sty or graphics.sty.}%
    \renewcommand\includegraphics[2][]{}%
  }%
  \providecommand\rotatebox[2]{#2}%
  \@ifundefined{ifGPcolor}{%
    \newif\ifGPcolor
    \GPcolortrue
  }{}%
  \@ifundefined{ifGPblacktext}{%
    \newif\ifGPblacktext
    \GPblacktextfalse
  }{}%
  % define a \g@addto@macro without @ in the name:
  \let\gplgaddtomacro\g@addto@macro
  % define empty templates for all commands taking text:
  \gdef\gplbacktext{}%
  \gdef\gplfronttext{}%
  \makeatother
  \ifGPblacktext
    % no textcolor at all
    \def\colorrgb#1{}%
    \def\colorgray#1{}%
  \else
    % gray or color?
    \ifGPcolor
      \def\colorrgb#1{\color[rgb]{#1}}%
      \def\colorgray#1{\color[gray]{#1}}%
      \expandafter\def\csname LTw\endcsname{\color{white}}%
      \expandafter\def\csname LTb\endcsname{\color{black}}%
      \expandafter\def\csname LTa\endcsname{\color{black}}%
      \expandafter\def\csname LT0\endcsname{\color[rgb]{1,0,0}}%
      \expandafter\def\csname LT1\endcsname{\color[rgb]{0,1,0}}%
      \expandafter\def\csname LT2\endcsname{\color[rgb]{0,0,1}}%
      \expandafter\def\csname LT3\endcsname{\color[rgb]{1,0,1}}%
      \expandafter\def\csname LT4\endcsname{\color[rgb]{0,1,1}}%
      \expandafter\def\csname LT5\endcsname{\color[rgb]{1,1,0}}%
      \expandafter\def\csname LT6\endcsname{\color[rgb]{0,0,0}}%
      \expandafter\def\csname LT7\endcsname{\color[rgb]{1,0.3,0}}%
      \expandafter\def\csname LT8\endcsname{\color[rgb]{0.5,0.5,0.5}}%
    \else
      % gray
      \def\colorrgb#1{\color{black}}%
      \def\colorgray#1{\color[gray]{#1}}%
      \expandafter\def\csname LTw\endcsname{\color{white}}%
      \expandafter\def\csname LTb\endcsname{\color{black}}%
      \expandafter\def\csname LTa\endcsname{\color{black}}%
      \expandafter\def\csname LT0\endcsname{\color{black}}%
      \expandafter\def\csname LT1\endcsname{\color{black}}%
      \expandafter\def\csname LT2\endcsname{\color{black}}%
      \expandafter\def\csname LT3\endcsname{\color{black}}%
      \expandafter\def\csname LT4\endcsname{\color{black}}%
      \expandafter\def\csname LT5\endcsname{\color{black}}%
      \expandafter\def\csname LT6\endcsname{\color{black}}%
      \expandafter\def\csname LT7\endcsname{\color{black}}%
      \expandafter\def\csname LT8\endcsname{\color{black}}%
    \fi
  \fi
  \setlength{\unitlength}{0.0500bp}%
  \begin{picture}(6480.00,1152.00)%
    \gplgaddtomacro\gplbacktext{%
    }%
    \gplgaddtomacro\gplfronttext{%
      \csname LTb\endcsname%
      \put(2537,1064){\makebox(0,0)[r]{\strut{}$B = 1$, $\kappa^{-1} = \num{0.1}$}}%
      \csname LTb\endcsname%
      \put(2537,778){\makebox(0,0)[r]{\strut{}$B = 1$, $\kappa^{-1} = \num{1.0}$}}%
      \csname LTb\endcsname%
      \put(2537,492){\makebox(0,0)[r]{\strut{}$B = 5$, $\kappa^{-1} = \num{0.1}$}}%
      \csname LTb\endcsname%
      \put(2537,206){\makebox(0,0)[r]{\strut{}$B = 5$, $\kappa^{-1} = \num{1.0}$}}%
      \csname LTb\endcsname%
      \put(5648,1064){\makebox(0,0)[r]{\strut{}DLVO, $\kappa^{-1} = \num{0.1}$}}%
      \csname LTb\endcsname%
      \put(5648,778){\makebox(0,0)[r]{\strut{}DLVO, $\kappa^{-1} = \num{1.0}$}}%
      \csname LTb\endcsname%
      \put(5648,492){\makebox(0,0)[r]{\strut{}HS, $\kappa^{-1} = \num{0.1}$}}%
    }%
    \gplbacktext
    \put(0,0){\includegraphics{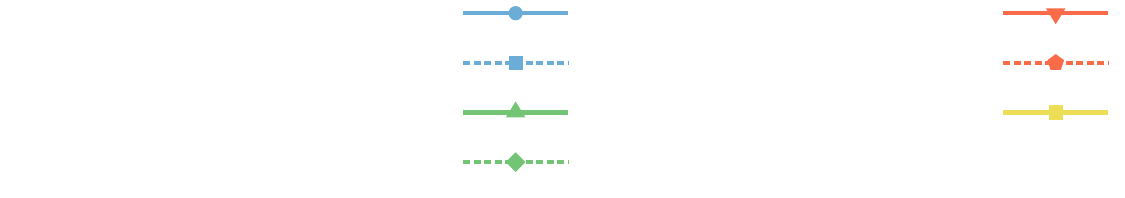}}%
    \gplfronttext
  \end{picture}%
\endgroup